\documentclass[twoside,twocolumn,9pt]{article}
\usepackage{extsizes}
\usepackage[super,sort&compress,comma]{natbib}
\usepackage[version=3]{mhchem}
\usepackage[left=1.5cm, right=1.5cm, top=1.785cm, bottom=2.0cm]{geometry}
\usepackage{balance}
\usepackage{times,mathptmx}
\usepackage{sectsty}
\usepackage{graphicx}
\usepackage{lastpage}
\usepackage[format=plain,justification=justified,singlelinecheck=false,font={stretch=1.125,small,sf},labelfont=bf,labelsep=space]{caption}
\usepackage{float}
\usepackage{fancyhdr}
\usepackage{fnpos}
\usepackage[english]{babel}
\usepackage[normalem]{ulem}
\addto{\captionsenglish}{%
  
}
\usepackage{array}
\usepackage{droidsans}
\usepackage{charter}
\usepackage[T1]{fontenc}
\usepackage[usenames,dvipsnames]{xcolor}
\usepackage{setspace}
\usepackage[compact]{titlesec}
\usepackage{hyperref}
\usepackage{amsmath,bm}
\usepackage{graphicx}
\usepackage{subfigure}
\usepackage{soul}
\usepackage{bbold}
\usepackage{color}

\usepackage{comment}

\newcommand{\beq}{\begin{equation}}
\newcommand{\eeq}{\end{equation}}
\newcommand{\vv}[1]{\left(\begin{array}{c}#1\end{array}\right)}

\usepackage{tikz}
\usetikzlibrary{arrows,shapes,positioning}
\usetikzlibrary{decorations.markings}
\tikzstyle arrowstyle=[scale=1]
\tikzstyle directed=[postaction={decorate,decoration={markings,
    mark=at position .65 with {\arrow[arrowstyle]{stealth}}}}]
\tikzstyle reverse directed=[postaction={decorate,decoration={markings,
    mark=at position .65 with {\arrowreversed[arrowstyle]{stealth};}}}]
\usetikzlibrary{calc}
\usetikzlibrary{patterns}

\usepackage{epstopdf}

\definecolor{cream}{RGB}{222,217,201}

\begin{document}

\pagestyle{fancy}
\thispagestyle{plain}
\fancypagestyle{plain}{

\fancyhead[C]{\includegraphics[width=18.5cm]{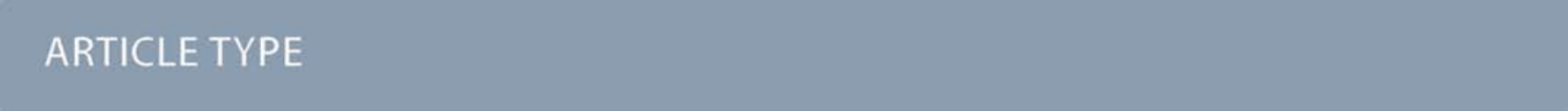}}
\fancyhead[L]{\hspace{0cm}\vspace{1.5cm}\includegraphics[height=30pt]{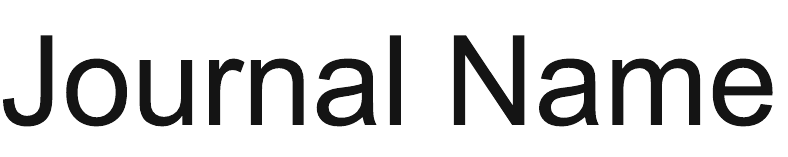}}
\fancyhead[R]{\hspace{0cm}\vspace{1.7cm}\includegraphics[height=55pt]{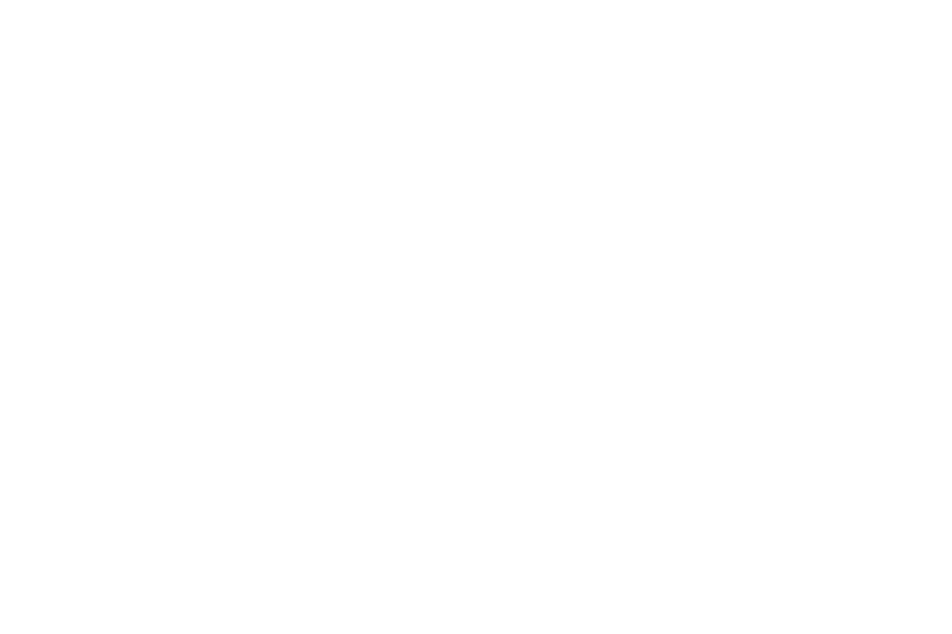}}
\renewcommand{\headrulewidth}{0pt}
}

\makeFNbottom
\makeatletter
\renewcommand\LARGE{\@setfontsize\LARGE{15pt}{17}}
\renewcommand\Large{\@setfontsize\Large{12pt}{14}}
\renewcommand\large{\@setfontsize\large{10pt}{12}}
\renewcommand\footnotesize{\@setfontsize\footnotesize{7pt}{10}}
\makeatother

\renewcommand{\thefootnote}{\fnsymbol{footnote}}
\renewcommand\footnoterule{\vspace*{1pt}%
\color{cream}\hrule width 3.5in height 0.4pt \color{black}\vspace*{5pt}}
\setcounter{secnumdepth}{5}

\makeatletter
\renewcommand\@biblabel[1]{#1}
\renewcommand\@makefntext[1]%
{\noindent\makebox[0pt][r]{\@thefnmark\,}#1}
\makeatother
\renewcommand{\figurename}{\small{Fig.}~}
\sectionfont{\sffamily\Large}
\subsectionfont{\normalsize}
\subsubsectionfont{\bf}
\setstretch{1.125} 
\setlength{\skip\footins}{0.8cm}
\setlength{\footnotesep}{0.25cm}
\setlength{\jot}{10pt}
\titlespacing*{\section}{0pt}{4pt}{4pt}
\titlespacing*{\subsection}{0pt}{15pt}{1pt}

\fancyfoot{}
\fancyfoot[LO,RE]{\vspace{-7.1pt}\includegraphics[height=9pt]{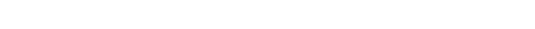}}
\fancyfoot[CO]{\vspace{-7.1pt}\hspace{13.2cm}\includegraphics{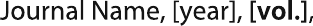}}
\fancyfoot[CE]{\vspace{-7.2pt}\hspace{-14.2cm}\includegraphics{RF}}
\fancyfoot[RO]{\footnotesize{\sffamily{1--\pageref{LastPage} ~\textbar  \hspace{2pt}\thepage}}}
\fancyfoot[LE]{\footnotesize{\sffamily{\thepage~\textbar\hspace{3.45cm} 1--\pageref{LastPage}}}}
\fancyhead{}
\renewcommand{\headrulewidth}{0pt}
\renewcommand{\footrulewidth}{0pt}
\setlength{\arrayrulewidth}{1pt}
\setlength{\columnsep}{6.5mm}
\setlength\bibsep{1pt}

\makeatletter
\newlength{\figrulesep}
\setlength{\figrulesep}{0.5\textfloatsep}

\newcommand{\topfigrule}{\vspace*{-1pt}%
\noindent{\color{cream}\rule[-\figrulesep]{\columnwidth}{1.5pt}} }

\newcommand{\botfigrule}{\vspace*{-2pt}%
\noindent{\color{cream}\rule[\figrulesep]{\columnwidth}{1.5pt}} }

\newcommand{\dblfigrule}{\vspace*{-1pt}%
\noindent{\color{cream}\rule[-\figrulesep]{\textwidth}{1.5pt}} }

\makeatother

\twocolumn[
  \begin{@twocolumnfalse}
\vspace{3cm}
\sffamily
\begin{tabular}{m{4.5cm} p{13.5cm} }


\includegraphics{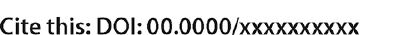} & \noindent\LARGE{\textbf{Directed force propagation in semiflexible networks}} \\
\vspace{0.3cm} & \vspace{0.3cm} \\

 & \noindent\large{Maximilian J.~Grill\textit{$^{a}$}, Jonathan Kernes\textit{$^{b}$}, Valentin M.~Slepukhin\textit{$^{b}$}, Wolfgang A.~Wall\textit{$^{a}$}, and Alex J.~Levine\textit{$^{b,c,d}$}} \\

\includegraphics{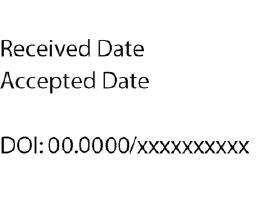} & \noindent\normalsize{We consider the propagation of tension along specific filament of a
semiflexible filament network in response to the application of a point force using a combination of numerical simulations and
analytic theory.  We find the distribution of force within the network is highly heterogenous, 
with a small number of fibers supporting a
significant fraction of the applied load over distances of multiple mesh sizes surrounding the point of force application.  We suggest that
these structures may be thought of as tensile force chains, whose structure we explore via simulation.  We develop 
self-consistent calculations of the point-force response function and introduce a transfer matrix approach 
to explore the decay of tension (into bending) energy
and the branching of tensile force chains in the network.}
\\

\end{tabular}

 \end{@twocolumnfalse} \vspace{0.6cm}

  ]

\renewcommand*\rmdefault{bch}\normalfont\upshape
\rmfamily
\section*{}
\vspace{-1cm}


\footnotetext{\textit{$^{a}$~Institute for Computational Mechanics, Technical University of Munich, 85748 Garching, Germany}}
\footnotetext{\textit{$^{b}$~Department of Physics \& Astronomy, University of California, Los Angeles. 90095  USA. }}
\footnotetext{\textit{$^{c}$~Department of Chemistry \& Biochemistry, University of California, Los Angeles. 90095  USA. }}
\footnotetext{\textit{$^{d}$~Department of Computational Medicine, University of California, Los Angeles. 90095  USA. }}




\section{Introduction}

The transmission of force through filamentous networks on the mesoscale is a complex problem that cannot be directly
addressed by appeals to continuum elasticity.  Understanding that, at sufficiently large length scales, a filamentous network
must act like a continuum elastic solid is
not helpful in predicting how that force is supported
at mesoscopic length scales in the network immediately surrounding the point of force application.
The complexities associated with this question are reasonably clear; they are
related both to the spatially heterogeneity and geometric complexity of filament interconnections and
the inherent nonlinearity of the filaments' force extension relations.
Biopolymer filaments are generically strongly strain hardening under tension, but quite soft under
compression, due to Euler buckling.

A well-known system that
combines both geometric heterogeneity of force-transmitting contacts and strong mechanical
nonlinearity is granular piles~\cite{liu1995force,Cates1998,Hurley2016}.
In a sand pile, one has a complex network of force-transmitting contacts that are elastically nonlinear -- the contacts
support large compressive loading and essentially no tensile loading.
These granular systems generically exhibit long and quite ramified force chains, spanning large numbers of intergrain contacts.
Although the bending stiffness of the filaments makes the analogy between filamentous
networks and granular media imperfect, one might expect similar force chains in such networks in the response to point forces.
This expectation seems to be supported by previous
simulations~\cite{head2005mechanical} of the point force response of Mikado networks and experiments on force-based interactions of
cells with the fibrous extracellular matrix (ECM)~\cite{Hart2013,rudnicki2013nonlinear}.
In the former, one sees the breakdown of the continuum elastic
response on scales much larger than the mesh size. In the latter, one observes intercellular force transmission over
long lengths, but only along particular, directed paths.
In the ECM experiments,
it is not clear if these long tracks of force transmission are a generic consequence of isotropic random networks,
or due, in part, to some filament anisotropy or heterogeneity either in the
form of stiffer filament bundles or spatial variations in the networks density.  Such heterogeneity may well be important.  In even
slightly anisotropic networks, previous work~\cite{Missel2010,foucard2015,majumdar2018mechanical}  has shown that
there exist long-ranged buckling scars forming in the network's response to even uniform shearing.
\begin{figure*}[htpb]
  \centering
  \begin{minipage}[t]{0.3\linewidth}%
    \begin{minipage}[t]{0.05\linewidth}%
      \vspace{-2cm}(A)
    \end{minipage}%
    \hfill
    \begin{minipage}[c]{0.92\linewidth}%
      \includegraphics[width=\linewidth]{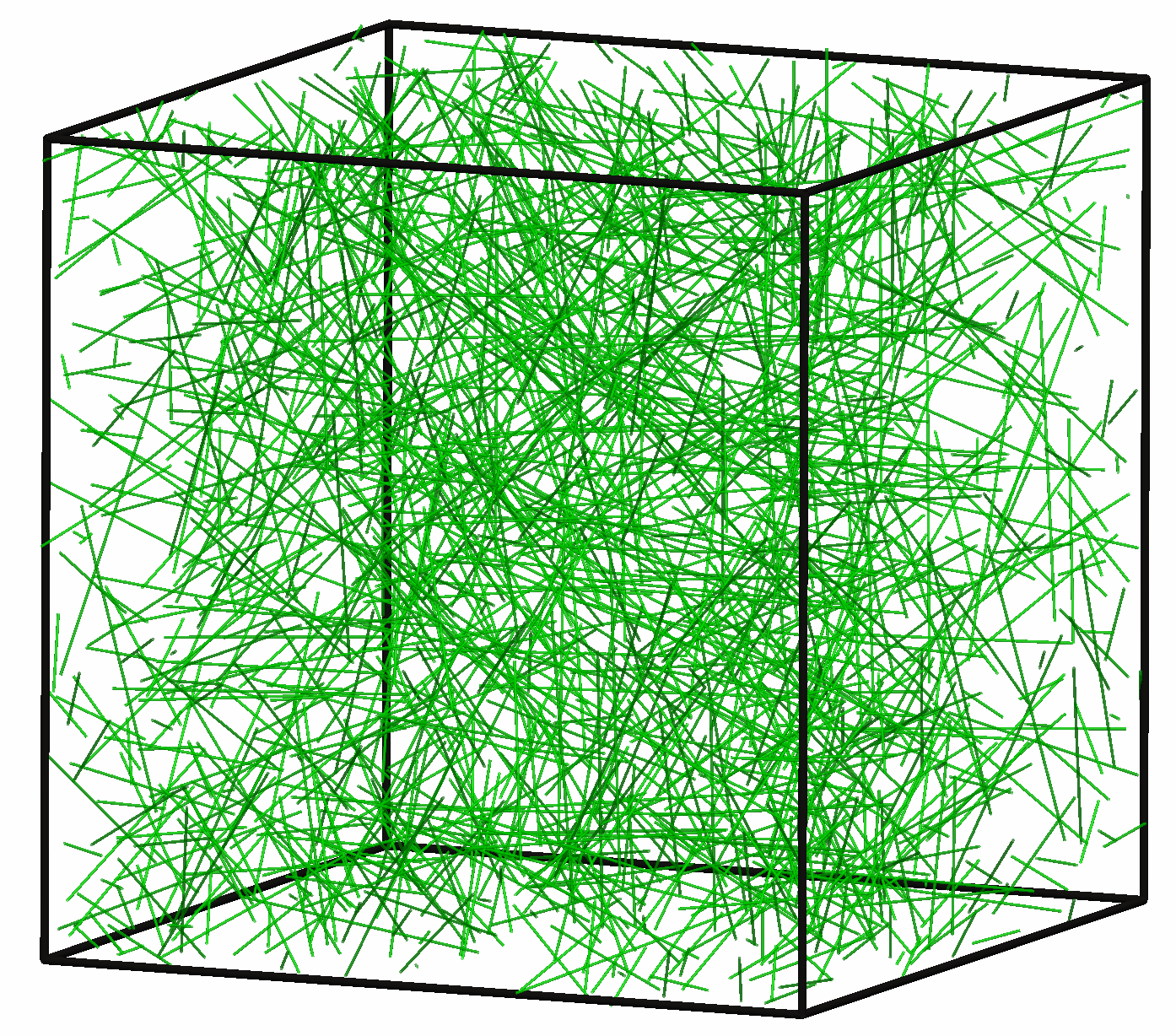}
    \end{minipage}%
  \end{minipage}
  \begin{minipage}[t]{0.39\linewidth}%
    \begin{minipage}[t]{0.05\linewidth}%
      \vspace{-2cm}(B)
    \end{minipage}%
    \hfill
    \begin{minipage}[c]{0.95\linewidth}%
      \includegraphics[width=\linewidth]{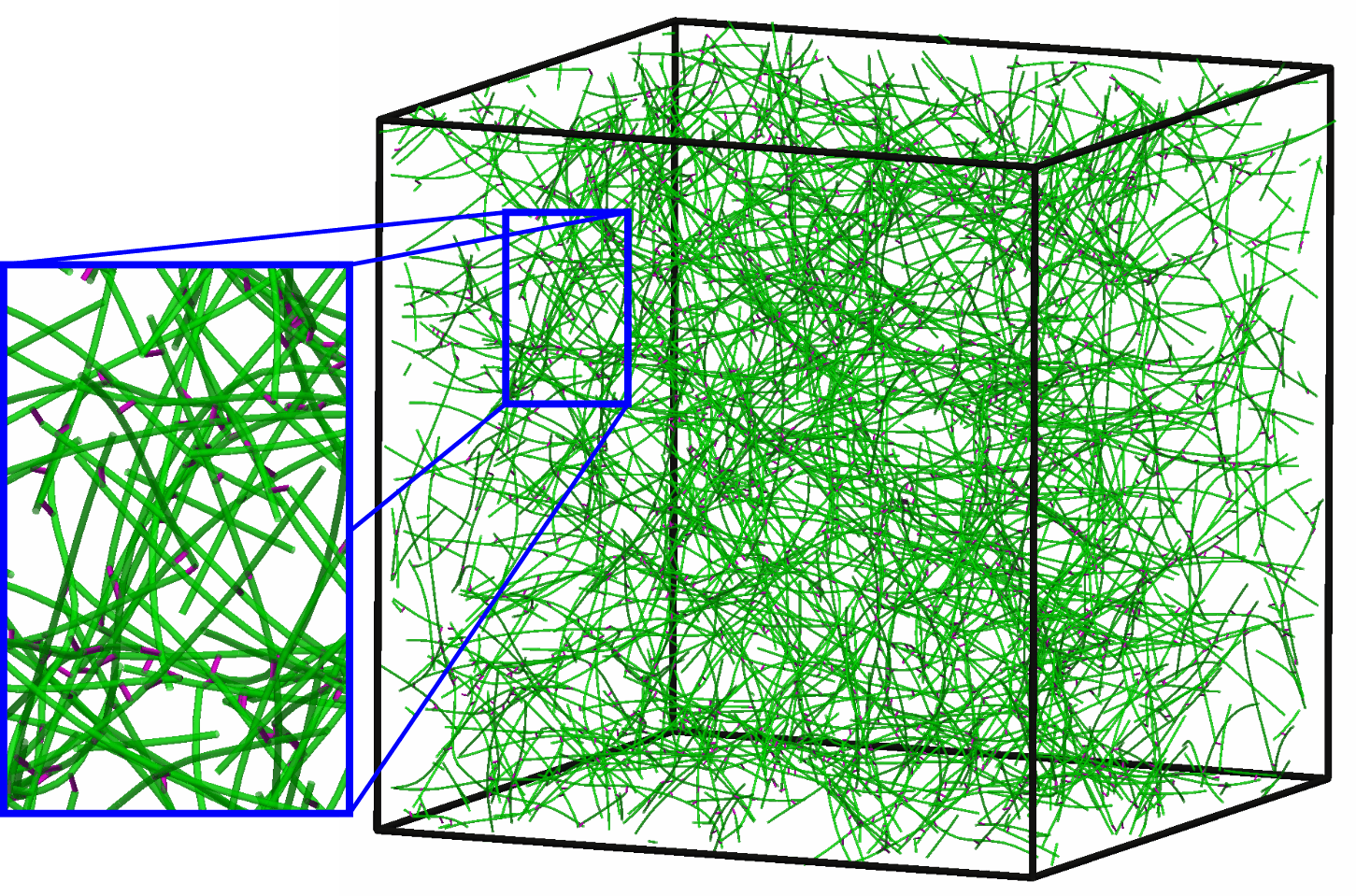}
    \end{minipage}%
  \end{minipage}
  \begin{minipage}[t]{0.3\linewidth}%
    \begin{minipage}[t]{0.08\linewidth}%
      \vspace{-2cm}(C)
    \end{minipage}%
    \hfill
    \begin{minipage}[c]{0.94\linewidth}%
      \includegraphics[width=\linewidth]{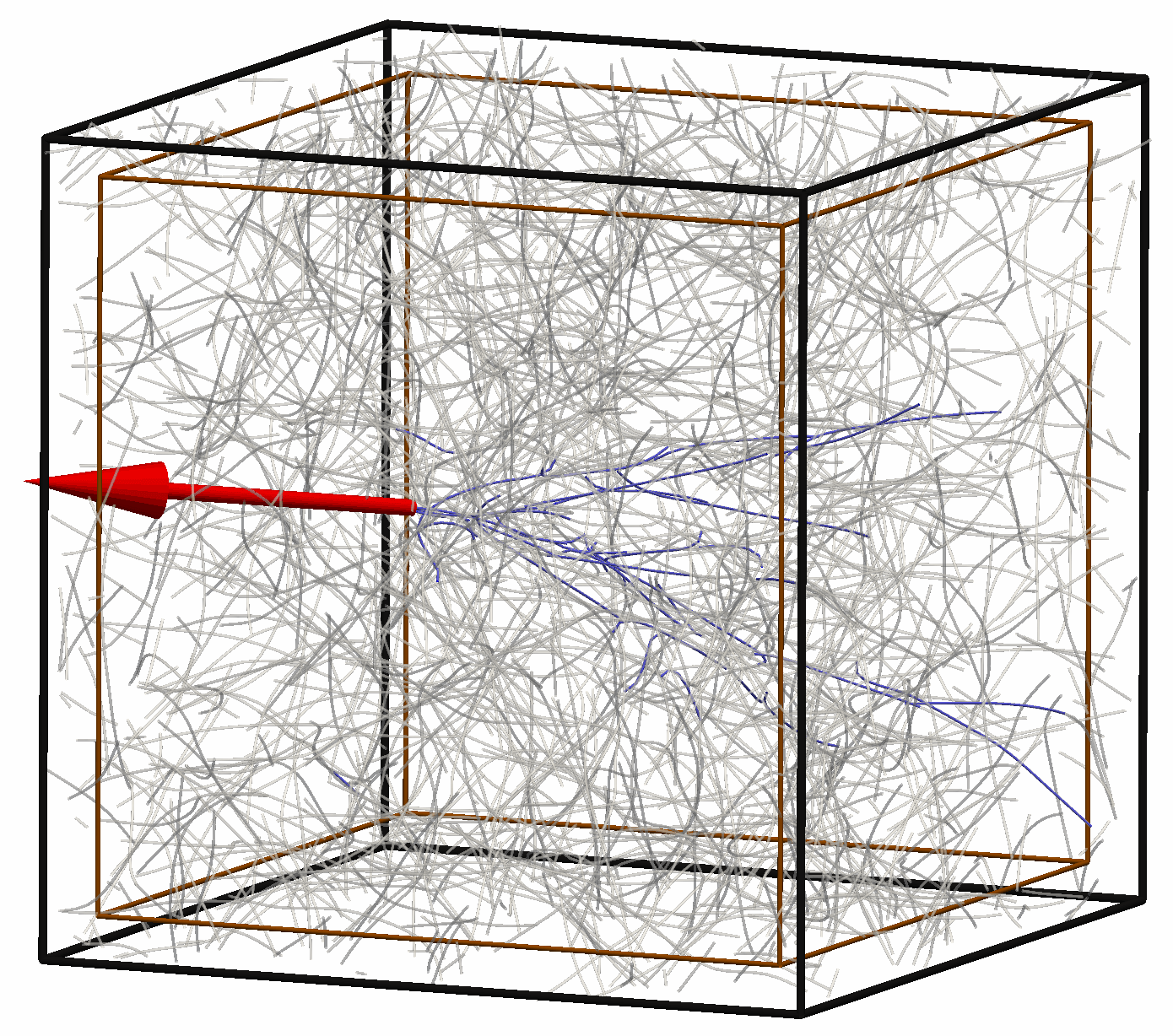}
    \end{minipage}%
  \end{minipage}\\
  \caption{(color online) Schematic of the setup and protocol of the numerical experiments.
           (A) Pre-processing: Random placement of straight filaments (green) and free linker molecules (not shown) in a cubic box with periodic boundaries.
           (B) Simulation of the network assembly driven by Brownian motion and random formation of cross links (pink).
           Subsequent equilibration of the system considering permanent cross links and zero temperature.
           (C) Simulation of the quasi-static network response to a point force (red arrow), applying zero displacement boundary conditions in the thin outermost shell of the network (between brown and black box).}
  \label{fig::simulation_experimental_setup}
\end{figure*}

Another consequence of the mesoscopic failure of continuum elasticity (by which we mean on scales greater several
mesh sizes) is that the collective point-force response of the network is remarkably heterogeneous.
Active microrheology experiments~\cite{kotlarchyk2011concentration,keating2017spatial}
have shown that the linear response of a bead embedded in the network to force
varies from location to location within the network by one order of magnitude.
Conversely, recent
experiments~\cite{Qingda2020} have shown that the response of the network to a force acting on
one filament is also highly heterogeneous.  The displacement of the surrounding network depends critically on which
filament is pulled, and in which direction.  These complexities of the
point force response are not attributable to large-scale spatial gradients in either the density of cross links or filaments.  Instead,
the experiments suggest that the inherent mesoscale structure of a
network, whose density and connectivity is, at least, statistically homogeneous or
self-averaging on long length scales, is responsible for these effects.

To better understand the peculiarities of the response of these seemingly simple systems, we examine in this manuscript
the question of force chains and the elastic response of isotropic and homogeneous filament networks to point forces.  These choices allow us to assess
the complexity of force propagation in the most simple form of a random filament network.  In this work,
we combine the results of large-scale, finite-element simulations with theoretical modeling
to determine how a force applied to a single point on a filament spreads out through the network.  We confine our studies
here entirely to the static response of the system.  We show
via simulation that tensile force chains exist in the vicinity of the point of force application.  We use
these simulations to further characterize both the spatial structure of the force chains and
the forces they carry, paying particular attention to how the force applied to a particular filament
bleeds off into the surrounding network through cross-linked junctions.  In addition, we
examine the collective response of the network to applied point forces by determining the point force response that is measured by low-frequency microrheology experiments in fibrous materials like the ECM.

We then turn to two theoretical approaches to the understanding force chains and the point
force response of the network, one that takes into account the small longitudinal compliance of the
constituent filaments and one that treats these as inextensible.  Finally, we compare
simulation and theory, as well as discuss our results in the broader context of intercellular force propagation in tissues in our summary. There
we address the question of whether one can account for the observations of both long force chains and the
dramatic spatial heterogeneity of the collective response of the network to
point forces within a statistically isotropic and homogeneous filament network.

\section{Simulations}
This section covers the large-scale, finite-element simulations we performed to study the point force response of randomly assembled 3D networks of semiflexible filaments.
After a brief description of the conducted numerical experiments and the underlying computational model in Sec.~\ref{sec::numerical_experiments} and \ref{sec::numerical_model}, respectively, we present the simulation results in Sec.~\ref{sec::simulation_results}.

\subsection{Computational experiments}\label{sec::numerical_experiments}
A schematic overview of the setup and protocol of the performed experiments is given in Fig.~\ref{fig::simulation_experimental_setup}.
In a pre-processing step, a number of straight filaments and free linker molecules are randomly distributed inside a cubic simulation box with periodic boundary conditions (Fig.~\ref{fig::simulation_experimental_setup}(A)).
We chose the filament concentration so that we end up with $834$ filaments in a box with edge length of $10 \mathrm{\mu m}$.
The number of linker molecules is chosen as $N_\ell=10^4$.
During the first simulation phase of $t = 1.5 \mathrm{s}$, the filaments experience
stochastic, thermal undulations, and cross links are
established, such that we obtain a random 3D network geometry (Fig.~\ref{fig::simulation_experimental_setup}(B)).
The final number of (doubly bound) cross links
in these assembled networks varies from 2998 to 3097 over the ten random realizations that have been considered
in our study. A subsequent relaxation phase of $1 \mathrm{s}$ allows the network to release some of the prestress that has been
trapped during the assembly in order to start the subsequent force application from an equilibrium state.
For this purpose, the thermal energy $k_\text{B} T$ and the binding and unbinding rates $k_\text{on}$ and $k_\text{off}$ of the linkers are set to zero after the assembly phase.

Finally, a point force is applied in order to investigate the quasi-static response of the network (Fig.~\ref{fig::simulation_experimental_setup}(C)).
The point of force application is chosen close to the box center in order to restrict the influence of the boundary conditions.
Specifically, the filament nodes in the outermost shell of the network sample
(the volume between the brown and black box in Fig.~\ref{fig::simulation_experimental_setup}(C)) are
pinned via zero displacement Dirichlet boundary conditions.
The thickness of this shell is chosen to be $0.5 \mathrm{\mu m}$, which corresponds to 5\% of the edge length.
The direction of the applied force is chosen either tangentially to the filament axis or transverse to it.
It is important to note, however, that the point force direction is kept constant, i.e., it will not follow the filament's deformation.
The force magnitude is increased linearly until it reaches its maximum~$F = 100 \mathrm{pN}$ after another $1 \mathrm{s}$ of simulation.
This is sufficiently slow to ensure a quasi-static response of the system. At the smallest forces recorded, we do observe viscous
effects dependent on the rate of force increase, which perturb our results for pulling forces around $1 \mathrm{pN}$.
Note that the third phase of the simulation, consisting of the actual force application, is run independently several times in order to
generate our complete data set, and to observe the influence of the direction of force application.
Starting from the identical equilibrated network configuration, the network is thus probed
along three mutually orthogonal axes with two directions each, which leads to six numerical
pulling experiments for each network geometry.

\subsection{Numerical model}\label{sec::numerical_model}
We employ the numerical model developed and applied in our previous work~\cite{Cyron2010,cyron2012,Cyron2013phasediagram,Mueller2014rheology,Mueller2015interpolatedcrosslinks,Maier2015,Kachan2016bundlingcasimir,Slepukhin2019}, describing semiflexible filaments via geometrically exact beam theory, subjected to Brownian dynamics.
Thermal excitations and the presence of cross-linker
molecules give rise to network self-assembly,
which produces isotropic and uniform random 3D networks, which are to be probed by applying a point force later on.
Further details including the parametrization of the model are given as follows.

\subsubsection{Filament model}
Each filament is modeled by nonlinear, geometrically exact, 3D Simo-Reissner beam theory and
discretized in space using beam finite elements.
In terms of the structural rigidity of the filament, we thus account for axial, torsional, bending,
and shear deformation.  All filaments are chosen to be initially straight with a length of
$L_0=4 \mathrm{\mu m}$ and persistence length~$L_p \approx 7 \mathrm{\mu m}$.
The geometrical and material parameters resemble F-actin, which is a key constituent of the cytoskeleton.
A complete specification is given by the cross-section area~$A=1.9 \times 10^{-7}\mathrm{\mu m}^2$, area
moment of inertia~$I = 2.85{\times}10^{-11}\mathrm{\mu m}^4$, polar moment
of inertia~$I_p = 5.7{\times}10^{-11}\mathrm{\mu m}^4$, Young's modulus~$E=10^{9} \mathrm{pN/\mu m^2}$,
and Poisson ratio~$\nu=0.3$.  By default, we discretized each filament with four beam finite elements of the
Hermitian Simo-Reissner type, which has been presented in our recent contribution~\cite{Meier2017b}.

As described above, once the networks have been created by the Brownian diffusive dynamics of
cross links and filaments, we explore force propagation in the network in zero temperature simulations.
Force propagation in these networks depends only on the relative size of two length scales,
the filament length and the distance between consecutive cross links along the filament.
As a result, our studies of force propagation under static loading apply equally well to
filament networks at all scales including cytoskeleton and ECM.

\subsubsection{Brownian dynamics}
To model the Brownian motion, we include viscous drag as well as thermal forces, each distributed along the
entire filament length as in previous work~\cite{cyron2012}.
Viscous forces and moments are computed assuming a quiescent background fluid and individual damping coefficients for translations parallel and perpendicular to the filament axis,  as well as rotation around the filament axis.
Thermal forces are determined from the stochastic Wiener process in accordance with the fluctuation-dissipation theorem.
Finally, an Implicit-Euler scheme is used to discretize in time and a Newton-Raphson algorithm solves the resulting nonlinear system of equations.
Further details on this simulation framework including all formulae can be found in Ref.~\cite{cyron2012}.

Here, temperature is set to $T{\,=\,}293\mathrm{K}$ and the dynamic viscosity of the quiescent background fluid to $\eta{\,=\,}10^{-3}\,\mathrm{Pa\,s}$.
The base time step size is chosen as~$\Delta t = 0.01 \mathrm{s}$, which is augmented by an adaptive time stepping scheme that reduces the time step size whenever necessary.

\subsubsection{Cross-link model}
Our numerical model tracks linker molecules explicitly as they switch between three possible states:  free, singly bound, or doubly bound.
All details on the linker model can be found in the original publication~\cite{Mueller2015interpolatedcrosslinks}.
Free linker molecules experience Brownian motion until eventually all binding criteria are met and they establish a first,
and later possibly a second, connection to a filament.
In the doubly bound state, {\em i.e.}, a cross-link spanning two filament binding spots on different filaments,
each linker is treated as an additional, very short
beam element, which can transmit forces and moments between the filaments.
The binding decision is made based on a given binding rate and a distance criterion that takes into
account the spatial extent and thus action range of the linker molecule.
Here, the length of the linker is chosen to be $L_\ell=0.1 \mathrm{\mu m}$ (with a tolerance of $\Delta L_\ell = 2 \mathrm{nm}$) and binding spots are assumed to be located equidistantly along the filament with a spacing of $\Delta s_\text{bs} = 0.1 \mathrm{\mu m}$.
During the assembly of the networks, the binding and unbinding rates are set to~$k_\text{on} = 10^5
\mathrm{s}^{-1}$ and $k_\text{off} = 0 \mathrm{s}^{-1}$ in order to speed up the network generation process.

\subsection{Simulation results}\label{sec::simulation_results}
All the simulations were performed by means of the parallel, multi-physics, in-house research code BACI.~\cite{BACI2020}
The following simulation results aim to characterize the collective point force response of the system, the resulting deformation
and the stress state inside the system. In particular, we explore how the applied external force propagates through the system.

\subsubsection{Force-displacement curves and effective spring constants}
%
\begin{figure*}[htpb]
  \centering
  \begin{minipage}[t]{0.9\linewidth}%
    \begin{minipage}[t]{0.05\linewidth}%
      \vspace{-2.5cm}(A)
    \end{minipage}%
    \begin{minipage}[c]{0.9\linewidth}%
      \includegraphics[width=\linewidth]{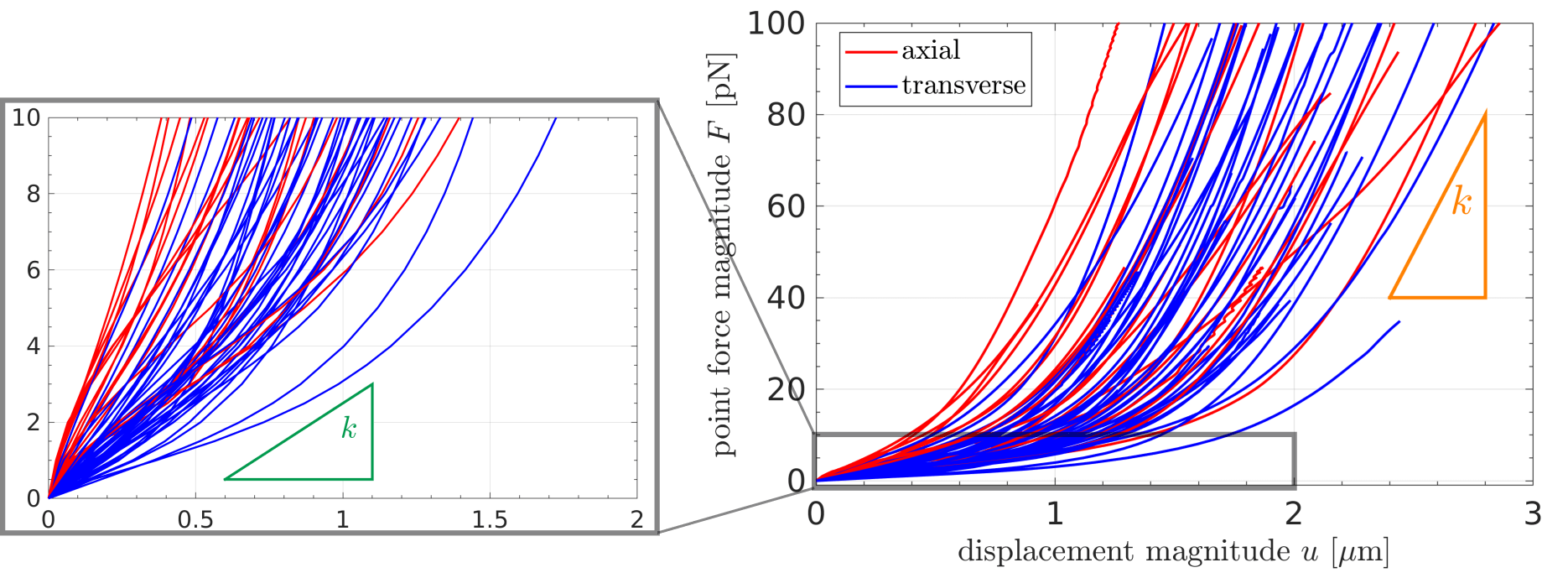}
    \end{minipage}%
  \end{minipage}\\\vspace{0.5cm}
  \begin{minipage}[t]{0.48\linewidth}%
    \begin{minipage}[t]{0.05\linewidth}%
      \vspace{-2.5cm}(B)
    \end{minipage}%
    \begin{minipage}[c]{0.9\linewidth}%
      \includegraphics[width=\linewidth]{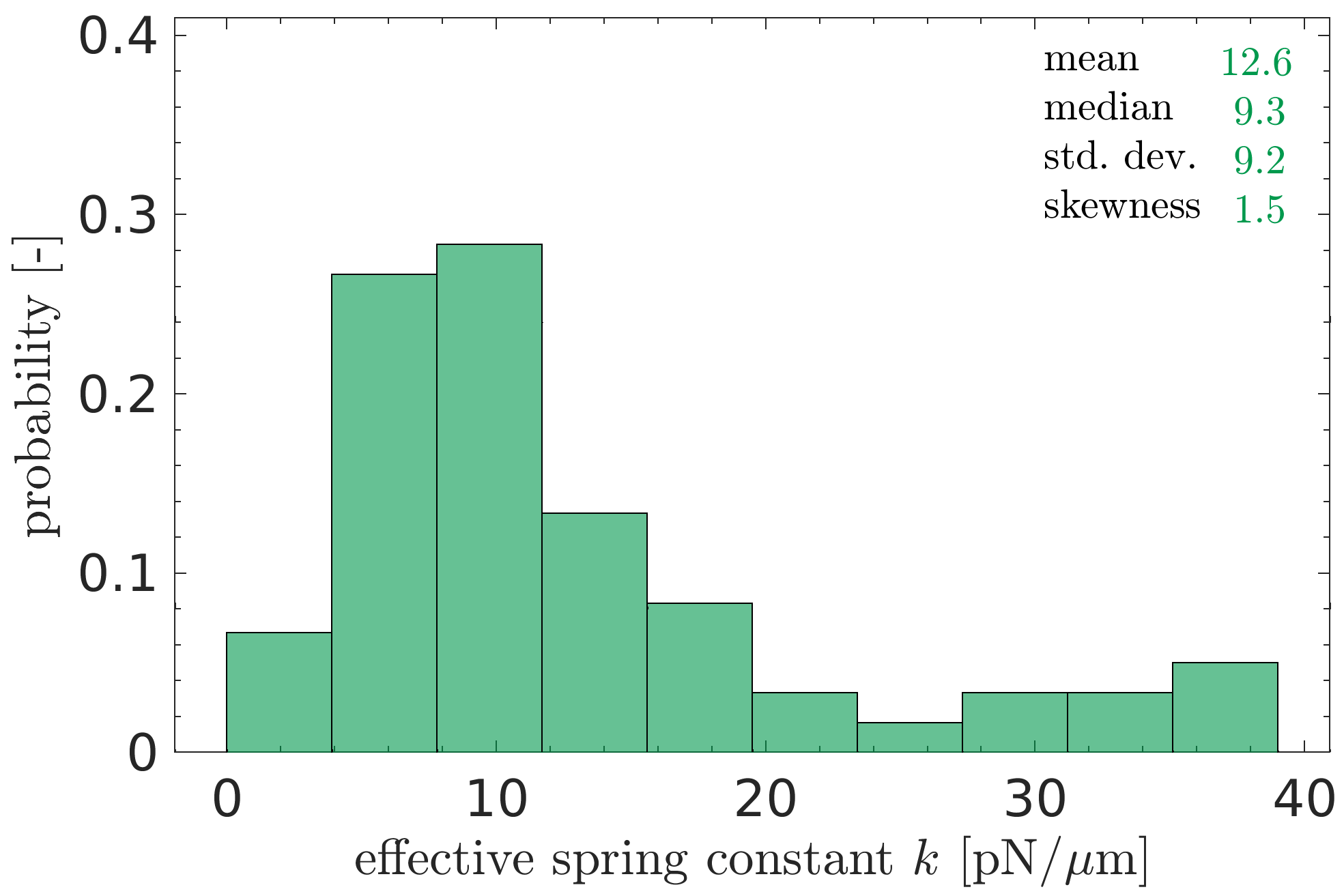}
    \end{minipage}%
  \end{minipage}%
  \begin{minipage}[c]{0.02\linewidth}%
  \rule{2pt}{5.5cm}
  \end{minipage}%
  \begin{minipage}[t]{0.48\linewidth}%
    \begin{minipage}[t]{0.05\linewidth}%
      \vspace{-2.5cm}(D)
    \end{minipage}%
    \begin{minipage}[c]{0.9\linewidth}%
      \includegraphics[width=\linewidth]{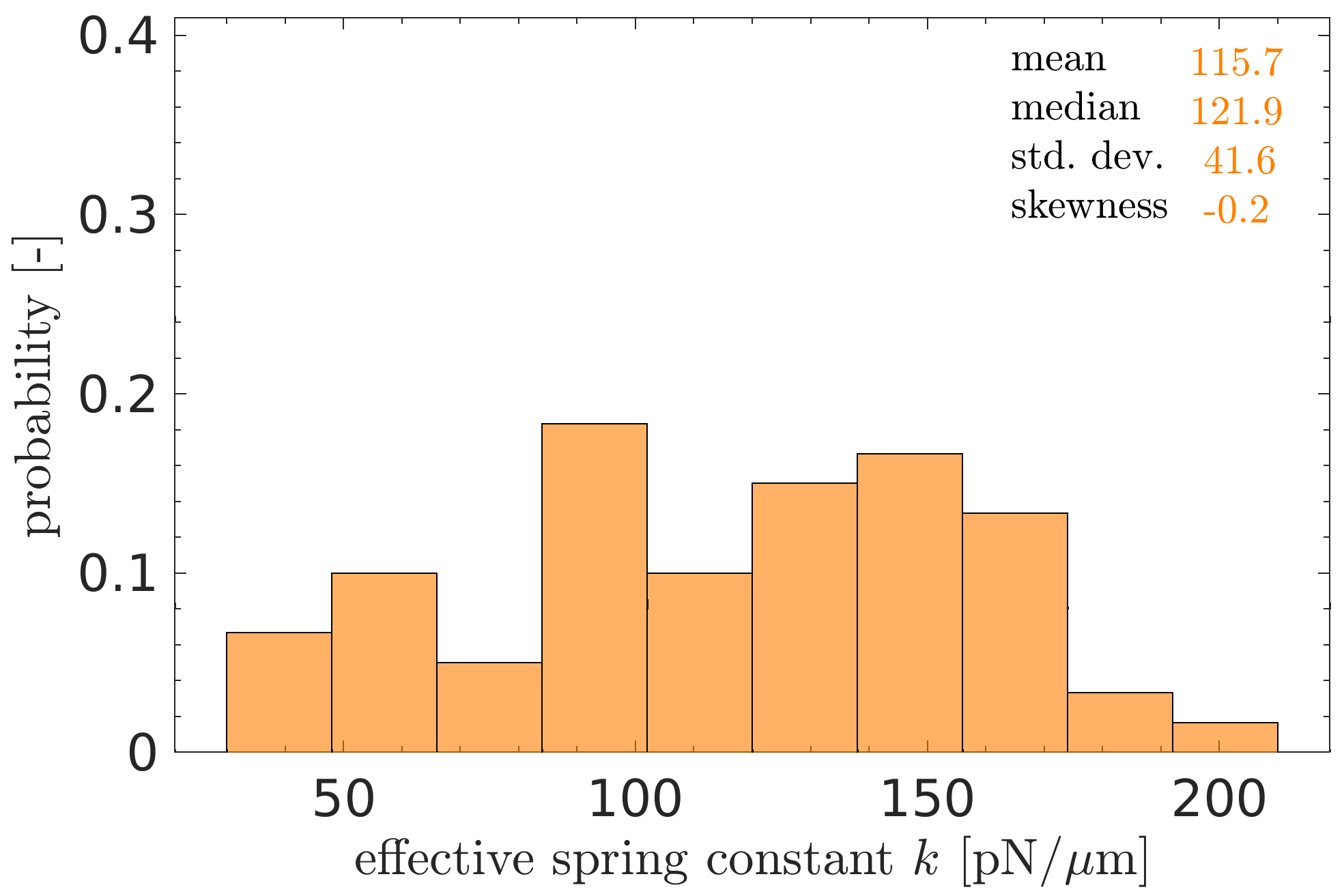}
    \end{minipage}%
  \end{minipage}\\\vspace{0.2cm}
  \begin{minipage}[t]{0.48\linewidth}%
    \begin{minipage}[t]{0.05\linewidth}%
      \vspace{-2.5cm}(C)
    \end{minipage}%
    \begin{minipage}[c]{0.9\linewidth}%
      \includegraphics[width=\linewidth]{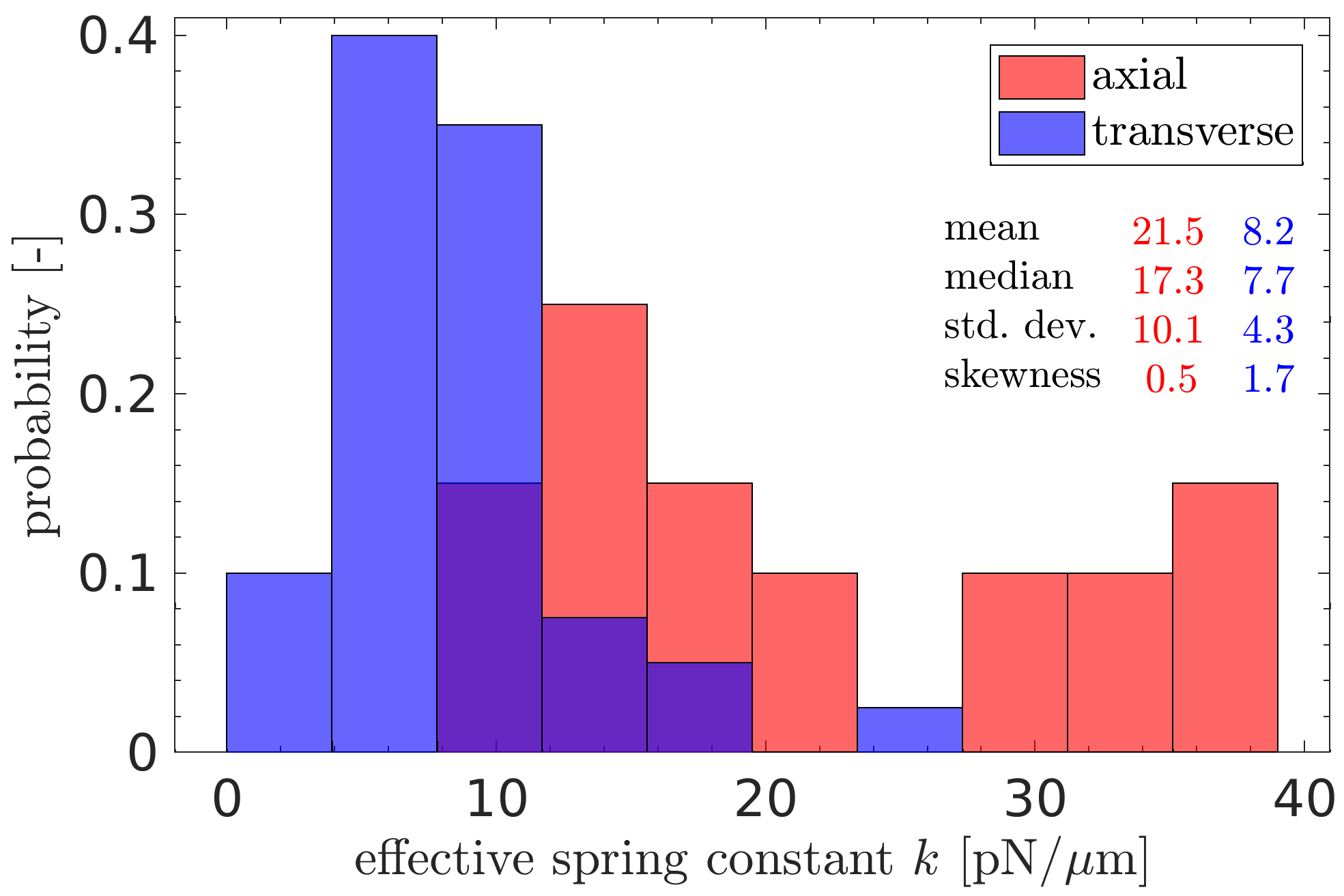}
    \end{minipage}%
  \end{minipage}%
  \begin{minipage}[c]{0.02\linewidth}%
  \rule{2pt}{5.5cm}
  \end{minipage}%
  \begin{minipage}[t]{0.48\linewidth}%
    \begin{minipage}[t]{0.05\linewidth}%
      \vspace{-2.5cm}(E)
    \end{minipage}%
    \begin{minipage}[c]{0.9\linewidth}%
      \includegraphics[width=\linewidth]{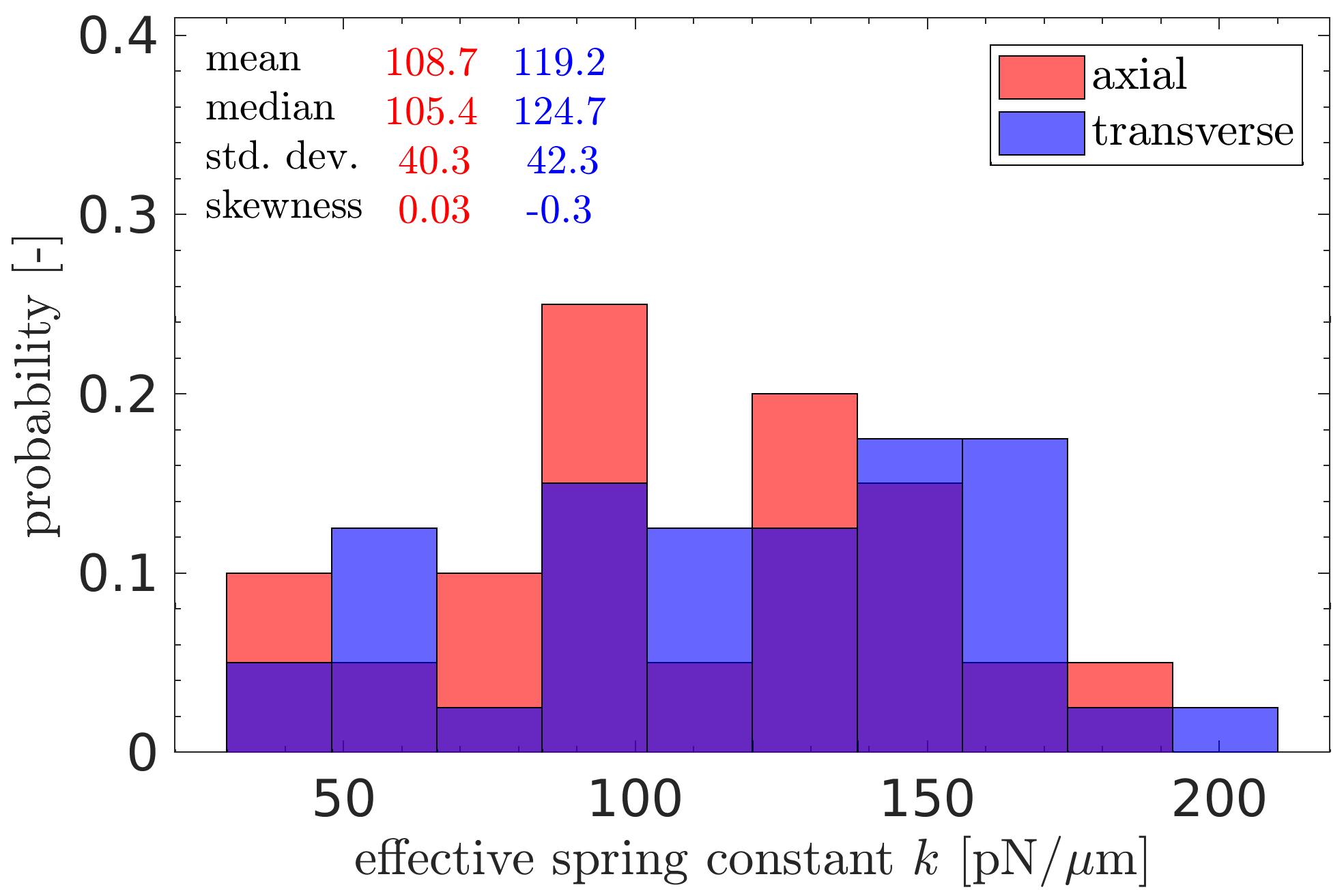}
    \end{minipage}%
  \end{minipage}\\
  \caption{(color online) (A) Force-displacement curves for 10 different network geometries with 2 axial (red) and 4 transverse (blue) pulling directions each, i.e., a total of 60 numerical pulling experiments.
           The inset on the left shows the magnified low force regime.
           The green and orange triangle indicates the calculation of the effective spring constants for the low and high force regime, respectively.
           (B) Histogram of effective spring constants for the low force regime and (C) the same data broken down into axial (red) and transverse (blue) pulling.
           (D) Histogram of effective spring constants for the high force regime and (E) the same data broken down into axial (red) and transverse (blue) pulling.}
  \label{fig::simulation_effective_spring_constants}
\end{figure*}
\begin{figure*}[htpb]
  \centering
  \begin{minipage}[c]{0.49\linewidth}%
    \begin{minipage}[t]{\linewidth}%
      \begin{minipage}[t]{0.06\linewidth}%
        \vspace{-0.7cm}(A)
      \end{minipage}%
      \begin{minipage}[c]{0.9\linewidth}%
        \includegraphics[width=\linewidth]{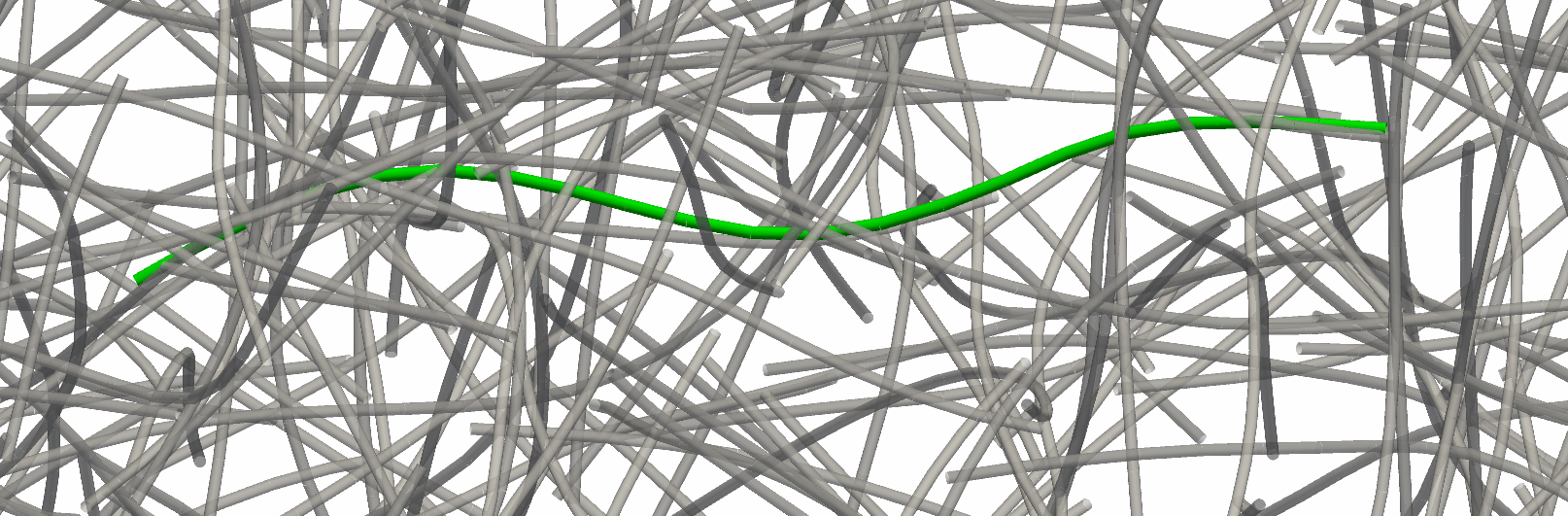}
      \end{minipage}%
    \end{minipage}\vspace{0.15cm}
    \begin{minipage}[t]{\linewidth}%
      \begin{minipage}[t]{0.06\linewidth}%
        \vspace{-0.7cm}(B)
      \end{minipage}%
      \begin{minipage}[c]{0.9\linewidth}%
        \includegraphics[width=\linewidth]{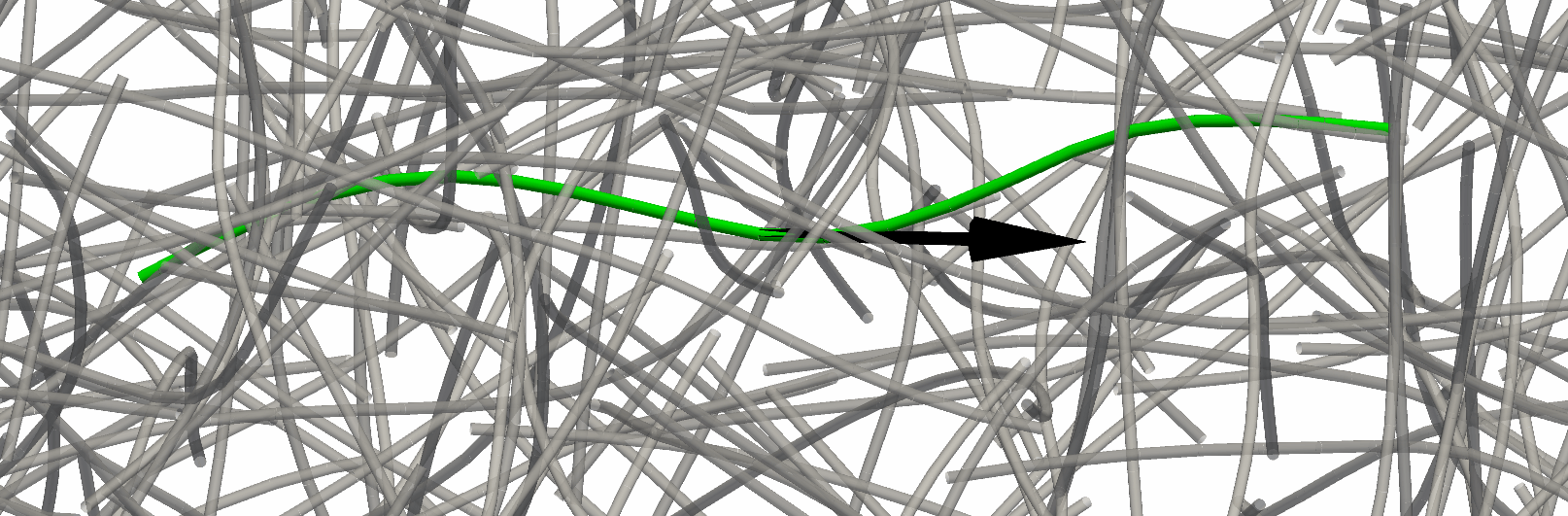}
      \end{minipage}%
    \end{minipage}\vspace{0.15cm}
    \begin{minipage}[t]{\linewidth}%
      \begin{minipage}[t]{0.06\linewidth}%
        \vspace{-0.7cm}(C)
      \end{minipage}%
      \begin{minipage}[c]{0.9\linewidth}%
        \includegraphics[width=\linewidth]{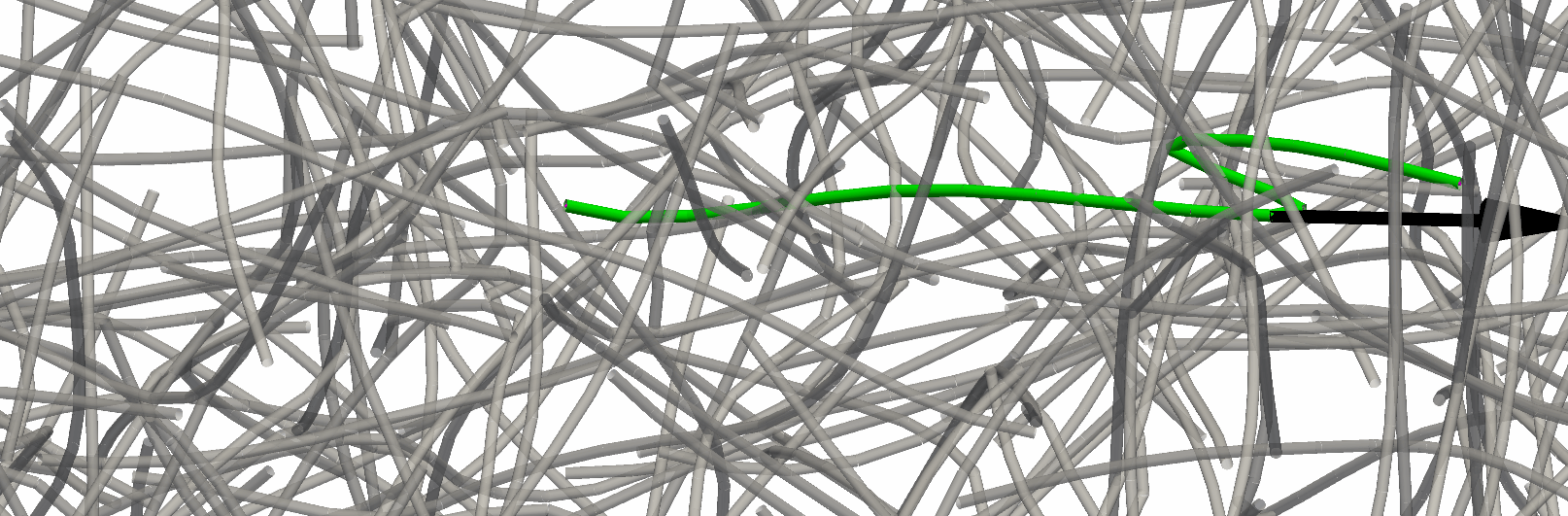}
      \end{minipage}%
    \end{minipage}%
  \end{minipage}%
  \begin{minipage}[c]{0.49\linewidth}%
    \begin{minipage}[t]{\linewidth}%
      \begin{minipage}[t]{0.06\linewidth}%
        \vspace{-1.5cm}(D)
      \end{minipage}%
      \begin{minipage}[c]{0.9\linewidth}%
        \includegraphics[width=\linewidth]{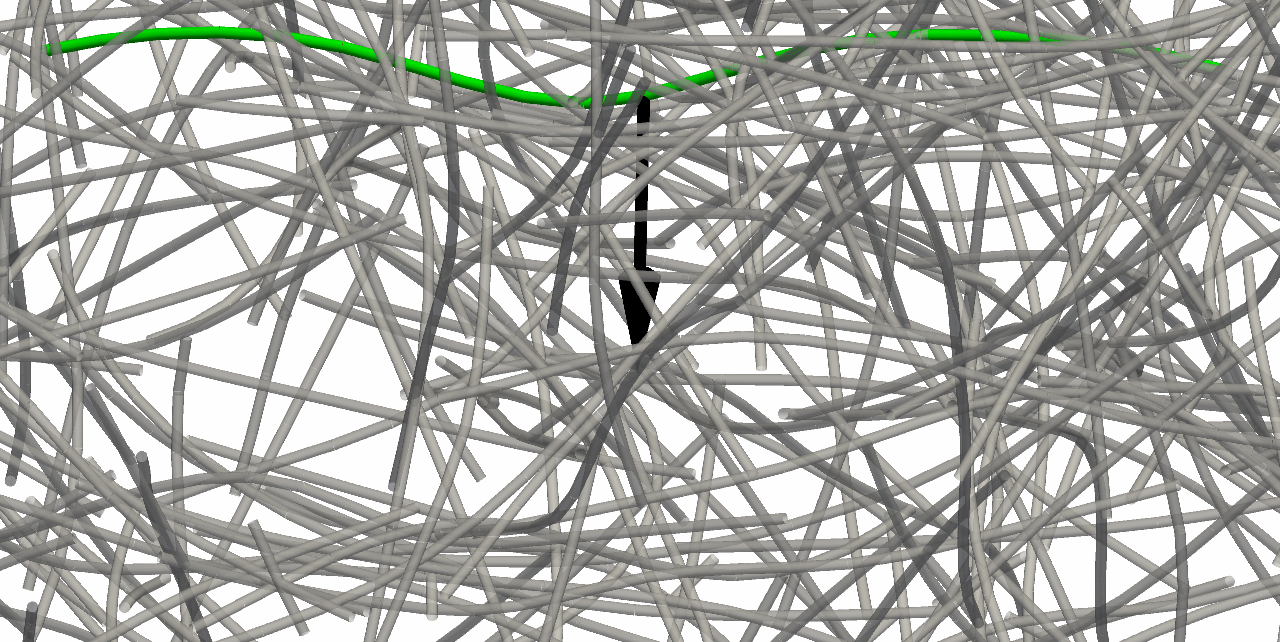}
      \end{minipage}%
    \end{minipage}\vspace{0.15cm}
    \begin{minipage}[t]{\linewidth}%
      \begin{minipage}[t]{0.06\linewidth}%
        \vspace{-1.5cm}(E)
      \end{minipage}%
      \begin{minipage}[c]{0.9\linewidth}%
        \includegraphics[width=\linewidth]{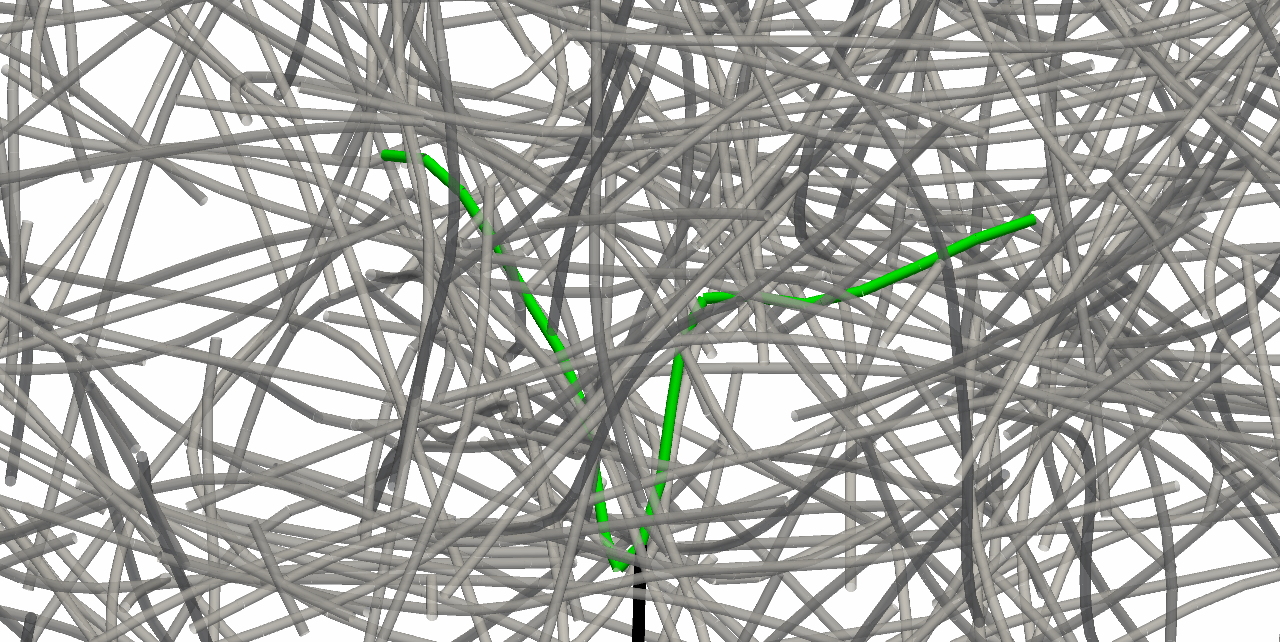}
      \end{minipage}%
    \end{minipage}%
  \end{minipage}%
  \caption{(Color online) Close-up views of the filament being pulled at (green) by a point force (black arrow).
           All other filaments in the network are shown in gray.
           (A) Equilibrated state, i.e., point force magnitude~$F=0$.
           (B) Axial loading at $F=1\mathrm{pN}$.
           (C) Axial loading at $F=100\mathrm{pN}$.
           (D) Transverse loading at $F=1\mathrm{pN}$.
           (E) Transverse loading at $F=100\mathrm{pN}$.}
  \label{fig::simulation_close-up_loaded_filament}
\end{figure*}

Fig.~\ref{fig::simulation_effective_spring_constants}(A) shows the measured force-displacement curves for 10 different network
geometries with two axial and four transverse pulling directions each, {\em i.e.}, for a total of sixty numerical pulling experiments.
The response from both the axially (red) and transversely (blue) applied point forces reveal a highly nonlinear, hyperelastic behavior.
Generally, we observe significantly smaller local stiffness in the regime of low
forces (see magnified part on the left of Fig.~\ref{fig::simulation_effective_spring_constants}(A)) as compared to the
high-force regime, which yields a pronounced strain hardening behavior.
To further characterize and investigate the system response, we compute the local,
effective spring constant as the slope of each curve,
both for the low and high force regime as indicated by the green and orange triangles, respectively.
Specifically, we use the first two data points of each curve to compute the slope at zero force and the
two data points with largest force values of each curve to compute the slope in the high force regime.

The resulting distributions of the effective spring constants in both regimes are shown in
Figs.~\ref{fig::simulation_effective_spring_constants}(B) and \ref{fig::simulation_effective_spring_constants}(D).
On average, the effective stiffness for high forces is approximately one order of magnitude higher than for low forces.
In addition, the shape of the distribution changes from a bell shape with fat right tail for low forces,
to a rather broad and uniform distribution for high forces.
Breaking this down into axial and transverse pulling experiments results in
the histograms shown in Fig.~\ref{fig::simulation_effective_spring_constants}(C) and (E).
In the low force regime (Fig.~\ref{fig::simulation_effective_spring_constants}(C)), this clearly
reveals a strong dependence on the pulling direction relative to the filament orientation.
On average, the stiffness values observed for tangentially applied point forces are a factor of 2.6 higher than for transverse point forces.
This is an expected result, because slender, semiflexible filaments typically have a significantly larger axial stiffness than
(effective) bending stiffness, making them much more compliant under transverse loading.
In the high force regime (Fig.~\ref{fig::simulation_effective_spring_constants}(E)), the picture is less clear and the
difference between both cases is reversed.
A close look at the deformed states of the network (see Fig.~\ref{fig::simulation_close-up_loaded_filament} for an example) suggests
a reason for this change.  When a filament is pulled transversely, it deforms as shown in Fig.~\ref{fig::simulation_close-up_loaded_filament} (E).
In effect, at high force the pulled filament (green), after becoming sharply bent, might be thought of as
two axially-tensed tensed filament halves, each of which may
generate its own tensile force chains and thus becoming stiffer to further pulling.  When the filament is pulled axially, however, this
effect is weaker -- see Fig.~\ref{fig::simulation_close-up_loaded_filament} (C). In that case, we surmise that the load is imperfectly
transferred to two tensile force chains, leading to a smaller increase in the collective stiffness of the system
to these large forces.  We speculate that the reversed effect of slightly higher mean effective stiffness in the
case of transverse loading might originate from the fact that the load is distributed more evenly on
both branches and thus involves a larger fraction of the entire network for high forces.

\subsubsection{Comparison of the resulting displacement field to the analytical solution for a linear elastic continuum}
%
\begin{figure*}[htb]
  \centering
  \begin{minipage}[t]{0.49\linewidth}%
    \begin{minipage}[t]{0.05\linewidth}%
      \vspace{-2.5cm}(A)
    \end{minipage}%
    \begin{minipage}[c]{0.9\linewidth}%
      \includegraphics[width=\linewidth]{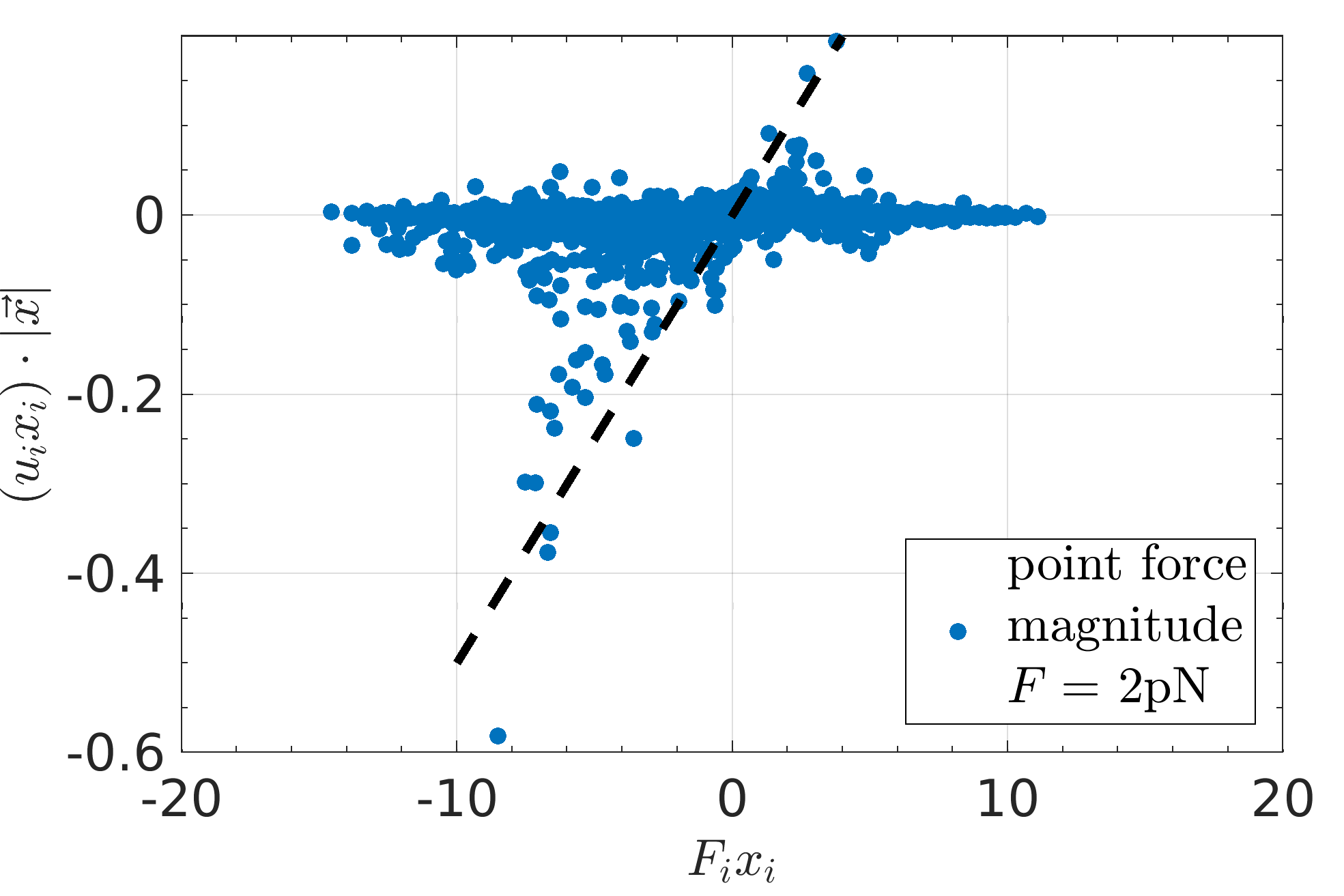}
    \end{minipage}%
  \end{minipage}%
  \begin{minipage}[t]{0.49\linewidth}%
    \begin{minipage}[t]{0.05\linewidth}%
      \vspace{-2.5cm}(B)
    \end{minipage}%
    \begin{minipage}[c]{0.9\linewidth}%
      \includegraphics[width=\linewidth]{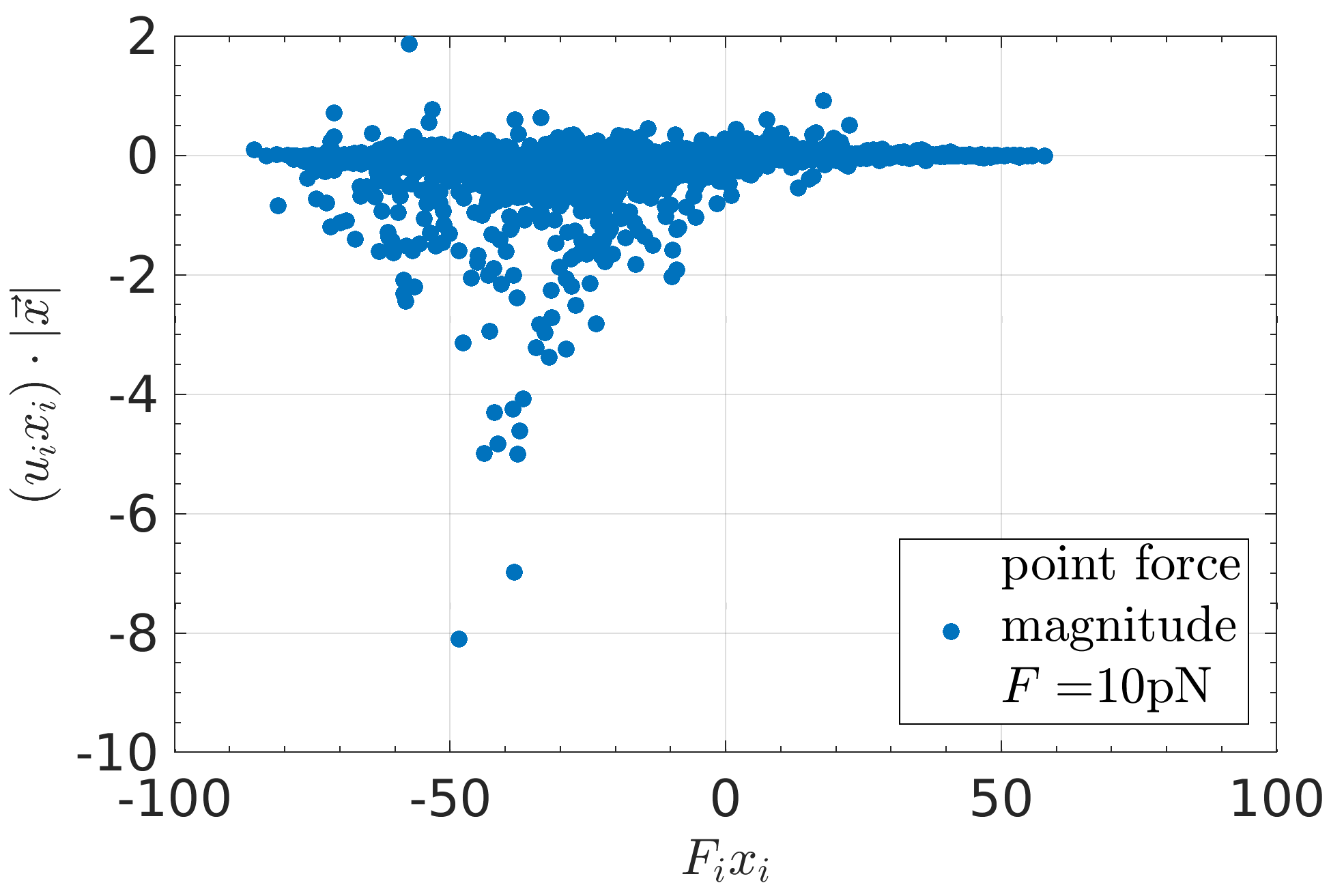}
    \end{minipage}%
  \end{minipage}\\\vspace{0.2cm}
    \begin{minipage}[t]{0.49\linewidth}%
    \begin{minipage}[t]{0.05\linewidth}%
      \vspace{-2.5cm}(C)
    \end{minipage}%
    \begin{minipage}[c]{0.9\linewidth}%
      \includegraphics[width=\linewidth]{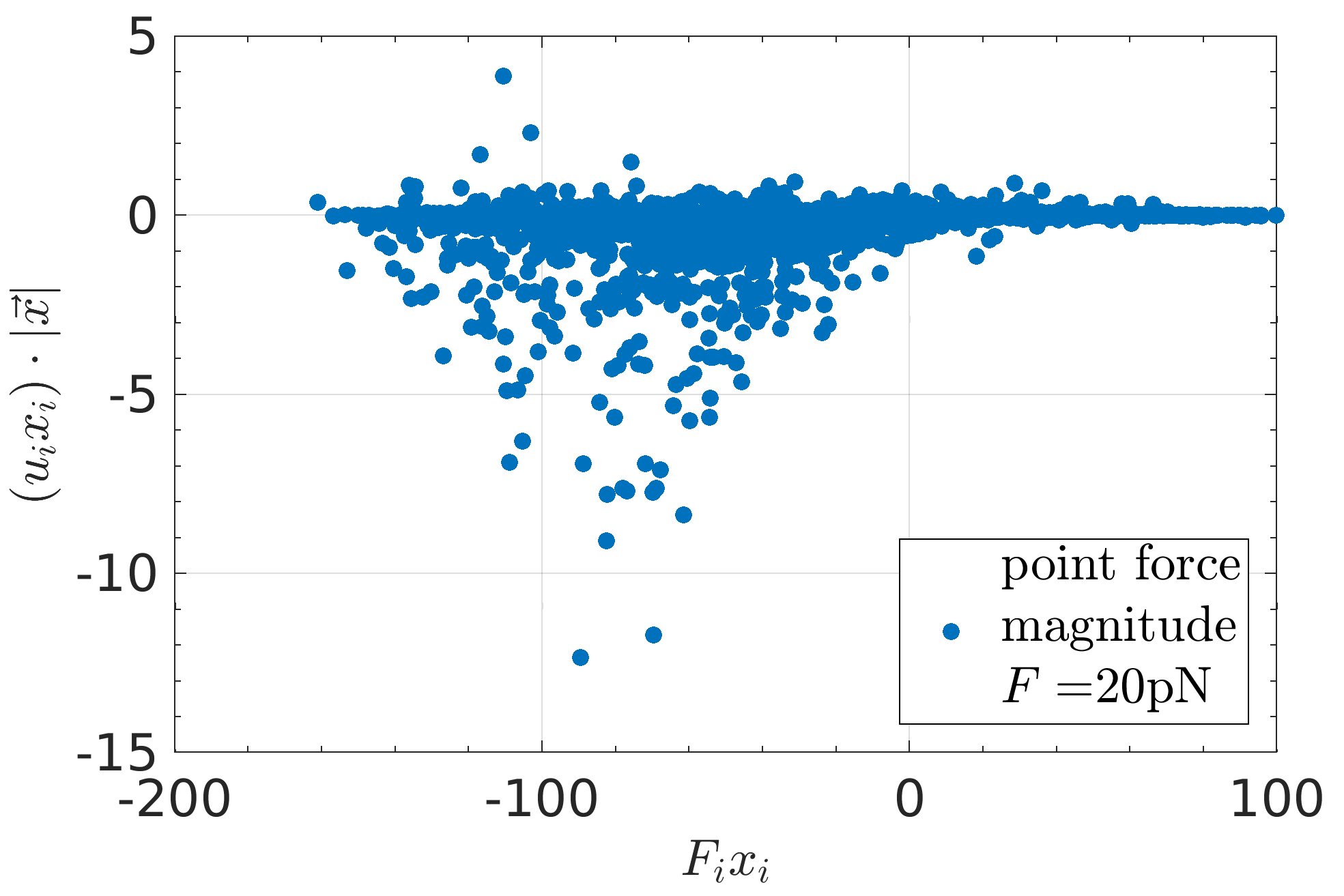}
    \end{minipage}%
  \end{minipage}%
  \begin{minipage}[t]{0.49\linewidth}%
    \begin{minipage}[t]{0.05\linewidth}%
      \vspace{-2.5cm}(D)
    \end{minipage}%
    \begin{minipage}[c]{0.9\linewidth}%
      \includegraphics[width=\linewidth]{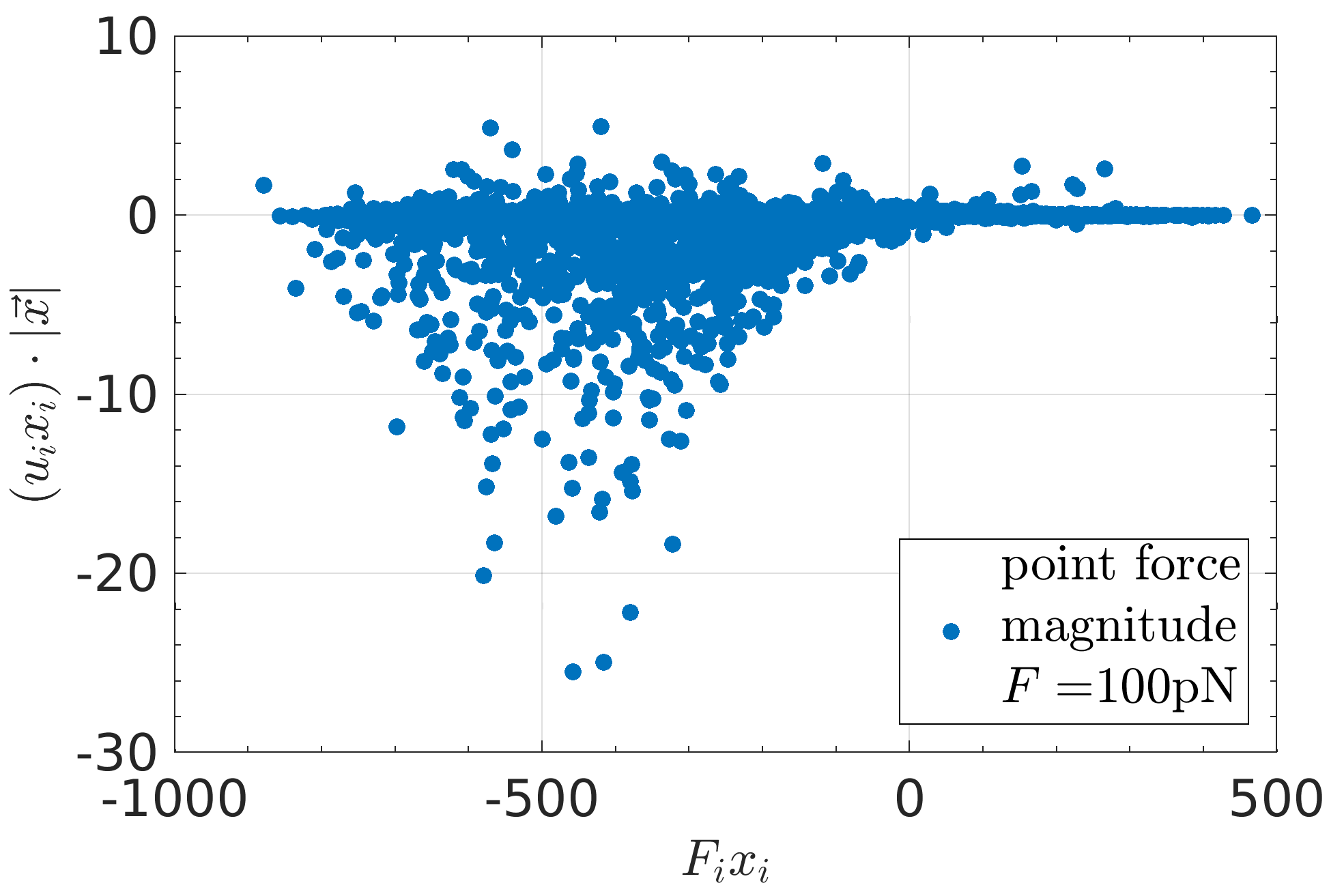}
    \end{minipage}%
  \end{minipage}%
  \caption{(color online) Scatter plots showing the correlation between the scalar products~$(u_i x_i)$ and $(F_i x_i)$, where $\vec{F}$ denotes the applied point force vector acting at the origin and $\vec{u}$ and $\vec{x}$ denote the displacement and position vector of a particular point.
           The data contains approx.~2000 nodes in the filament network for one (axial) pulling experiment at a point force magnitude of (A) $F=2\mathrm{pN}$, (B) $F=10\mathrm{pN}$, (C) $F=20\mathrm{pN}$, and (D) $F=100\mathrm{pN}$.}
  \label{fig::simulation_comparison_to_linear_elastic_continuum}
\end{figure*}
In this section, we characterize the resulting displacement field in the filament network, and compare it to the
analytical solution for an infinite homogeneous, isotropic, linear elastic continuum. For an isotropic elastic continuum, the
 displacement $u_i$ of each point $x_i$ of the medium with Young's modulus $E$ and Poisson ratio $\nu$ under the action point force
 $F_i$ at the origin obeys the equation~\cite{Landau1986}
\begin{equation}\label{eq::continuous_medium_point_force_solution}
(u_i x_i) |\vec{x} | = \frac{1+\nu}{2 \pi E} (F_i x_i).
\end{equation}

Fig.~\ref{fig::simulation_comparison_to_linear_elastic_continuum} characterizes the resulting displacement field in the network at a point force magnitude of (A) $F=2\mathrm{pN}$, (B) $F=10\mathrm{pN}$, (C) $F=20\mathrm{pN}$, and (D) $F=100\mathrm{pN}$.
Each scatter plot contains the data of ~2000 nodes in the filament network. These plots show the correlation
between the scalar products~$(u_i x_i)$ and $(F_i x_i)$, where $\vec{F}$ denotes the applied point force vector acting at the origin
and $\vec{u}$ and $\vec{x}$ denote the displacement and position vector of a particular point in the network.
Whereas the analytical solution for a homogeneous, isotropic, linear elastic continuum would be a straight line
with the slope equal to $(1+\nu)/(2 \pi E)$ (see Eq.~\ref{eq::continuous_medium_point_force_solution}), the simulation results for the filament network show an entirely different behavior.

Irrespective of the force's magnitude, there is a high concentration of points around zero displacement, indicating that a
large fraction of the network is barely influenced by the applied point force.
The remaining data points become distributed more widely upon increasing the force magnitude. Only for the smallest force
value of $F=2\mathrm{pN}$ (Fig.~\ref{fig::simulation_comparison_to_linear_elastic_continuum}(A)) can one find a linear feature
consistent with the solution from continuum elasticity. We show this as a guide to the eye with a black, dashed line.
Note also the asymmetry of the data with respect to the origin, which again increases with the force magnitude. It may be
explained by the strong asymmetry of the filaments to bear tensile versus compressive loads.
Altogether, one may conclude -- on the mesoscale considered here -- the point force response of a semiflexible filament
network is conceptually different to the response of a continuous elastic medium, even in the regime of small forces. We now consider
other ways to characterize the distribution of forces within the network.

\subsubsection{Distribution of axial force in the network}
\begin{figure}[htpb]
  \centering
  \includegraphics[width=\linewidth]{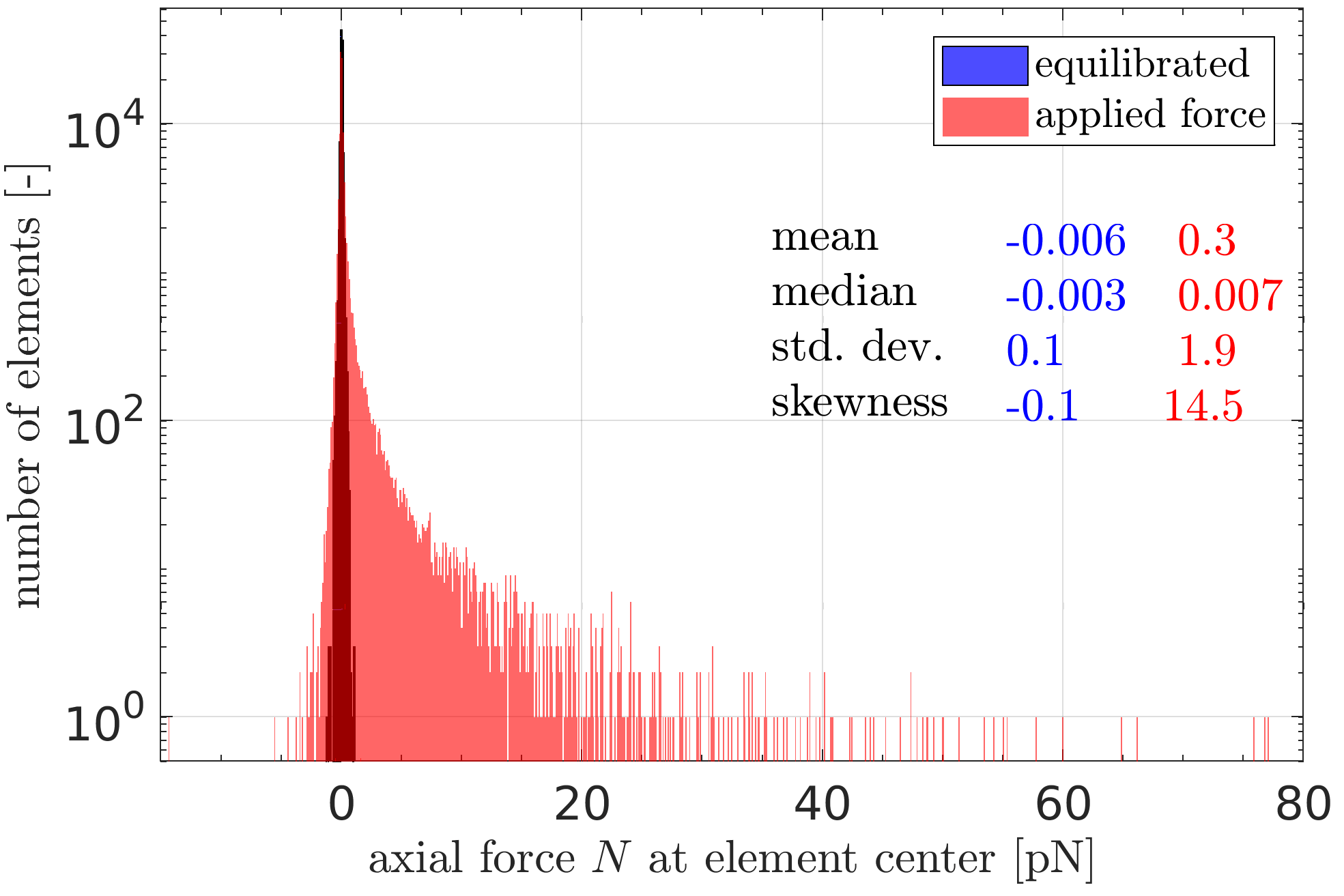}
  \caption{(color online) Histogram of axial forces at the element center in the entire network for 30 numerical pulling experiments and an applied point force magnitude of $F=100 \mathrm{pN}$.
          Note the logarithmic scale on the vertical axis.}
  \label{fig::simulation_histogram_axial_force_ele_centers}
\end{figure}

%
To characterize the stress state of the entire filament network as a result of point force application, we report the
distribution of axial force measured at the center of each finite element.
Fig.~\ref{fig::simulation_histogram_axial_force_ele_centers} compares, on a semi-log scale,
the distribution obtained after the equilibration phase (blue)
with the one obtained for an applied point force magnitude of $F=100 \mathrm{pN}$ (red).
In order to obtain more data points in the long high-force tail of the distribution, we have included
data from thirty numerical pulling experiments.

As expected, the reference distribution of axial forces in the equilibrated state has a sharp peak near
zero force, with mean value and standard deviation $-0.006 \pm 0.1 \mathrm{pN}$.
Applying the point force significantly broadens and shifts the distribution towards tensile, {\em i.e.}, positive force
values with mean value and standard deviation $0.3 \pm 1.9 \mathrm{pN}$.
Also, the skewness increases by more than two orders of magnitude, reflecting the well-known,
strong asymmetry between tensile and compressive force transmission in semiflexible filaments.
The median, however, of that distribution changes only very little
from $-0.003 \mathrm{pN}$ to $0.007 \mathrm{pN}$ when compared to the significant increase of the mean value.
This indicates, once again, that the vast majority of the filaments in the network remain almost unaffected by the applied point force.
Together with the long tail of the distribution, it provides evidence for the existence and importance of tensile force chains.

\subsubsection{Force propagation via force chains}

%
\begin{figure*}[htb]
  \centering
  \begin{minipage}[t]{0.32\linewidth}%
    \begin{minipage}[t]{0.08\linewidth}%
      \vspace{-2.5cm}(A)
    \end{minipage}%
    \begin{minipage}[c]{0.9\linewidth}%
      \includegraphics[width=\linewidth]{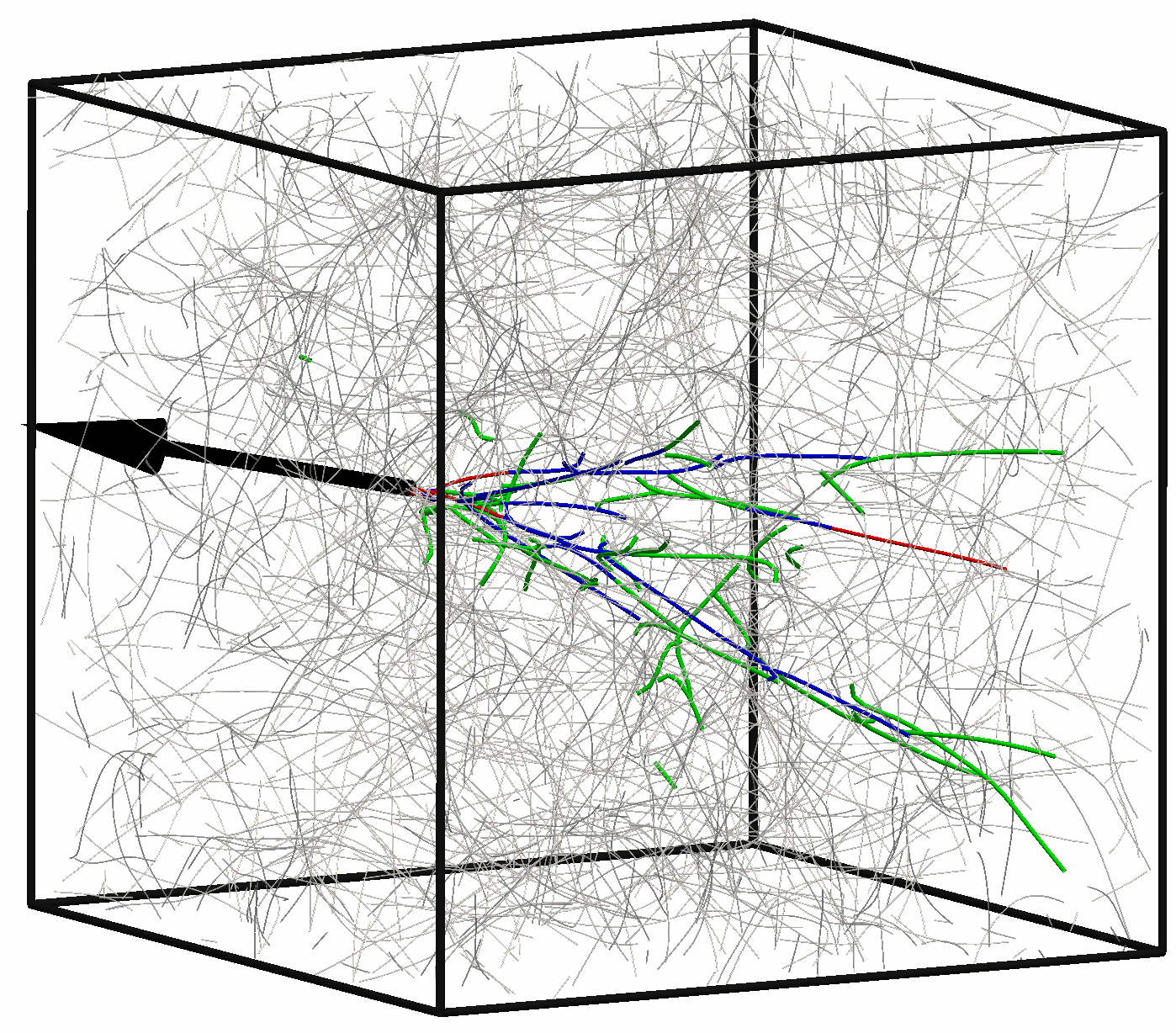}
    \end{minipage}%
  \end{minipage}
  \begin{minipage}[t]{0.32\linewidth}%
    \begin{minipage}[t]{0.08\linewidth}%
      \vspace{-2.5cm}(B)
    \end{minipage}%
    \begin{minipage}[c]{0.9\linewidth}%
      \includegraphics[width=\linewidth]{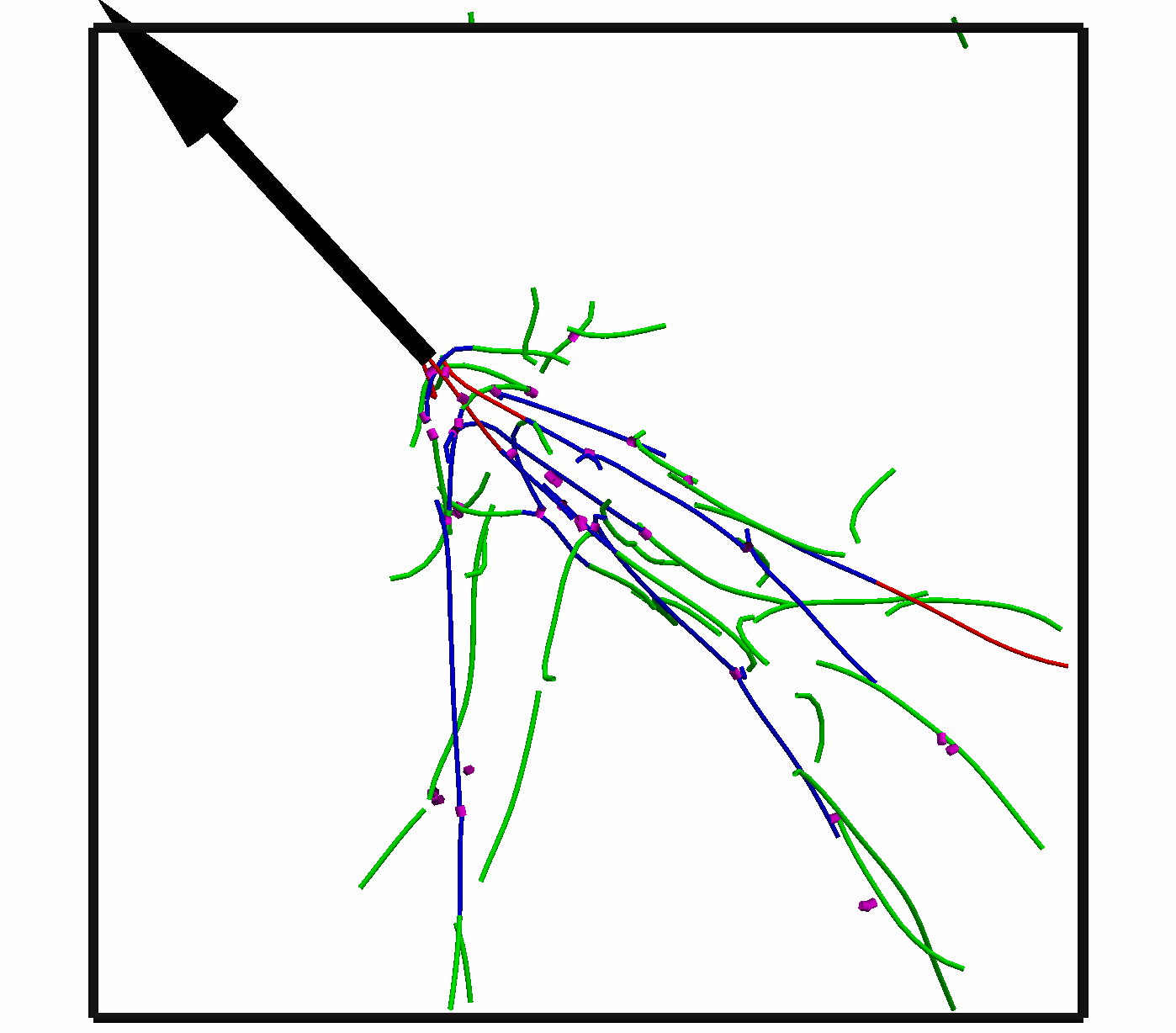}
    \end{minipage}%
  \end{minipage}
  \begin{minipage}[t]{0.32\linewidth}%
    \begin{minipage}[t]{0.08\linewidth}%
      \vspace{-2.5cm}(C)
    \end{minipage}%
    \begin{minipage}[c]{0.9\linewidth}%
      \includegraphics[width=\linewidth]{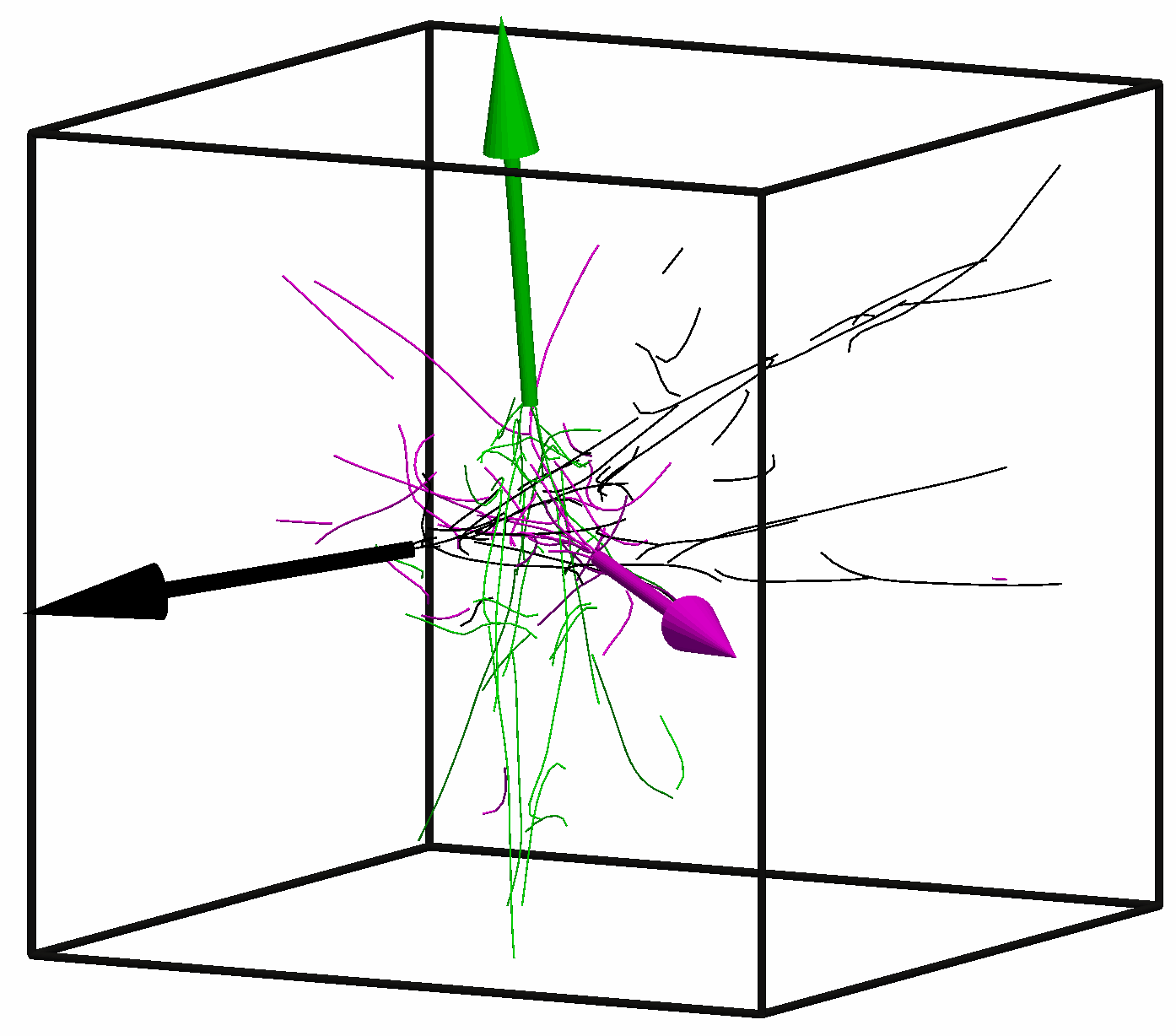}
    \end{minipage}%
  \end{minipage}\\\vspace{0.5cm}
  \begin{minipage}[t]{0.45\linewidth}%
    \begin{minipage}[t]{0.07\linewidth}%
      \vspace{-2.5cm}(D)
    \end{minipage}%
    \begin{minipage}[c]{0.95\linewidth}%
      \includegraphics[width=\linewidth]{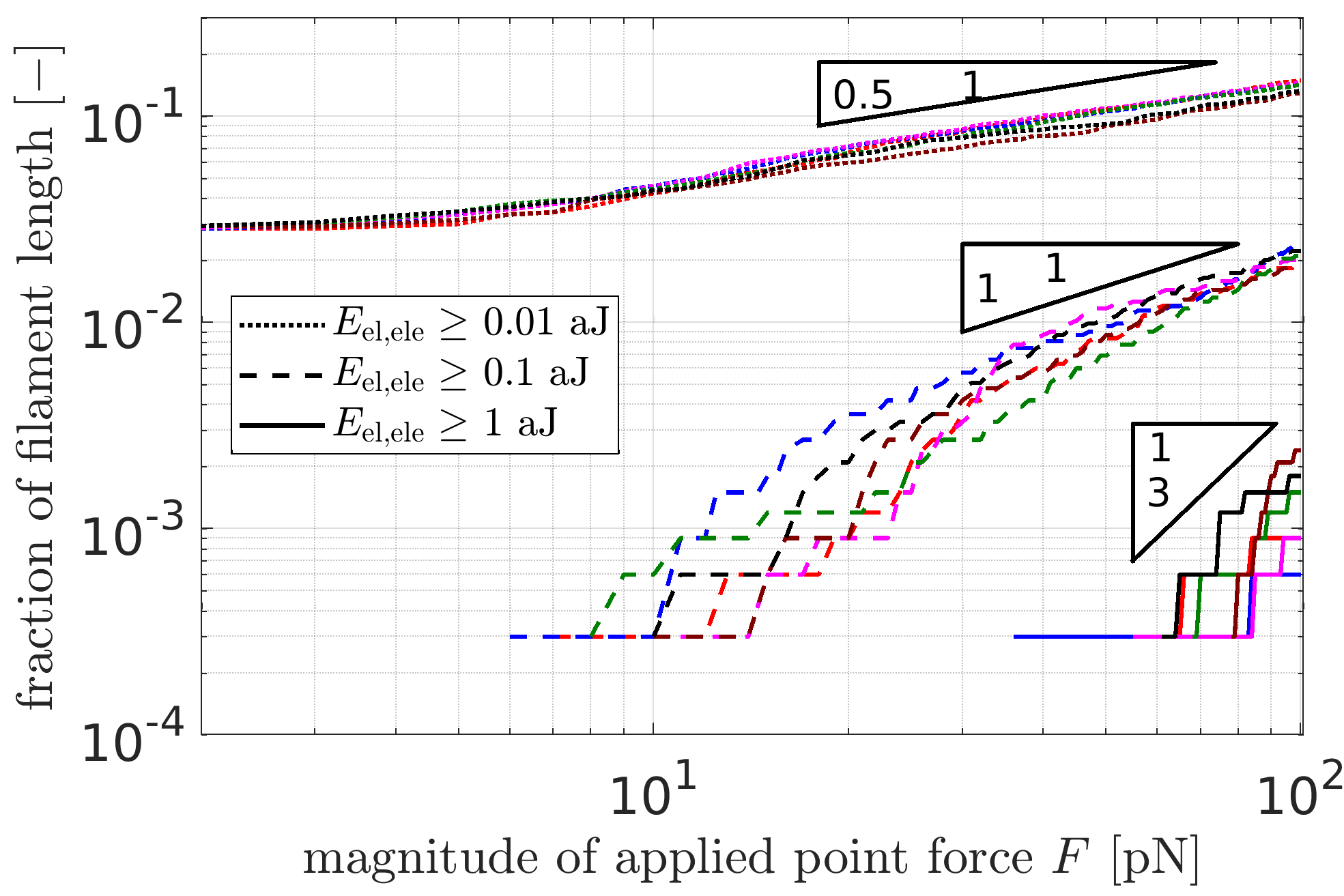}
    \end{minipage}%
  \end{minipage}\hspace{0.5cm}
  \begin{minipage}[t]{0.47\linewidth}%
    \begin{minipage}[t]{0.07\linewidth}%
      \vspace{-2.5cm}(E)
    \end{minipage}%
    \begin{minipage}[c]{0.95\linewidth}%
      \includegraphics[width=\linewidth]{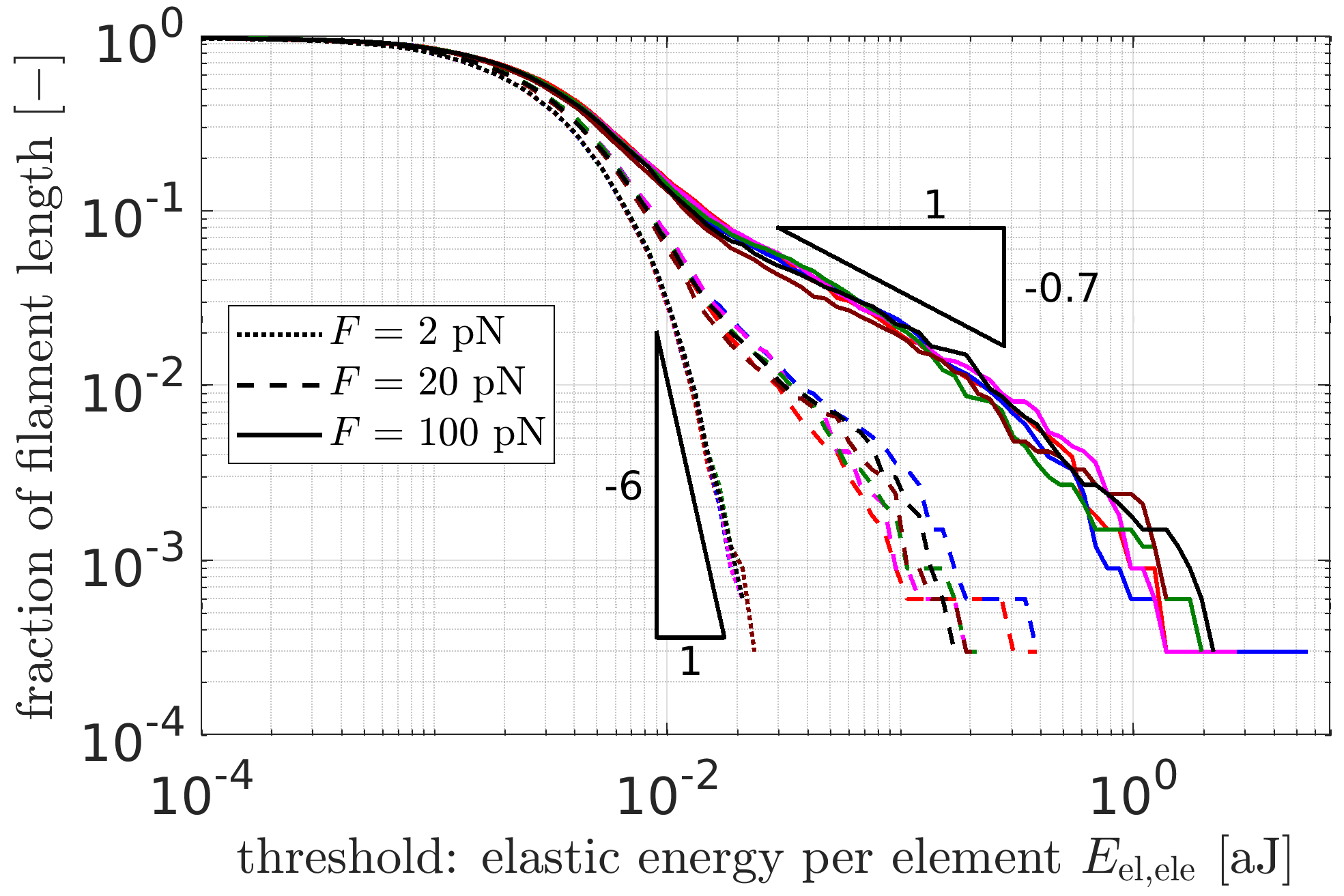}
    \end{minipage}%
  \end{minipage}
  \caption{(color online) Analysis of force chains.
           (A) Force chains resulting for an elastic energy threshold (per finite element) of $0.75 \mathrm{aJ}$ (red), $0.25 \mathrm{aJ}$ (blue), and $0.075 \mathrm{aJ}$ (green) among all other filaments in the network (gray, thin lines).
           (B) Top view hiding all other filaments for clarity and showing also all cross links with a force magnitude above $8 \mathrm{pN}$ (pink).
           (C) Overlay of three different pulling directions and the resulting force chains (for an elastic energy threshold of $0.1 \mathrm{aJ}$) in one color each.
           The simulation snapshots in (A)-(C) show the state for the highest considered point force magnitude~$F=100\mathrm{pN}$.
           (D) Fraction of filament length exceeding a certain elastic energy threshold (per finite element) over the magnitude of the applied point force for three different threshold values.
           Colors indicate 6 different pulling experiments.
           (E) Fraction of filament length exceeding a certain elastic energy threshold (per finite element) over the threshold value for three different force magnitudes.
           Again, colors indicate 6 different pulling experiments.}
  \label{fig::simulation_forcechains}
\end{figure*}

Fig.~\ref{fig::simulation_forcechains}(A) shows the concept of force chains to be used for the characterization of force
propagation in filament networks.
Filament segments with an elastic energy per finite element above $0.75 \mathrm{aJ}$, $0.25 \mathrm{aJ}$,
and $0.075 \mathrm{aJ}$ are highlighted in red, blue, and green respectively, while all other filaments are depicted as thin, gray lines.
Choosing and varying this threshold value allows one to track force transmission from the point of force application to the
support at the boundaries.
The resulting force-chain structures are typically connected, and span cone-shaped subregions of the network volume.
To get a clearer picture, all other untensed filaments are hidden in Fig.~\ref{fig::simulation_forcechains}(B).
Cross links supporting a force greater than a threshold of $8 \mathrm{pN}$ are also shown in pink to demonstrate that the tensile force chains
pass between filaments via such highly loaded cross links.   Finally, Fig.~\ref{fig::simulation_forcechains}(C) shows an overlay of three different pulling directions and the resulting force chains distinguished by color.
All the simulation snapshots shown in Fig.~\ref{fig::simulation_forcechains}(A)-(C) correspond to the highest point force magnitude~$F=100\mathrm{pN}$ used in the numerical experiments.

Since the threshold value is fundamentally arbitrary, it is helpful to look at changes of force-chain-related quantities as function
of  that threshold.  Fig.~\ref{fig::simulation_forcechains}(D) shows a double-logarithmic plot of the fraction of filament length making up the
force chains as a function of the applied point force magnitude for three different threshold values (dotted, dashed, and solid lines).
The colors indicate six pulling experiments with different point force directions.
The smallest threshold value $0.01 \mathrm{aJ}$ is chosen such that it is exceeded by (a small fraction of) filaments already for small point force magnitudes (dotted lines).
We observe that the corresponding filament length fraction
approximately increases with the square root of the point force magnitude in the high force regime.
A second threshold of $0.1 \mathrm{aJ}$ is exceeded only for intermediate to high point force magnitudes and the corresponding filament length fraction seems to increase linearly for high force values.
The highest threshold value of $1 \mathrm{aJ}$ is exceeded only for very high point force magnitudes and only in a very small fraction of the filament length.

Fig.~\ref{fig::simulation_forcechains}(E) gives the complementary picture, where the fraction of filament length is plotted
as a function of the threshold value for three different values of the applied point force magnitude (dotted, dashed, and solid lines).
Again, the colors indicate six pulling experiments with different point force directions.
Naturally, as the elastic energy threshold approaches zero, the entire network is activated, and so all curves collapse to unity.
For the smallest chosen force magnitude $2 \mathrm{pN}$, we observe a rapid decrease of the filament fraction decaying
approximately $\sim E^{-6}$ at large $E$.
This behavior changes for higher applied force magnitudes, where the fraction of filament length falls off much slower,
especially in the regime of intermediate threshold values.
For the largest threshold values $E$ observed for a given force magnitude, there seems to be a similar behavior with $\sim E^{-6}$ decay for all three force magnitudes considered here.

\subsubsection{Decay of axial tension along force chains}
%
\begin{figure*}[htpb]
  \centering
  \begin{minipage}[t]{\linewidth}%
    \begin{minipage}[t]{0.04\linewidth}%
      \vspace{-1.5cm}(A)
    \end{minipage}%
    \begin{minipage}[c]{0.945\linewidth}%
      \includegraphics[width=\linewidth]{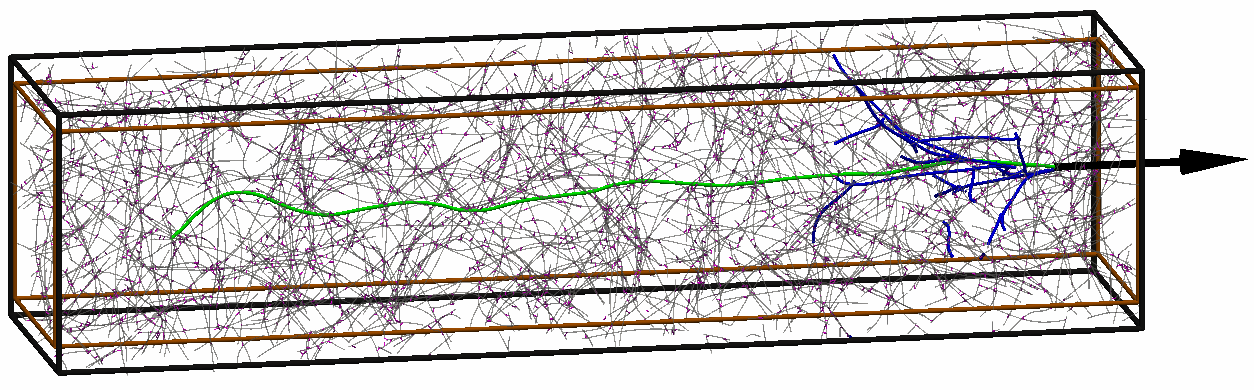}
    \end{minipage}%
  \end{minipage}\\\vspace{0.2cm}
  \begin{minipage}[c]{0.495\linewidth}%
    \begin{minipage}[t]{\linewidth}%
      \begin{minipage}[t]{0.06\linewidth}%
        \vspace{-1cm}(B)
      \end{minipage}%
      \begin{minipage}[c]{0.95\linewidth}%
        \includegraphics[width=\linewidth]{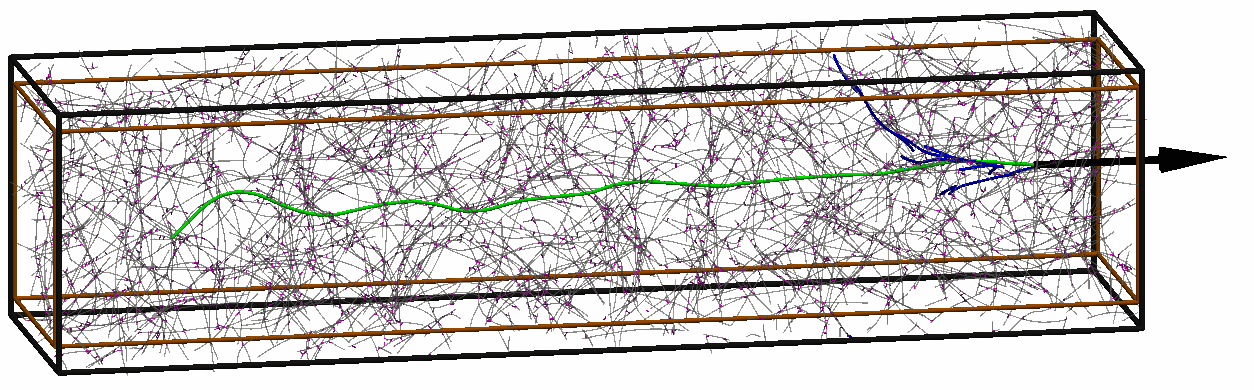}
      \end{minipage}%
    \end{minipage}\\\vspace{0.1cm}
    \begin{minipage}[t]{\linewidth}%
      \begin{minipage}[t]{0.06\linewidth}%
        \vspace{-1cm}(C)
      \end{minipage}%
      \begin{minipage}[c]{0.95\linewidth}%
        \includegraphics[width=\linewidth]{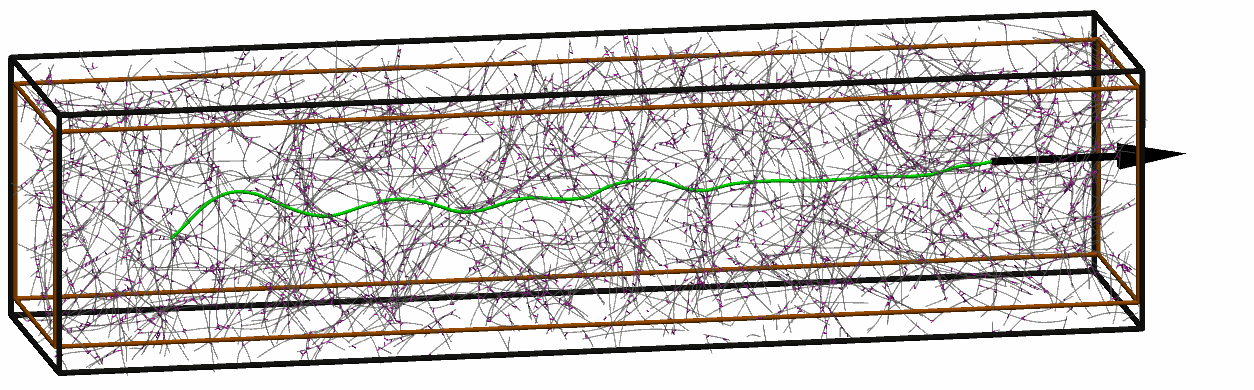}
      \end{minipage}%
    \end{minipage}%
  \end{minipage}%
  \hfill
  \begin{minipage}[c]{0.49\linewidth}%
    \begin{minipage}[t]{0.051\linewidth}%
      \vspace{-2.5cm}(D)
    \end{minipage}%
    \begin{minipage}[c]{0.95\linewidth}%
      \includegraphics[width=\linewidth]{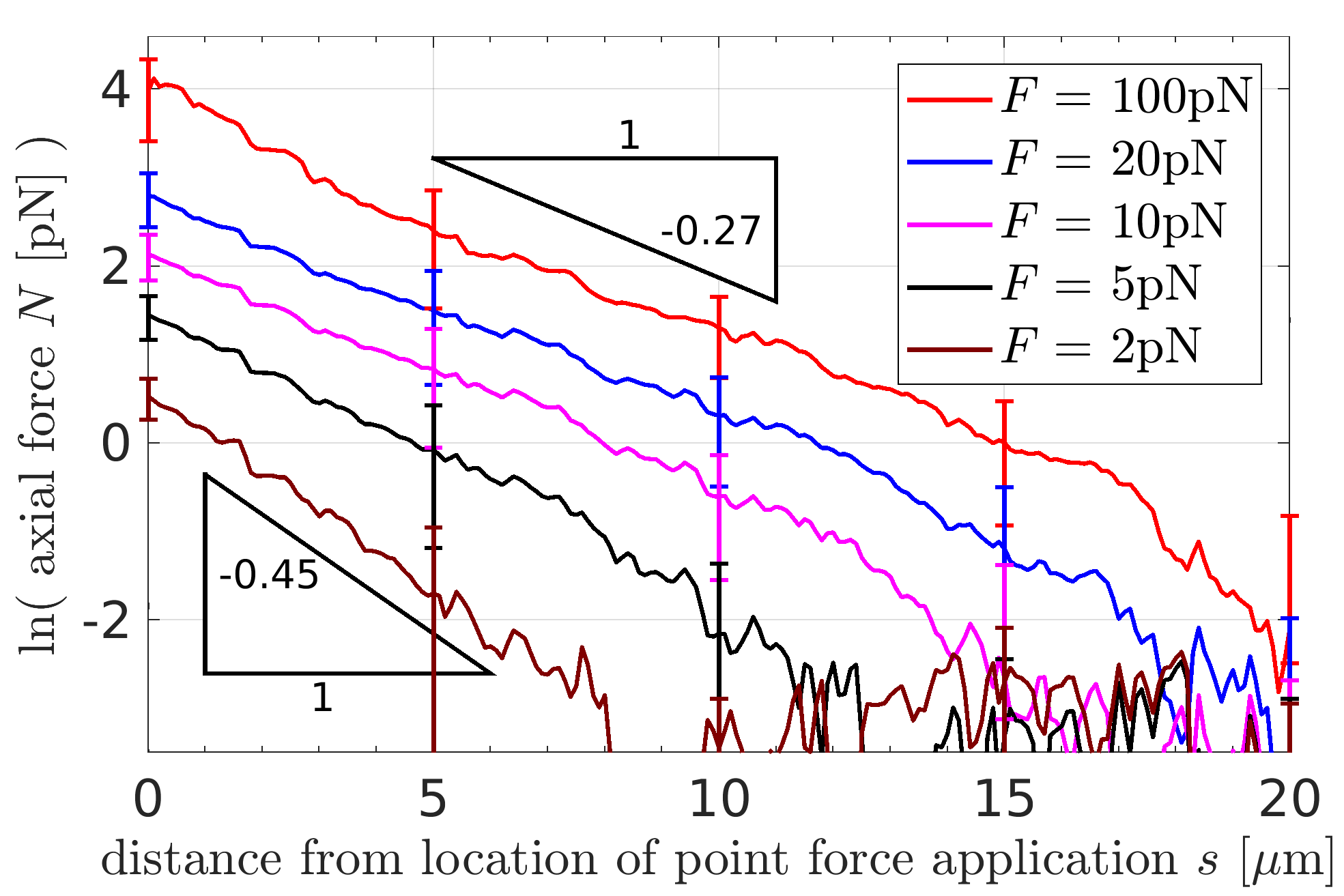}
    \end{minipage}%
  \end{minipage}\\\vspace{0.4cm}
  \begin{minipage}[t]{0.8\linewidth}%
    \begin{minipage}[t]{0.05\linewidth}%
      \vspace{-1.5cm}(E)
    \end{minipage}%
    \begin{minipage}[c]{0.95\linewidth}%
      \includegraphics[width=\linewidth]{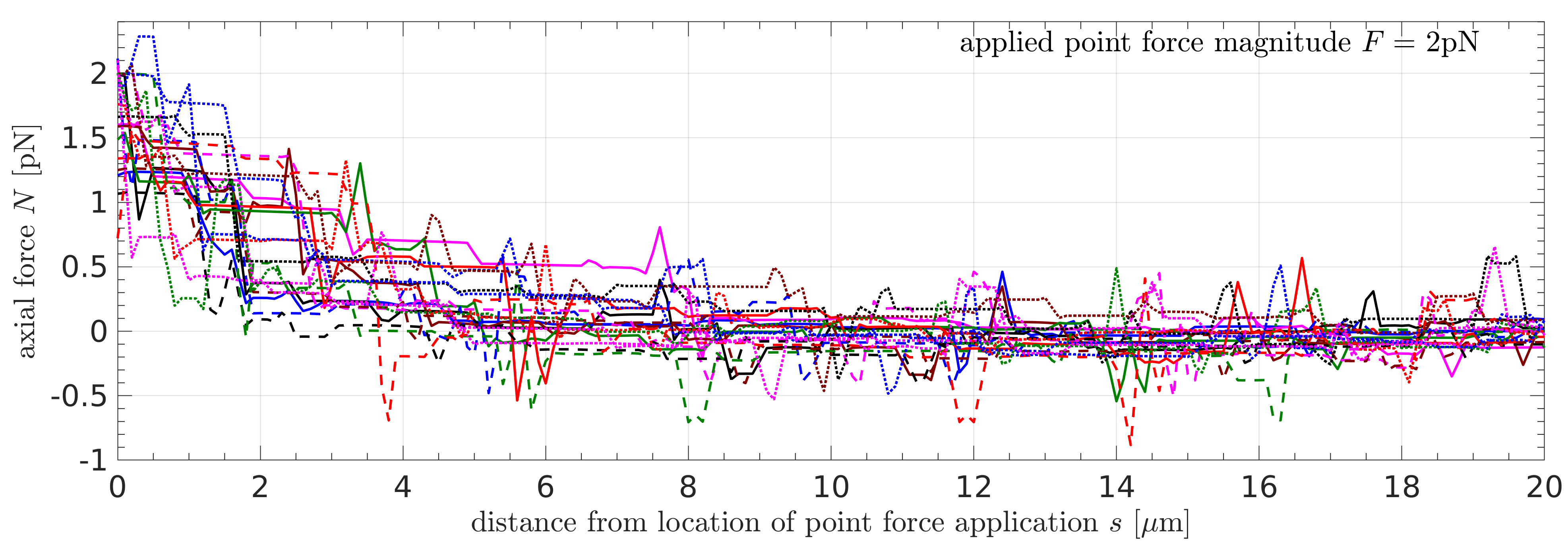}
    \end{minipage}%
  \end{minipage}\\
  \begin{minipage}[t]{0.8\linewidth}%
    \begin{minipage}[t]{0.05\linewidth}%
      \vspace{-1.5cm}(F)
    \end{minipage}%
    \begin{minipage}[c]{0.95\linewidth}%
      \includegraphics[width=\linewidth]{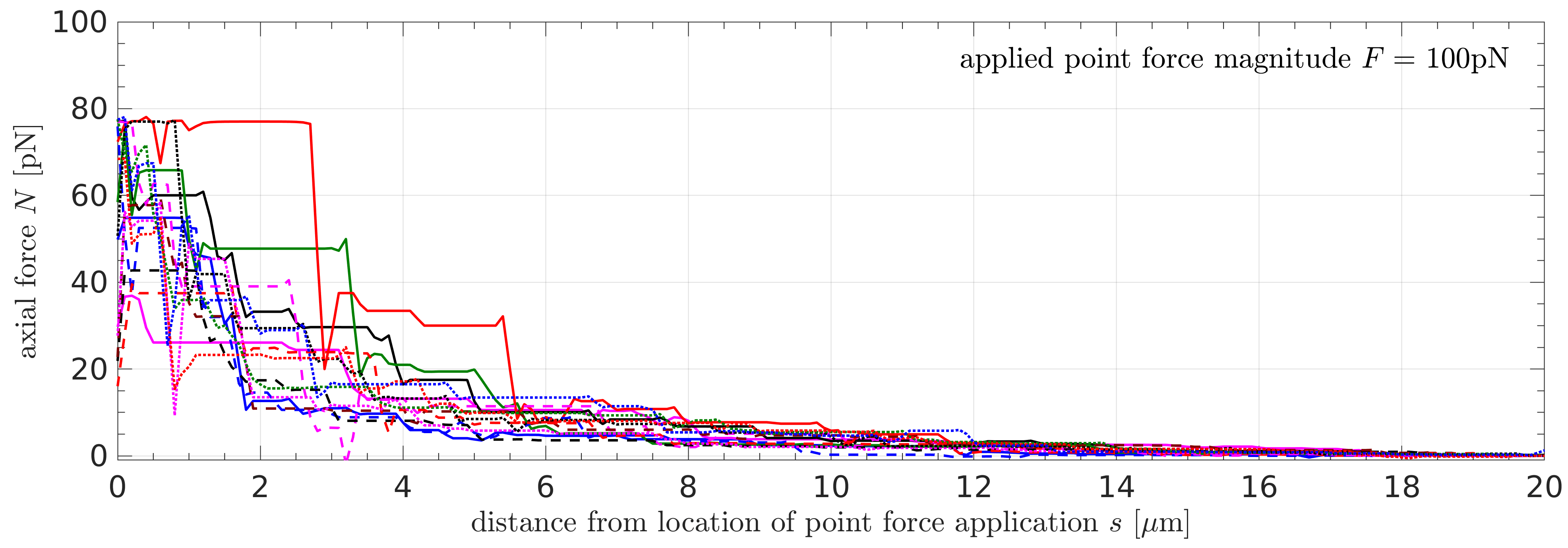}
    \end{minipage}%
  \end{minipage}%
  \caption{(color online) Simulation snapshots for an applied point force of magnitude (A) $F=100 \mathrm{pN}$, (B) $F=50 \mathrm{pN}$, or (C) $F=2 \mathrm{pN}$.
           The central, long filament is highlighted in green in the middle of the network of all other filaments (thinner, gray lines) and cross links (pink).
           All filaments in the force chains for an elastic energy threshold (per finite element) $E_\mathrm{el,ele}=0.1\mathrm{aJ}$ are highlighted in blue.
           The black arrow indicates the applied point force and the thin shell outside the brown box represents the region where filaments are pinned.
           (D) Semi-logarithmic plot of the axial force~$N$ along the central filament for 5 different point force magnitudes~$F$, using the average over 20 pulling experiments.
           The error bars indicate the standard deviation at five exemplarily chosen, equidistant points along the filament.
           The curves for all individual pulling experiments at a point force magnitude of~$F=2 \mathrm{pN}$ and $F=100 \mathrm{pN}$ are shown in (E) and (F), respectively.
           Each pulling experiment is indicated by a different combination of color and line style. Performing a linear fit to the 
           mean tension decay in (C), we obtain decay lengths 
           $\xi=2.24,2.99,3.55,3.80,\,$and $3.67 \mu\text{m}$, in order of increasing applied force.}
  \label{fig::simulation_box25x06x06}
\end{figure*}
For a detailed study of the decay of the axial force along the filaments and force chains, we use
a slightly modified setup, which is shown in Fig.~\ref{fig::simulation_box25x06x06}(A).
In addition to all the filaments with (initial, stress-free) length~$L_0=4\mathrm{\mu m}$ (gray), one
long filament with~$L_0=20\mathrm{\mu m}$ (green) is placed along the axis of an elongated
simulation box of size~$25 \times 6 \times 6 \mathrm{\mu m}$.
The central filament is discretized with 100 beam finite elements to ensure a fine spatial discretization.
Instead of the filament midpoint, the point force is now applied to one of the filament
endpoints and we only consider tangential tensile loading.
All other parts of the setup and protocol of the numerical experiments as described in
Sec.~\ref{sec::numerical_experiments} and \ref{sec::numerical_model} remain unchanged.
In particular, once again 10 different network geometries have been generated by
random initial placement of the straight filaments, simulating the dynamic assembly driven by
Brownian motion, and a subsequent equilibration simulation.

Fig.~\ref{fig::simulation_box25x06x06}(A), (B), and (C) show simulation snapshots of the same
pulling experiment at three different point force magnitudes of $F=100 \mathrm{pN}$, $F=50 \mathrm{pN}$, and $F=2 \mathrm{pN}$, respectively.
All filaments in the force chains for an elastic energy threshold (per finite element) $E_\mathrm{el,ele}=0.1\mathrm{aJ}$ are highlighted in blue.
This already reveals that the perturbation of the network in form of the applied point force is transmitted
along a few paths of cross-linked filaments in the vicinity of the location of force application.
From the persistent wavy form of the left half of the central filament even for very high point
force magnitudes, it becomes obvious that the perturbation is absorbed quite rapidly.
Note in this respect that the force chains reaching from the central filament to the pinned
boundaries typically include a couple of cross links and thus different filaments such that the
pathological edge case of one filament reaching from the central filament to the pinned
boundary is typically not observed in this simulation setup.

For a quantitative analysis of the tension decay along the filaments, we look at the axial force along the central filament.
The semi-logarithmic plot in Fig.~\ref{fig::simulation_box25x06x06}(D) shows the mean
values over all twenty numerical pulling experiments obtained for 5 different point force magnitudes.
In addition, the data for all twenty individual realizations at the lowest ($F=2 \mathrm{pN}$) and highest ($F=100 \mathrm{pN}$)
force magnitude is plotted (with linear scale on the vertical axis) in Fig.~\ref{fig::simulation_box25x06x06}(E) and (F), respectively.
Most importantly, Fig.~\ref{fig::simulation_box25x06x06}(D) reveals an approximately exponential decay of the
average axial tension with increasing distance from the location of force application -- more or less irrespective
of the applied force magnitude.
Second, the slope of the curves and therefore also the characteristic decay length, varies slightly over the five
different applied force magnitudes with the fastest decay being observed for the smallest force magnitude.
Finally, from looking at the data for each individual realization in Fig.~\ref{fig::simulation_box25x06x06}(E) and (F), a
characteristic step-wise decay behavior is revealed.
This can be explained by the fact that tensile force is transmitted to
other filaments at the (randomly distributed) discrete locations of cross links along the filament axis.
To conclude, this analysis confirms the rapid
absorption of tension along the filaments in a quantitative manner, and reveals an exponential decay,
which is consistent with the theoretical calculations presented in the following section.

\section{Analytical models}
We adopt a Mikado model of the semiflexible filament network~\cite{head2003distinct,head2003deformation,head2005mechanical}.
Inextensible filaments are placed one by one in a cube of volume $L^3$, truncating at the boundaries. Intersecting
filaments are then cross-linked together, creating a collection of filament {\it segments} that form the network.  The lower coordination number of our model differs from those of
previous studies on force chains on lattices~\cite{liu1995force,coppersmith1996model}.  We study this model in two dimensions.  We believe that this choice is
not important for our analysis of  force balance at a given cross-linked node in the network. Dimensionality is important when discussing the isostaticity condition
in the network~\cite{Souslov2009,Mao2010,kane2014topological}. It is also likely to be important in its effect on how branching force chains
interact with each other.  In three dimensions, we expect them to more rarely intersect than in two.
We briefly comment on these points in the conclusions.
\begin{figure}[htbp]
\centering
\includegraphics[width=0.9\linewidth]{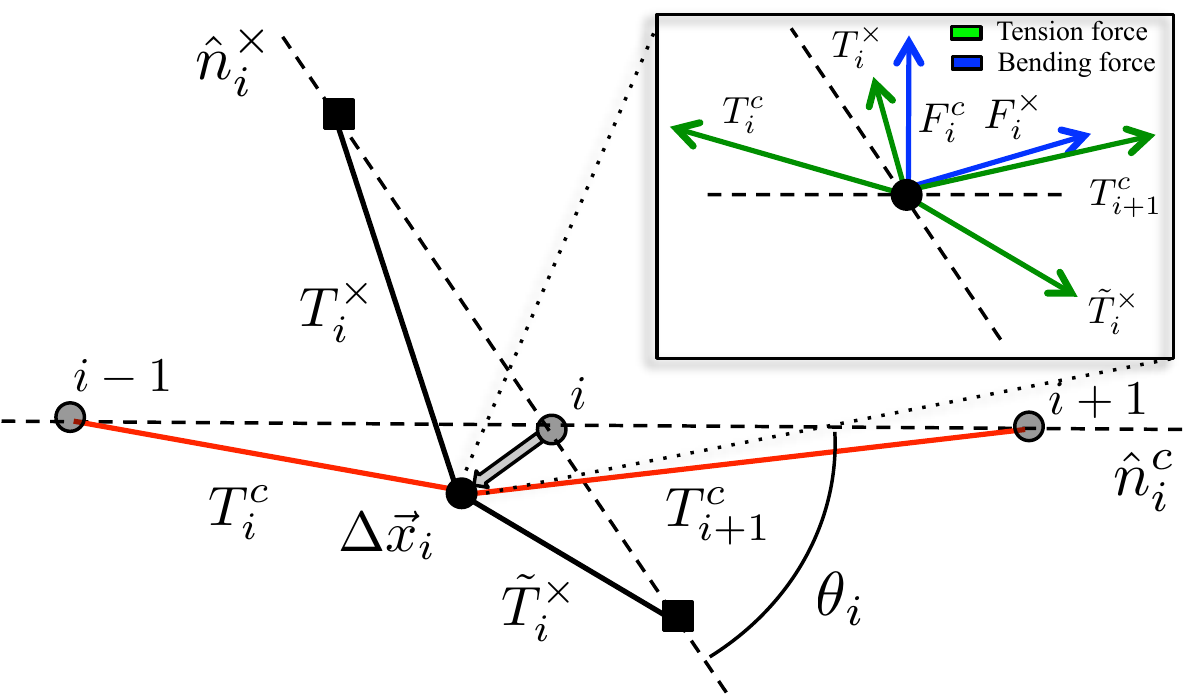}
\caption{(color online) schematic representation of forces and displacements at a single node.
Top: The central filament, pointing in the $\hat n_i^c$ direction, is aligned along the x-axis. The crossing filament, in the direction $\hat n_i^\times$ makes an angle $\theta_i$ with the central filament. The network is assumed pulled in the $-\hat x$ direction, leading to two incoming tensions $T_i^c$ and $T_i^\times$, and two outgoing tensions $T_{i+1}^c,\,\tilde T_i^\times$ at a node. In the self-consistent model of sec.~\ref{sec:self-consistent-theory}, displacements (wide gray arrow) $\Delta \vec x_i$ are with respect to the undeformed state. In the accordion model of sec.~\ref{sec: accordion}, $\Delta \vec x_i$ is relative to neighboring nodes -- see Fig.~\ref{fig: accordion displacement} for further details. In the inset we show the free-body diagram for forces at node $x_i$. Small displacements are subject to a linear restoring force, modeled by a collection of springs, leading to two bending forces (blue arrows), and four tensions (green arrows).}
\label{fig: FBD}
\end{figure}

The mechanics of each filament is controlled by its bending rigidity $\kappa$ and its longitudinal modulus $\mu$.  In thermalized networks of
effectively inextensible filaments, these two elastic constants are actually connected.  The filaments bend and therefore contract under thermal
fluctuations, leading to an entropic  effective longitudinal spring constant $k_{\rm entropic} \sim \kappa^2/(k_\text{B} T \ell^3)~\cite{mackintosh1995}$.
Here we treat the longitudinal compliance $\mu$ as a simple phenomenological constant.  For a large number of regular filament networks, one can
compute the collective elastic constants in terms of these filament-level elastic quantities~\cite{mao2013effective,mao2013elasticity}, as well as
for random elastic gels~\cite{mackintosh1995}.

The Hamiltonian density for a filament segment (directed along $\hat x$ axis) spanning two cross linked sites on that filament is given by
\beq
\label{eq:H}
\frac{\delta \mathcal{H}}{\delta s} = \frac{\mu}{2} \frac{\delta \ell}{\delta s} + \frac{\kappa}{2} \frac{\delta \theta}{\delta s},
\eeq
where $\delta \ell(s)$ is extensional deformation of the filament as a function of arclength $s$, and $\delta \theta(s)$ the change in angle of
local filament tangent with respect to the $x$-axis. The elastic moduli $\mu$ and $\kappa$ set a length scale
\beq
\label{eq: l bend}
\ell_\text{bend} = \sqrt{\kappa/\mu},
\eeq
governing the competition between bending and stretching in the network. The length $\ell_\text{bend}$ sets a
tension absorption length, to be confirmed later.  Incidentally, $\ell_\text{bend}$ also represents the crossover length for
ensemble-averaged semiflexible networks to shift from non-affine to affine elasticity~\cite{head2003deformation,head2005mechanical}.
The lowest energy modes of the Mikado model involve filament bending.  In one limit where this bending energy is taken to be zero, these
result in so-called {\em floppy modes}~\cite{Heussinger2006}, or zero-energy deformations of the network.

Force propagation within the network depends strongly on the boundary conditions imposed at the ends of the filaments.
To illustrate this point, consider a situation where, in the undeformed state, the
central filament (along the $x$ axis) is crossed by a number of filaments normal to that central filament. If
one were to pull on the left end of the central inextensible filament, while leaving the right end free, one would expect that the displacement of the network
would resemble a sort of ``bow and arrow'' configuration, in which each crossing filament bends and that bending transfers the tensile loading
on the central filament to these crossing ones.  The displacements of each of these cross-linked points
would be equal but nonzero.  On the other hand,
if one were to pin the right end of the central filament, these displacement would all vanish and the tensile load would be perfectly transmitted along the
central filament. Another way that the connection to the boundary can play a role is in the excitation of topologically protected surface modes of the
network~\cite{zhou2018}.  We do not consider such surface states here.  In our calculations, we assume that the filament is not directly pinned to the
boundary, except where explicitly noted. As long as the tension applied to the filament in question has been transferred to the rest of the network before
coupling to the boundary, we expect that the effect of the boundary condition should be small.

\subsection{Self-consistent theory of tension propagation within linear response}
\label{sec:self-consistent-theory}

We begin by treating the filaments as linear elastic elements, but accounting for their different responses to longitudinal and bending deformation.
The bending rigidity of the filaments eliminates the {\em floppy modes} of the network.  We now apply a tension
$\tau$ along a particular filament within the network at an arbitrary point of the filament.  The point where this force is applied is displaced by
$\Delta x$ along the $\hat{x}$ axis.  In the following, we use a self-consistent approach to compute the collective linear response of the network to this force by
computing
\begin{equation}
\label{k-parallel-def}
k_\parallel = \frac{ \tau}{\Delta x}.
\end{equation}
The self-consistency condition is invoked by demanding that the effective spring constant of the particular filament to which we apply the
force, called the  \textit{central} filament hereafter, is equal to the one of all other filaments cross-linked to it.  The validity of this
approximation rests on the assumption
that the tension in the crossing filaments are all selected from the same distribution, and that there are no correlations in those tensions.  The former
seems reasonable for a statistically homogeneous network. The latter is not obviously valid, especially if there are a large number of closed paths, or
loops, in the network along which tensions may propagate.  Thus, we do not
expect the self-consistent approach to remain valid for the case of regular lattices, where
such loops abound.

Now, we consider force balance at one node on the central filament labeled $i$ -- see Fig.~\ref{fig: FBD}.
As shown in the inset of that figure, there are six forces acting on the node on the central filament, shown in blue and green.  
We treat these forces as being
linearly related to the displacement of the nodes.  Specifically, the longitudinal springs associated with the extension of the crossing
filaments (shown in black) have an effective Hookean spring constant $k_{\parallel}$, which takes into account both the longitudinal
compliance of that filament and the displacement of other nodes in the network.  This spring constant will be determined self-consistently in
the following calculation.  This Hookean spring generates the forces  $T^\times_i$ and $\tilde T^\times_i $ shown in green in the 
inset of Fig.~\ref{fig: FBD}.

In addition to this force, there is a force associated with the bending of the filament crossing the central one at node $i$.  This spring constant,
$k_{\perp} \sim \kappa / \ell_c^3$, is proportional to the bending modulus of the filaments and inversely proportional to the distance between consecutive
cross links along a given filament, $\ell_c$~\cite{Frey:03}.   The movement of
the neighboring nodes can be taken into account by a diminishing of this constant: $k_\perp\rightarrow  \epsilon k_\perp$ for $ 0 < \epsilon < 1$.
This bending spring generates the force $F_i^\times$ (shown in blue).  The central filament
may also bend,  generating a displacement of node $i$ in the vertical direction $ \Delta y_i $. Using the same $k_\perp$,
this produces the force $F_i^c = k_\perp  \Delta y_i $.  We expect that the bending modulus of the central filament will not be
affected by the motion of the surrounding nodes, so that $k_\perp >\epsilon k_\perp$ is fixed.

We now apply tension $T_{i}^c$ to the filament segment to the left of the node and in the left direction.
Writing down the force-balance conditions along the $X$ and $Y$ axes, we obtain:
\begin{eqnarray}
-T_i^c + T_{i + 1}^c  + (\tilde T_i^\times - T_i^\times)  \cos \theta_i + F_i^\times \sin \theta_i = 0,
\\
(T_i^\times - \tilde T_i^\times) \sin \theta_i - F_i^\times \cos \theta_i - F_i^c = 0.
\end{eqnarray}
The forces, written in terms of displacements of node $i$, are given by:
\begin{eqnarray}
F_i^c = - k_\perp \Delta y_i,
\\
\tilde T_i^\times - T_i^\times   = - k_\parallel ( \Delta x_i  \cos \theta_i - \Delta y_i \sin \theta_i ),
\\
 F_i^\times = - \epsilon k_\perp ( \Delta x_i \sin \theta_i + \Delta y_i \cos \theta_i ).
\end{eqnarray}
After some algebra we obtain
\begin{equation}
 T_{i}^c - T_{i+1}^c     =   \frac{   (\epsilon   k_\parallel k_\perp    + ( k_\parallel   \cos^2 \theta_i +  \epsilon k_\perp \sin^2 \theta_i)  k_\perp     }{( k_\parallel \sin^2 \theta_i     + \epsilon k_\perp  \cos^2 \theta_i +  k_\perp )  } \Delta x_i.
\label{eq:force_balance}
\end{equation}

To connect the displacement of the $i^{\rm th}$ and $(i+1)^{\rm th}$ nodes (counting to the left),  we include the Hookean extensibility of
the central filament segment between these nodes to write
\begin{equation}
T_i^c = k_s( \Delta x_{i-1} - \Delta x_{i } ).
\label{eq:Hookes}
\end{equation}
For many biopolymers, at low forces the longitudinal compliance is dominated by the pulling out of thermally generated undulatory modes. In
that case, the Hookean spring constant introduced above in Eq.~\ref{eq:Hookes} can be related to the bending modulus
and temperature via $k_s = 6 \kappa^2  / k_{\rm B}T \ell_c^4 $~\cite{mackintosh1995}.
However, nothing in the following analysis requires this, and the calculation applies equally well to athermal systems.  We can solve Eqs.~\ref{eq:force_balance} and \ref{eq:Hookes}
numerically to find the self-consistent solution for $k_\parallel(k_s)$.

\begin{figure}
 \includegraphics[width=0.96\linewidth]{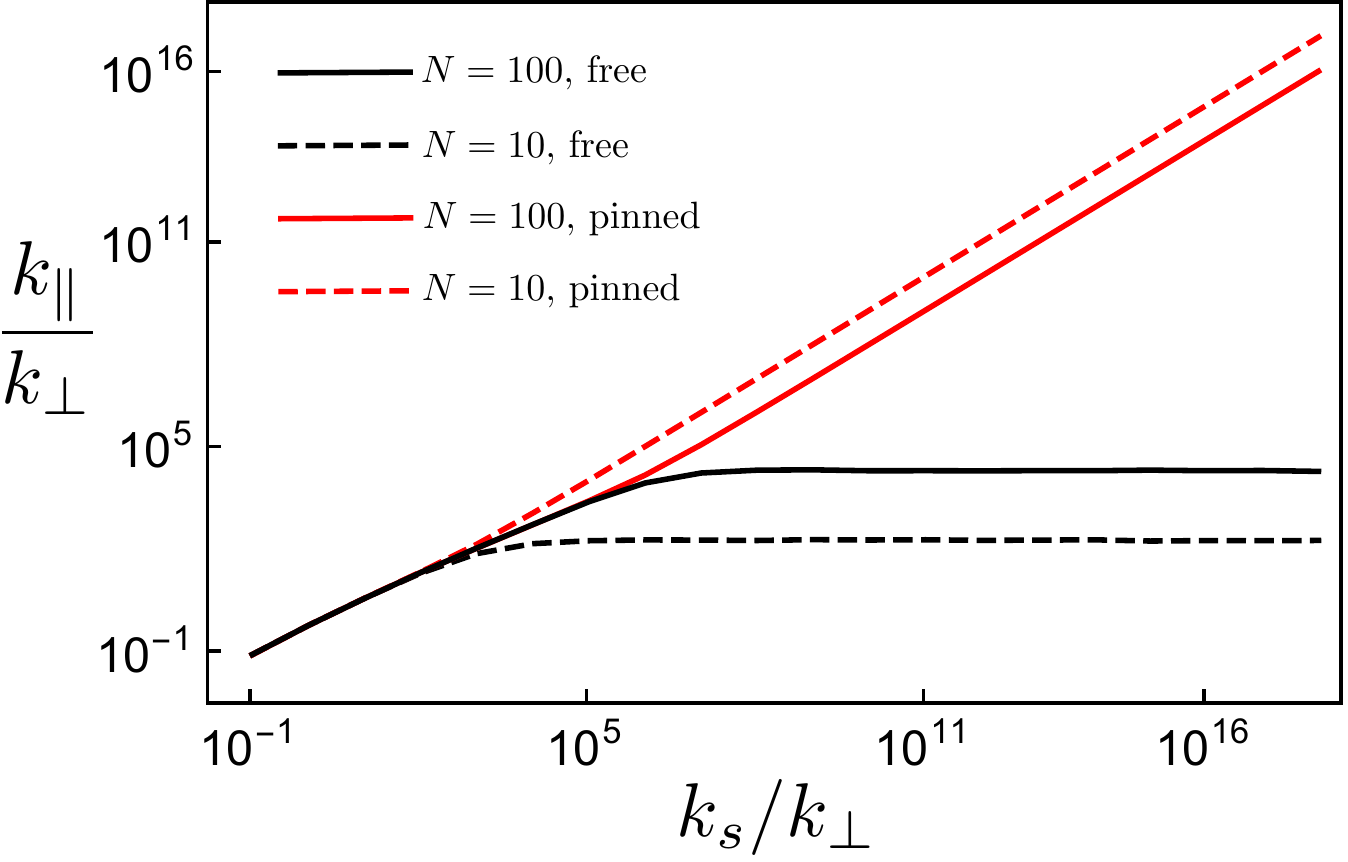}
 \caption{(color online) Numerical solution of the self-consistent Eqs.~\ref{eq:force_balance} and~\ref{eq:Hookes} for both free (black) and pinned (red) boundary conditions, for a total number of crossing filaments $N = 100$ (solid curves) and $N = 10 $ (dashed curves). We set $\epsilon = 1 $. For small longitudinal spring constants 
 $k_s$ the behavior of $k_\parallel$ is independent of the number of cross links $N$ and the boundary conditions. The transition to 
 a regime where boundary condition affect the result occurs at higher $k_{s}$ for larger $N$.}
 \label{fgr:kpar-ks}
\end{figure}

This solution is shown in Fig.~\ref{fgr:kpar-ks}, where we plot
the effective longitudinal response $k_{\parallel}$ of the network as a function of the segment
longitudinal spring for different numbers of cross links $N$ and for pinned (solid) and free (dashed) boundary conditions.  
Both spring constants are scaled by the underlying bending spring constant of the network $k_{\perp}$.  On the log-log
plot, we observe that for small longitudinal spring constants, the collective longitudinal spring constant of the network grows 
$\sim k_{s}^{3/4}$.  In this region the behavior of $k_\parallel$ does not depend on the number of cross links or 
the boundary conditions on the filament. For larger $k_{s}$ we observe the transition of $k_\parallel$  to a plateau for free end boundary 
condition. There is  linear growth in the case of a pinned end boundary condition. This transition occurs at a
characteristic spring constant that depends on the number $N$ of crossing filaments, as shown in 
Fig.~\ref{fgr:kpar-ks}. 

In addition to this numerical solution of the self-consistent equations, there are two particular cases that admit a straight-forward analytic solution.
It is instructive to look at them to directly observe in detail how tension propagates along the central filament.  The first case applies to
perfectly inextensible filaments, $k_s \rightarrow \infty$. The second is a scale-free solution, {\em i.e.}, one that is independent of the total number
of filaments cross linked to the central one, $N$.  We refer to this solution as a critical point.  The case of the inextensible filament
leads to a constant $\Delta x_i = \Delta x $ for all $i$. Summing Eq.~\ref{eq:force_balance} for all $i$ we obtain
\begin{equation}
 T_{0}^c - T_{N }^c     =   \Delta x \sum_{i=1}^N \frac{   (   \epsilon k_\parallel k_\perp    + ( k_\parallel   \cos^2 \theta_i + \epsilon  k_\perp \sin^2 \theta_i)  k_\perp     }{( k_\parallel \sin^2 \theta_i     + \epsilon k_\perp  \cos^2 \theta_i +  k_\perp )  }.
\label{eq:force_balance2}
\end{equation}
Imposing self-consistency for $k_\parallel$ forces that spring constant to satisfy
\begin{equation}
 k_\parallel     =   \left \langle \sum_{i=1}^N \frac{   (   \epsilon k_\parallel k_\perp    + ( k_\parallel   \cos^2 \theta_i +  \epsilon k_\perp \sin^2 \theta_i)  k_\perp     }{( k_\parallel \sin^2 \theta_i     + \epsilon k_\perp  \cos^2 \theta_i +  k_\perp )  }  \right \rangle.
\label{eq:force_balance3}
\end{equation}
Note  that the random angle $\theta_i$ at each node is assumed to be uncorrelated with the other angles , so each average is independent of the
others. Thus, we find
\begin{equation}
k_\parallel  = \frac{N}{2 \pi} \int_0^{2 \pi} d \theta \frac{\epsilon k_{\perp} k_{\parallel} + k_{\perp} (k_{\parallel} \cos^2 \theta + \epsilon k_{\perp} \sin^2 \theta ) }{k_{\parallel} \sin^2 \theta +\epsilon  k_{\perp} \cos^2 \theta + k_{\perp} }.
\end{equation}

After performing the integral and solving the self-consistent equation we obtain
\begin{eqnarray}
k_{\parallel } &=& \frac{1}{2}\left(-  2 k_{\perp} N + k_{\perp} N^2   +  \epsilon  k_{\perp} N^2\right) \\ \nonumber
&+& \frac{1}{2}\left(\sqrt{( 2 k_{\perp} N - k_{\perp} N^2   - \epsilon  k_{\perp} N^2  )^2  + 4 N^2 \epsilon k_{\perp} k_{\perp} } \right).
\end{eqnarray}
In the limit of large $N$, this simplifies to
\begin{equation}
k_{\parallel } =  ( k_{\perp}    + \epsilon   k_{\perp} ) N^2.
\end{equation}
The effective spring constant is proportional to $N^2$.

The other case that allows for an analytical solution results from the assumption that the displacements decay exponentially from
node to node,  {\em i.e.}, $\Delta x_{i+1} = (1/q)  \Delta x_{i} $ for all $i$ and for $q>1$. Substituting this ansatz into
Eqs.~\ref{eq:force_balance},~\ref{eq:Hookes} gives, after some algebra,
\begin{equation}
\left(q  - 1 \right) -  \left(1  - \frac{1}{q} \right)  =\frac{1}{k_s}  \frac{ \epsilon k_{\perp} k_{\parallel} + k_{\perp} (k_{\parallel} \cos^2 \theta_i + \epsilon k_{\perp} \sin^2 \theta_i ) }{k_{\parallel} \sin^2 \theta_i +\epsilon  k_{\perp} \cos^2 \theta_i + k_{\perp} }.
\end{equation}
This is consistent with our assumptions if, and only if, the right hand side of the above equation is also independent of $i$. This requires that the
dependence on the random angles $\theta_i$ vanishes. The necessary condition for this is $\epsilon k_\perp = k_\parallel$, which leads to:
\begin{equation}
q^2 - \left( 2 + \epsilon \frac{k_\perp}{k_s} \right) q + 1 = 0.
\end{equation}
The solution of this quadratic equation defines the allowed values of  $q$ consistent with our assumption of an exponential decay of displacements
and the spring constant relation $k_\parallel = \epsilon k_\perp $. Depending whether we apply tension (a) between nodes or (b) directly on the node,
we get (a) $k_\parallel = T_{0}^c / \Delta x_{ 1}  $, or (b)  $k_\parallel = T_{ 1}^c / \Delta x_N $. Thus, in the first case (a)
\begin{equation}
k_\parallel^{(a)} =   k_s \left(1  - \frac{1}{q} \right),
\end{equation}
and in the second case (b)
\begin{equation}
k_\parallel^{(b)} =   k_s(q  - 1 ).
\end{equation}
We obtain two equations for roots of the polynomial $q_{1,2}$. One for case (a)
\begin{equation}
q^2_a -\left[2 + \left(1  - \frac{1}{q_a} \right)  \right] q_a + 1 = 0,
\end{equation}
and one for case (b)
\begin{equation}
q^2_b -\left(2 + q_b  - 1 \right) q_b + 1 = 0.
\end{equation}
The first (a) gives us the roots
\begin{equation}
q_a = 1,2.
\end{equation}
The second (b) gives us a linear equation for $q_{b}$, having the single root
\begin{equation}
q_b  = 1.
\end{equation}
These roots are now independent of the number of crossing filaments, which means that the decay of tension and displacement along the
central filament is independent of filament length at this critical point.

The solution $q = 1$ corresponds to the previously considered case of inextensible filaments, which have equal displacements at each node. The
other solution, $q = 2$, is consistent with the assumed exponential decay of those displacements. At every node, the magnitude of the tension
falls by a factor of 2, i.e., $T_{i+1}^c =  T_i^c / 2$. Substituting this tension relation into the force balance equation for the $\hat x$ 
direction, we obtain
\begin{equation}
   (\tilde T_i^\times - T_i^\times)  \cos \theta_i + F_i^\times \sin \theta_i = T_i^c/2,
\end{equation}
which means that the tensile force exactly splits between the crossing filament and the central filament at each node.
Using the mean-field approximation, we can infer that the displacement of node $i-1$ on the central filament is equal to the displacement of
the other nodes on the  $i^{\rm th}$
crossing filament that neighbor the central one. This is due to the fact that both of these nodes have half of the tensile load $T_{i+1}$.
By continuity, we assume this division of the tension and  displacement is, at least approximately, true when the system is not precisely at this
critical point $k_s = 2 \epsilon k_\perp $.  This is supported by our numerical results. This observation leads us to
 introduce what we call the accordion approximation, which assumes this relation between displacements remains exactly valid.  We use this
 approximation to study the problem of force propagation in the vicinity of the critical point in the following section.

\subsection{Accordion model}
\label{sec: accordion}
We treat the filaments as identical linear elastic elements. There are two spring constants, which should be derivable from the
derived from the longitudinal modulus and bending moduli~\cite{mao2013effective} of the filaments and the where $\ell_c$ is the average
spacing between consecutive crosslinks along the same filament.  In terms of these quantities, one finds
\beq
\label{eq: k defs}
k_\parallel' = \mu/\ell_c, \qquad k_\perp' = \kappa/\ell_c^3.
\eeq
This result may not hold for our zero temperature simulations of bent filaments.  We return to this point in the conclusions where we
quantitatively compare our theoretical predictions with our numerical results.
The longitudinal and transverse springs ($k'_\parallel$ and $k'_\perp$) attached to each node are
located along the blue and green arrows respectively in the inset of Fig.~\ref{fig: FBD}.

\begin{figure}[htbp]
\centering
\includegraphics[width=0.9\linewidth]{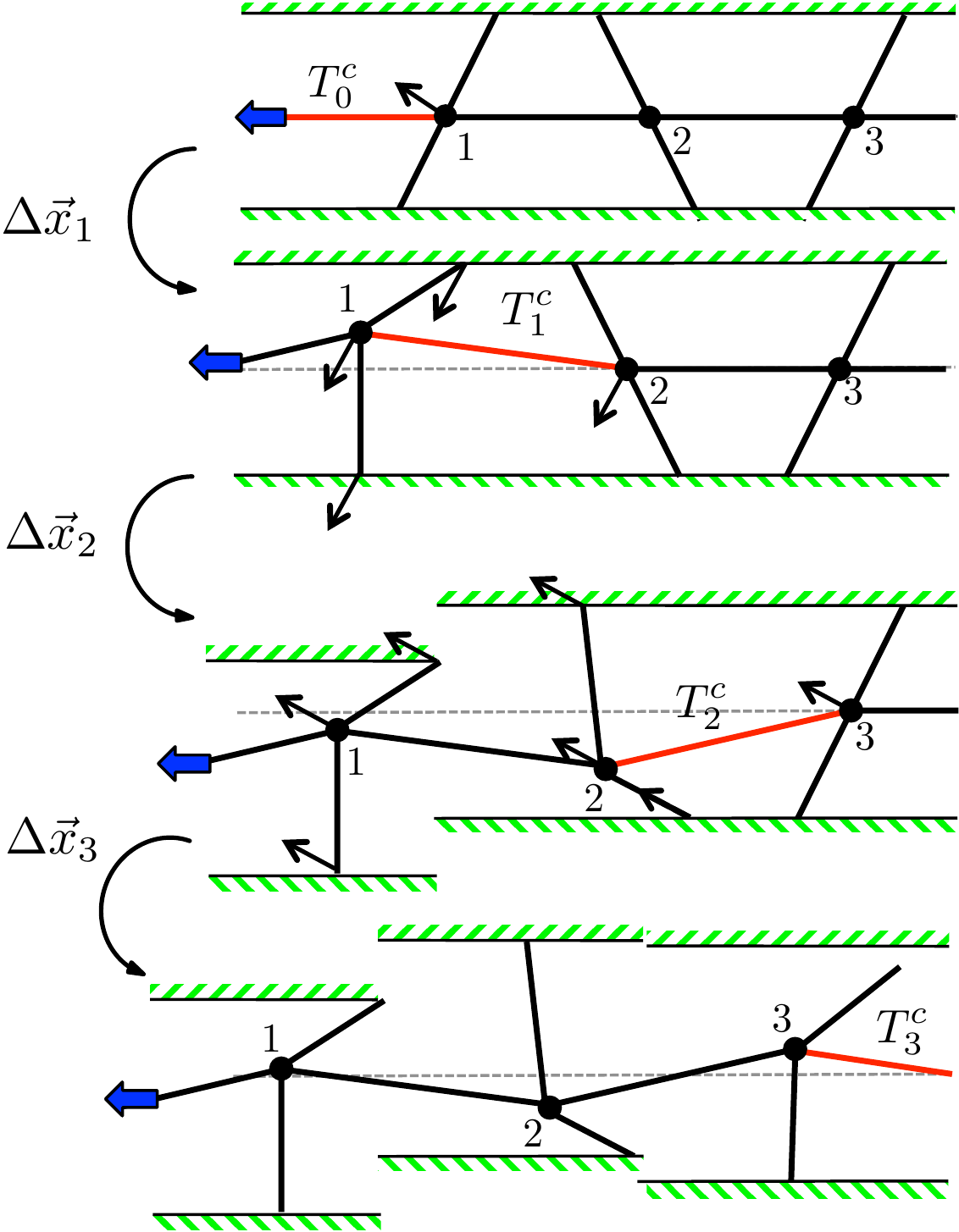}
\caption{(color online) Schematic of the accordion model deformation. The network is initially undeformed, and oriented with the central filament (solid black line) running horizontal. For reference, the position of the central filament in the undeformed network is shown via dashed, gray line. Hashed green boundaries represent the network in the transverse direction, black circles represent crosslinks, and solid black lines represent filament segments. Filaments under unbalanced tension are shown in red. In the top panel, a tension $T_0$ (blue thick arrow) is applied at the leftmost filament segment causing force imbalance on the first node. This node is allowed to come to mechanical equilibrium by displacing an amount $\Delta \vec x_1$. This puts a tension $T_1$ on its neighboring filament segment. Node two is allowed to come to mechanical equilibrium by simultaneously displacing itself, node one, node one's crossing filament segment, and the network transverse to node one, by an amount $\Delta \vec x_2$. This procedure maintains force balance between node one and its crossing filament, while putting filament segment three under tension $T_3$. The displacement procedure is iterated until we reach the pinned boundary.}
\label{fig: accordion displacement}
\end{figure}

We consider the pulling scheme depicted in Fig.~\ref{fig: accordion displacement}, where at the far left end of the central filament
we apply a leftward tension $T_{0}^c$ (wide blue arrow). The far right end of the filament is pinned to the anchored face. The central filament is oriented along the
$\hat x$-axis. We also fix the lateral extent of the network in the direction transverse to the central filament.

Pulling the central filament creates a force imbalance on the first node, located at $\vec x_1$. This node moves in order to reestablish mechanical
equilibrium. For the moment, we assume that all nodes to the right of this node are stationary.  The
movement of node one has now, in turn, created a force imbalance on its neighboring node located at $\vec x_2$. In
keeping with the accordion approximation, we move all points in the
network to the left of node two and its crossing filament by the {\it same} displacement $\Delta \vec x_2$, such that node two is
now in mechanical equilibrium. Since node one, the filament crossing the central one at node one, and node two have all displaced equally, node one remains
in mechanical equilibrium. The result of this procedure is the third panel in Fig.~\ref{fig: accordion displacement}. There is
now a force imbalance at node three. We now apply the same displacement procedure, this time collectively
moving nodes one and two, and their crossing filaments in addition to node three. The method is then iterated
until we reach the pinned face at the boundary (assumed to be infinitely far away).

This procedure treats the network is as type of layered solid, with the layers having normals along the direction of the central filament.  Each layer displaces uniformly, but that displacement varies from layer to layer.  In this way, the  deformation is reminiscent of the extension of an accordion, from which we take the name of our approach.

By using an effective transverse spring constant $k_\perp$, we have incorporated bending by approximating the bending energy contribution of a fiber as the sum of the bending energies of the cusps created by node displacement. For $\theta_i$ the local tangent formed between the displaced nodes $i$ and $i-1$, a cusp has energy $U_\text{cusp} = \frac{\kappa}{2}(\theta_{i+1}-\theta_i)^2$~\cite{mao2013effective}. Bending couples nearest neighbor normal displacements along the central filament. Our iterative accordion model cannot account for this, and misses bending energy reductions when adjacent nodes have similar transverse displacements. Since the applied force is directed in the longitudinal direction though, we expect this to be negligible, with the main bending contributions coming from crossed fibers.

The propagation of tension down the central fiber can now be described in terms of transfer matrices that relate the incoming to the outgoing tensions.
We first consider the displacements of a single cross-linker as shown in Fig.~\ref{fig: FBD}, subject to two
tensions $T_i^c$ and $T_i^\times$ applied from the left. The crossing filament is restricted to angles of intersection $\theta_i < \pi/2$,
at which point the tension would have to change sign. The system Hamilton is given by the sum over nodes~\cite{mao2013effective}
\beq
\label{eq: H}
H = \frac{1}{2}\sum_{i=1}^{N_c}\sum_{s=\{c,\times\}} \left[k'_\parallel ( \hat n_{i}^s \cdot \Delta \vec x_i)^2,
+k'_\perp ( \hat n_{i}^s \times \Delta \vec x_{i})^2 + 2 T_i^s (\hat n_i^s \cdot \Delta \vec x_i)\right].
\eeq
with accordion model displacements $\Delta \vec x_i$ as defined in Fig.~\ref{fig: accordion displacement}. The $i$ summation is over all nodes. Since the central filament is assumed straight along the $x$-axis in the undeformed state, for every node $i$ we may replace $\hat n_i^c = \hat x$.


Force balance determines the displacement $\Delta \vec x_i$ in terms of incoming tensions $T_i^c$ and $T_i^\times$. In terms of $\Delta \vec x_i$, the two outgoing tensions must satisfy:
$\tilde T_{i+1}^c = k'_\parallel \hat x \cdot \Delta \vec x_i$ and $\tilde T_i^\times = k'_\parallel \hat n_i^\times \cdot \Delta \vec x_i$. Solving for the displacement $\Delta \vec x_i$ allows us to express the outgoing tensions in terms of the incoming tensions. Since there is a linear relation between these pairs of forces, we may introduce a
transfer matrix converting incoming tensions into outgoing ones.  This is the key simplification of the accordion method. The transfer matrix is given
by
\beq
\label{eq: M schematic}
\vv{T_{i+1}^c \\ \tilde T_i^\times} = \mathbf{M}(\theta_i) \vv{T_i^c \\ T_i^\times},
\eeq
as a function of crossing angle $\theta_i$.  Because these angles are random, the transfer matrix
$\mathbf{M}$ is, itself, a random variable. We assume that the network is isotropic, so that the distribution of these crossing angles is uniform and uncorrelated from node to node.  We write
\beq
\label{eq: P theta}
P(\theta_1,\ldots, \theta_N) = \left(\frac{2}{\pi}\right)^N.
\eeq
Expressing the transfer matrix in terms of the random angle, we obtain
\beq
\label{eq: M def}
\mathbf{M}(\theta)=
\left(
\begin{array}{cc}
 \frac{(1-\alpha^2) \sin ^2\theta+2 \alpha^2 }{(1-\alpha^2 )^2 \sin ^2\theta+4 \alpha^2 } & \frac{2 \alpha^2  \cos \theta }{(1-\alpha^2 )^2 \sin ^2\theta+4 \alpha^2 } \\
 \frac{2 \alpha^2  \cos \theta }{(1-\alpha^2)^2 \sin ^2\theta +4 \alpha^2 } & \frac{(1-\alpha^2) \sin ^2\theta +2 \alpha^2 }{(1-\alpha^2)^2 \sin ^2\theta +4 \alpha^2 } \\
\end{array}
\right),
\eeq
where we have defined the ratio $\alpha$ of bending to stretching in the system:
\beq
\label{eq: alpha}
\alpha = \sqrt{k_\perp/k_\parallel}.
\eeq
If we consider the relations in Eq.~\ref{eq: k defs} to be valid, $\alpha = \ell_\text{bend}/\ell_c$ as well.

If there are no loops, which would allow forces transferred away from the central filament to return to it,
then we note that $T_i^\times=0$. Due to the fact that the total force must split at network junctions and that loops require
multiple junctions (at least two) we expect that the tension contribution from loops will be subdominant.  If we ignore the loop
contribution and set  $T_i^\times=0$, then $M_{11}(\theta_i)$ gives the fraction of
tension $T_{i+1}^c/T_i^c$ that remains on the central fiber after a crossing.

Tension propagation along the central filament is now determined by the repeated multiplication
of the transfer matrix component $M_{11}(\theta)$.
We define the tension fraction remaining on the central filament after the $n_{c}^{\rm th}$ cross link
\beq
\label{eq: tension fraction def}
\tau_{n_c} = T_{n_c}^c/T_0^c.
\eeq
We find that
\beq
\label{eq: tau nc def}
\tau_{n_c} = \prod_{i=1}^{n_c}M_{11}(\theta_i).
\eeq
For any given central filament, the tension in each segment $\tau_{n_c}$ is also a random variable. Given the probability
distribution of angles $\theta_i$ factorizes, the moments of segment tension over an ensemble of central filaments are
computable by a single contour integration.
We find the average
\beq
\label{eq: tau1 avg}
\langle \tau_1 \rangle = \frac{1}{1+\alpha},
\eeq
and the variance
\beq
\label{eq: tau1 var}
\langle \tau_1^2\rangle-\langle \tau_1\rangle^2 =\frac{\alpha(1-\alpha)^2}{4 \left(1+\alpha\right)^2 (1+\alpha^2)}.
\eeq
The angled brackets denote angular averages.

We observe that the tension on the central filament decays exponentially. In the limit of small bending
as compared to stretching, where $\alpha \ll 1$, that decay length (measured in units of number of cross link nodes)
is approximately $1/\alpha$.  For filament systems, this limit of small $\alpha$ is likely to be valid.  We can convert that
decay per cross link to a decay per distance $x$ along the filament, by approximating $n_{c} = x/\ell_{c}$.
In that case, and in the limit of small $\alpha$, we may write
\beq
\label{eq: tau large x}
\langle \tau_{n_c} \rangle \xrightarrow[\alpha \ll 1]{x\gg \ell_c} e^{- x/\xi},
\eeq
where we have defined the tension propagation length
\beq
\label{eq: xi}
\xi = \frac{\ell_{c}}{\alpha}.
\eeq
This can be viewed as penetration length for tension in a network, before it is absorbed by bending of crossed filaments.

\begin{figure}
\centering
\includegraphics[width=0.9\linewidth]{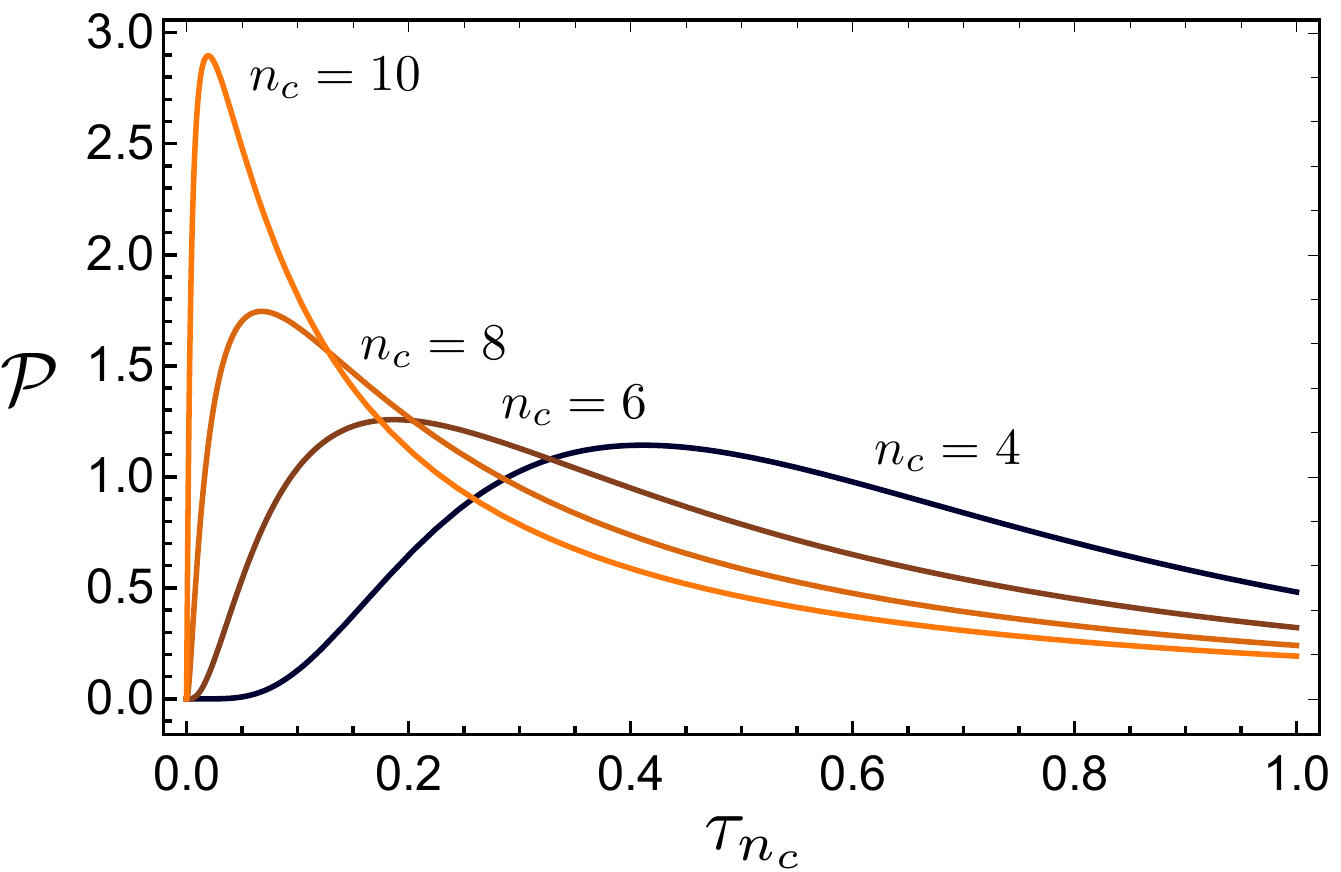}
\includegraphics[width=0.9\linewidth]{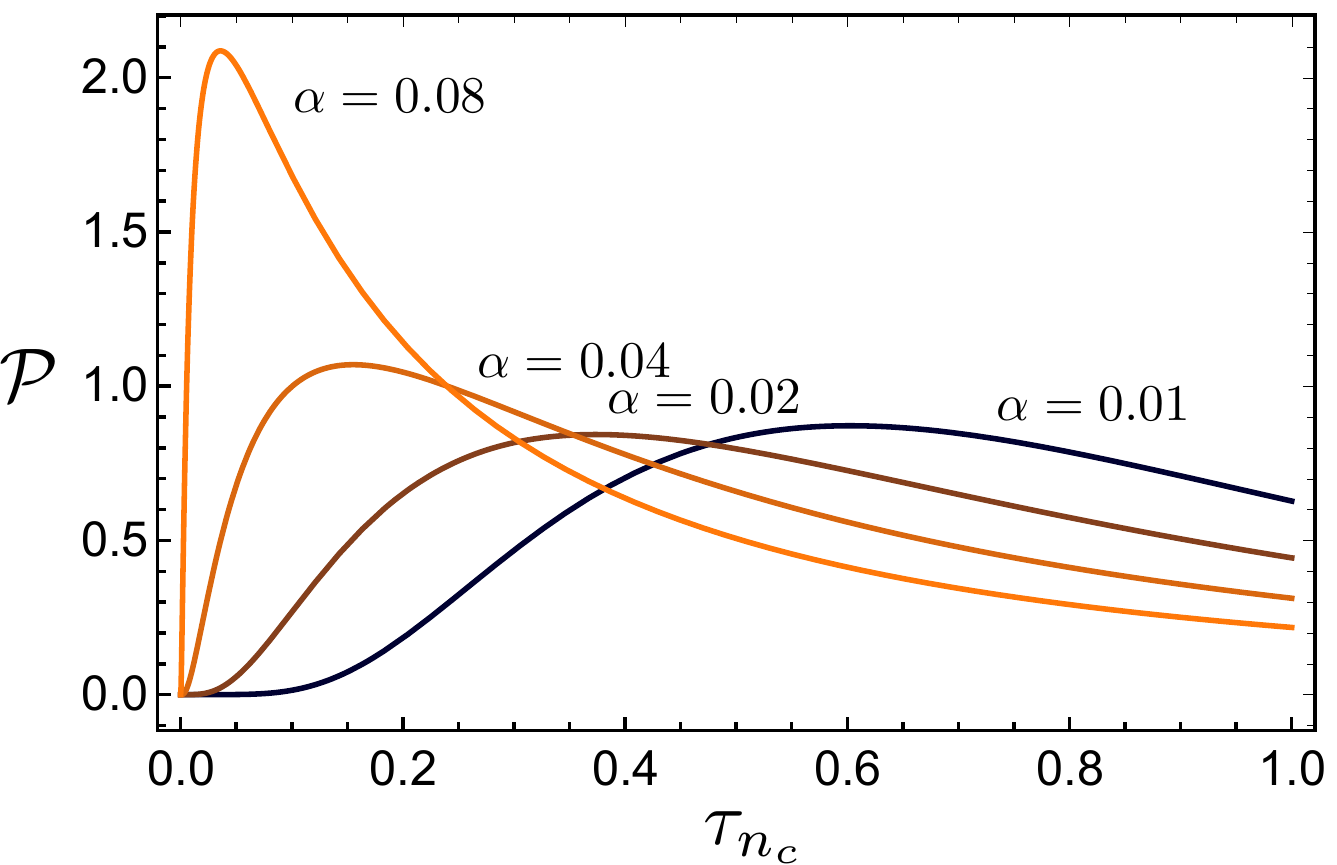}
\caption{(color online) probability distribution of the tension fraction $\tau_{n_c}$ along the central filament.
Top: we fix $\alpha=0.1$ and vary the number $N_c$ of crossings. Bottom: we fix $n_c=10$ and vary the bending to
stretching ratio $\alpha$. As the number of crossings ($\alpha$) increases, the peak of the distribution tends toward zero,
indicating that most of the tension has been absorbed into the network via bending.}
\label{fig: lognormal}
\end{figure}

More information on the distribution of tension along the central filament can be gleaned by taking the logarithm of Eq.~\ref{eq: tau nc def}.
We find that $\ln \tau_{n_c}$ is given by the sum of $n_c$ independent random variables $ \ln M_{11}(\theta_i)$. Applying the central
limit theorem to that sum of random variables yields a log-normal distribution of tension.  Specifically, the probability of observing
$\ln \tau $ at the $n_{c}^{\rm th}$ cross link
is governed by a normal distribution with mean $\tilde \mu = n_c \langle \ln M_{11}(\theta)\rangle$
and variance $\sigma^2 = n_c \langle (\ln M_{11}(\theta)-\tilde \mu/n_c)^2 \rangle $. We can then convert
back to a distribution for tension itself via a change of variables. We find the distribution
\beq
\mathcal{P}_{n_c \gg 1}(\tau) = \frac{\text{exp}\left[- \left(\ln \tau - \tilde  \mu\right)^2/2\sigma^2\right] }{\sigma \tau \sqrt{2\pi}} ,
\eeq
where we have indicated that $n_c \gg 1$ must hold in order to make use of the central limit theorem.
The average $\tilde \mu$ can be computed analytically using the identity $ \int_0^{\pi/2} \ln(1+ a \sin^2(\theta)) d\theta = \pi \ln(\frac{1+\sqrt{1+a}}{2})$~\cite{gradshteyn2014table}. We obtain
\beq
\tilde \mu =\langle \ln M_{11}(\theta) \rangle = 2 \ln \left(\frac{\sqrt{\alpha^2 +1}+\sqrt{2} \alpha}{(\alpha+1)^2}\right).
\eeq
The variance $\sigma^2$, on the other hand, does not have a simple representation in terms of known functions. We numerically compute it, and plot the corresponding distributions for ${\mathcal {P}}_{n_c}(\tau)$ as a function of both $n_c$ and $\alpha$ -- see Fig.~\ref{fig: lognormal}.

Though $\tau_{n_c}$ is by definition less than unity, the probability distributions possess long tails extending beyond one. This is a consequence of our use of the central limit theorem to infer the full probability distribution from its moments. A similar issue arises in the study of spin glasses~\cite{mezard1987spin} and directed paths in random media~\cite{kardar2007statistical}.

As the number of crossings $n_c$ increases, the upper panel of Fig.~\ref{fig: lognormal} shows that the tension distribution exhibits an increasingly narrow peak at smaller values of tension. This is a consequence of tension decay, whereby at long distances the network has absorbed most of the tensile load into bending. At fixed $n_c$, the lower panel of Fig.~\ref{fig: lognormal} shows that a similar effect occurs with increasing $\alpha$. Larger $\alpha$ leads to more bending absorption, and hence a peak of the distribution at smaller tensions.

\begin{figure}
\centering
\includegraphics[scale=0.7]{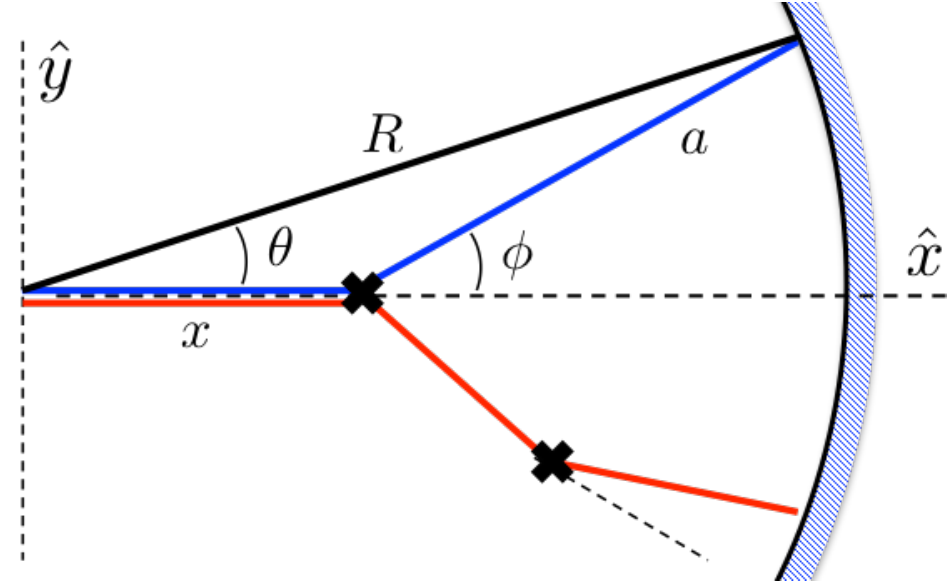}
\caption{Off-axis branching of tension, for one (blue line) and two (red line) scattering events. The black semi-arc on the right denotes a tension detector at $R$ away from the site of point force application at the origin. Averaging over network configurations amounts to a sum over path configurations. These are organized in terms of the number of scattering events, where tension changes direction.}
\label{fig: optical paths}
\end{figure}

We now turn to an analysis of force branching to investigate force-bearing pathways that diverge from the central filament. At a given node, the central filament with incoming tension $T_i$ will distribute a fraction of that
tension, defined as the tension fraction $\bar \tau_i(\theta)=\tilde T_{i}^\times/T_i^c$, along the crossing filament making an angle $\theta_i$ with it. This tension fraction is given by a multiplicative factor
$\bar \tau(\theta_i)=M_{21}(\theta_i)$. As expected, this has a vanishing probability to split tension into the direction perpendicular to the central filament.

The matrix element $M_{21}$ is invariant under inversion of $\alpha$:
\beq
M_{21}(\theta,\alpha) = M_{21}(\theta,1/\alpha).
\eeq
This suggests that branching effects depend only on the ratio of bending to stretching in a network. A network that is weak to bending and stiff to stretching could exhibit the same branching characteristics as a network that is stiff to bending and weak to stretching.

We consider the following thought experiment. Having applied a tension to the central filament, we measure the tension on other filaments
passing through a thin spherical shell of radius $R$. One can then repeat this measurement on multiple realizations of the network to
arrive at an ensemble average of this tension propagation in a particular (solid) angle. Assuming that this average must be azimuthally
symmetric (about the central filament) we define the ensemble-averaged angular distribution of tension branching as $T_R(\theta)$.

In order to calculate $T_R(\theta)$, we consider the process shown in Fig.~\ref{fig: optical paths}. Ensemble averaging of the network
makes the position of nodes at any location equally likely, so the average is equivalent to taking a sum over all possible paths for
tension to leave from the origin and reach the shell. Paths can be categorized by the number of ``scattering" events
(the black x's in Fig.~\ref{fig: optical paths}), where the tension path changes direction. For shell radii on the order of a few $\ell_{c}$,
these paths consist of a small number of straight line segments with average length $\ell_{c}$.
The problem simplifies if we look in the far field,  $R \gg \ell_c$, where we may replace an analysis of discrete scattering
events at the randomly placed nodes with decay of tension as a function of distance along the central filament -- see
Eq.~\ref{eq: tau large x}. In that case, each path is given by the exponentially decaying tension propagators $e^{- |\bf{r}|/\xi}$,
connected by scattering events
where the tension is reduced by a factor of $M_{21}(\theta_i)$, for angular deviation $\theta_i$ from the unscattered direction.

In the single scattering approximation, we include only the effects of one scatterer. Here, each path can be confined to a plane. We can
extend our results to three dimensions by computing the tension at fixed azimuthal angle in one plane, then rotating. For more
than one scattering event, one must take into account the three dimensional nature of the network. Neglecting contributions from
zero scattering, ({\em i.e.}~tension decay along the central filament), using the geometry/notation of Fig.~\ref{fig: optical paths}, we
find the single scattering tension
\beq
T_R(\theta) = \int_0^{R \cos \theta} dx e^{-x/\xi} M_{21}[\phi(x)] e^{-a/\xi}.
\eeq
To normalize, we divide by the ensemble averaged total tension on all filaments crossing through the spherical shell. The normalized tension is now
\beq
\overline{T}_R(\theta) = \frac{T_R(\theta)}{\mathcal{N}},
\eeq
with normalization
\beq
\mathcal{N} = \frac{2}{\pi} \int_0^{\pi/2} T_R(\theta) d \theta.
\eeq

The tension may be expressed in terms of the dimensionless ratio $R/\xi$, which is the number of tension decay lengths we are away from the origin. This is accomplished via change of integration variables $x \to x R \cos \theta$. Expressing $M_{21}[\phi(x)]$ in terms of $x$ using the geometry of Fig.~\ref{fig: optical paths}, we obtain the integral equation
\begin{eqnarray}
\overline{T}_{\frac{R}{\xi}}(\theta) &=& \frac{\cos \theta}{\mathcal{N}} \int_0^1 \bigg[ \frac{2 \alpha ^2 (1-x) \cos \theta \sqrt{1+x(x-2) \cos ^2\theta}}{\left(1-\alpha ^2\right)^2 \sin ^2\theta+4 \alpha ^2 \left(x(x-2) \cos ^2\theta+1\right)} \nonumber \\
&\times& e^{-(R/\xi) x} e^{- (R/\xi)\sqrt{1+x(x-2)\cos^2\theta}} \bigg ]dx .
\end{eqnarray}

\begin{figure}[h]
\centering
\includegraphics[width=0.9\linewidth]{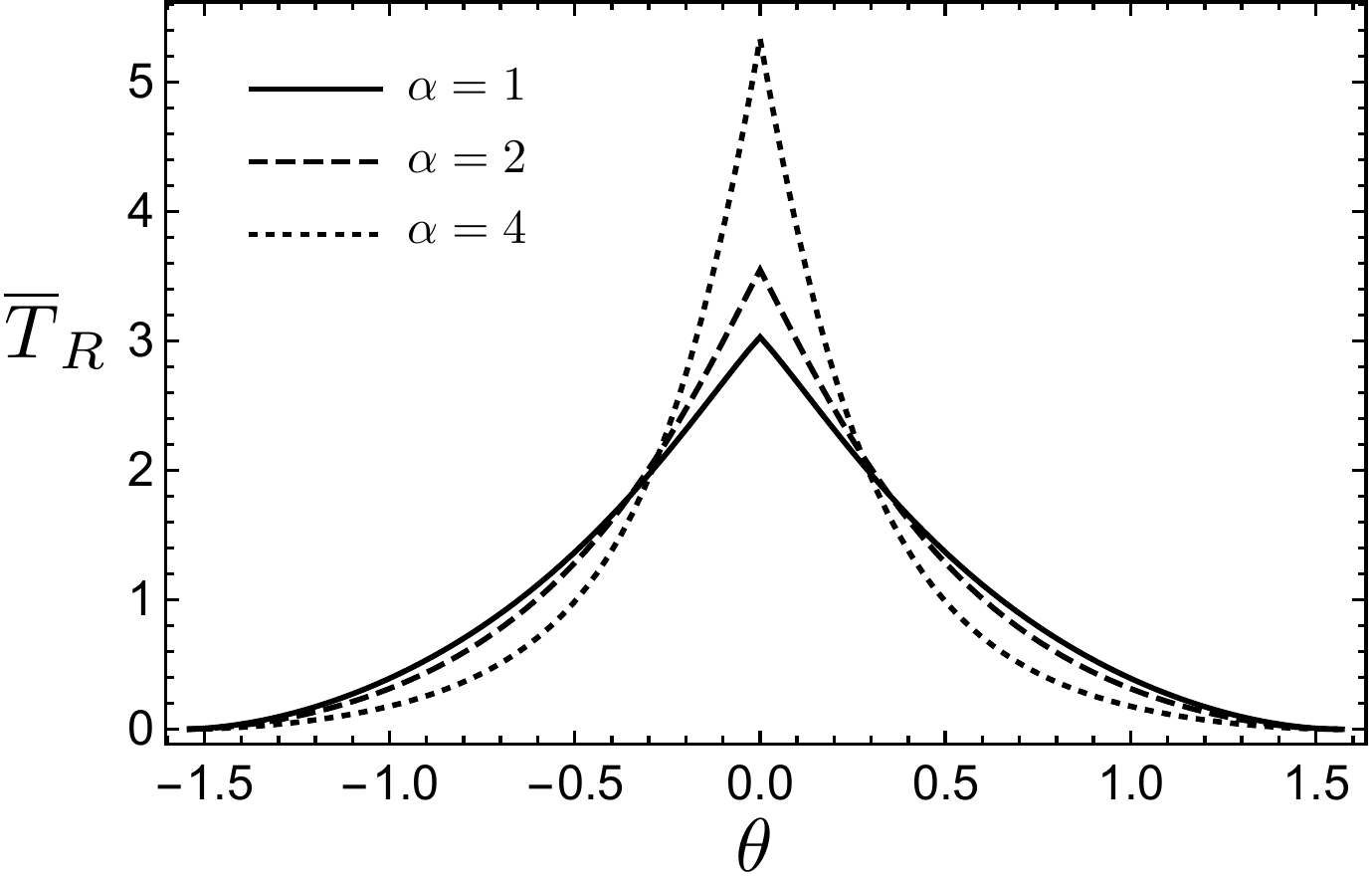}
\includegraphics[width=0.9\linewidth]{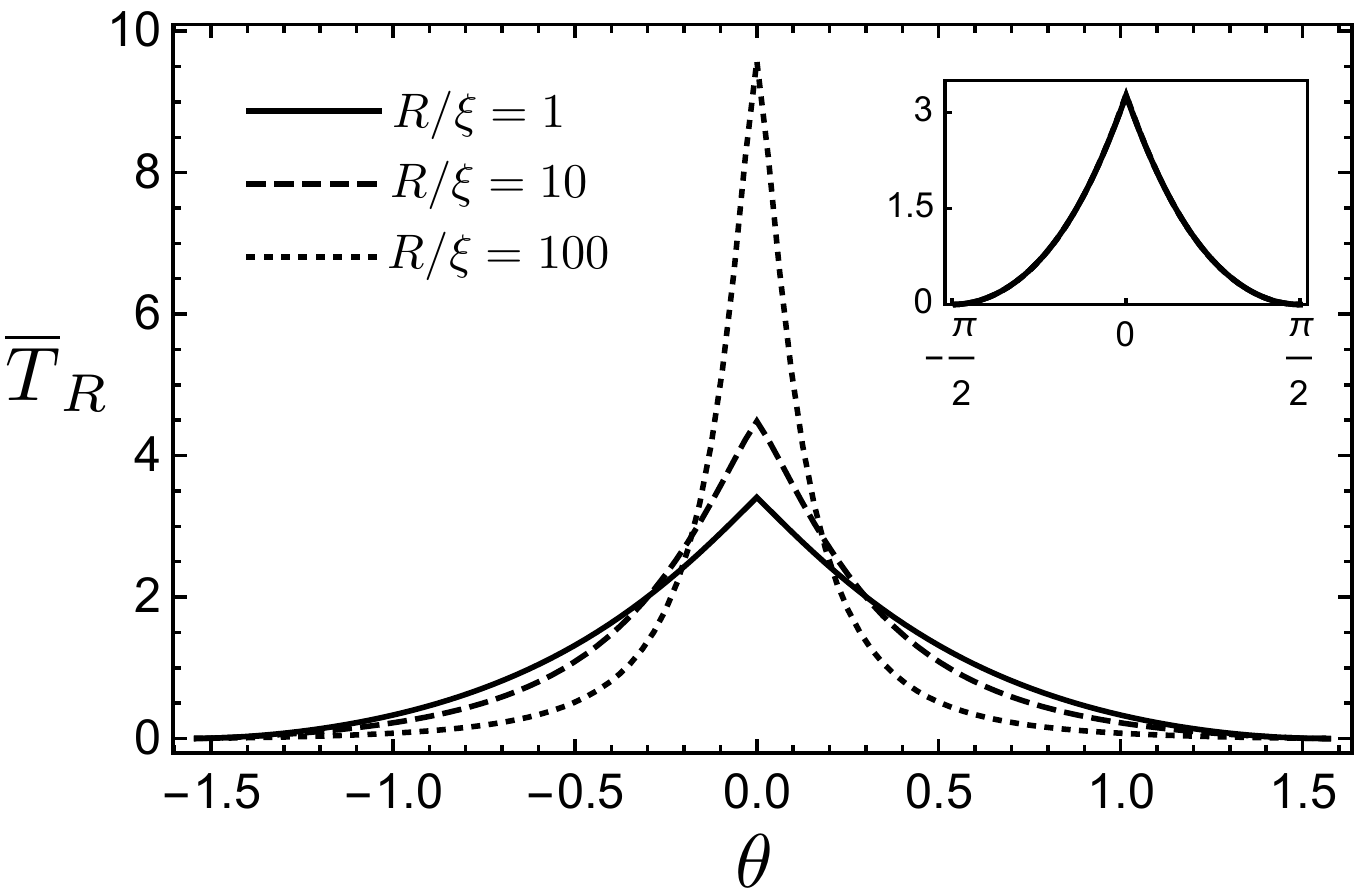}
\caption{Ensemble averaged tension on filaments located a distance $R$ away and at azimuthal angle $\theta$ with respect to a point force applied at the origin. Tension is normalized by the total tension on all filaments passing through a shell of radius R. Top: we fix $R/\xi =2$ and vary the ratio of bending to stretching $\alpha$. Bottom: we fix $\alpha=0.5$ and vary the distance from the point force $R/\xi$. The distribution of tension is widest at $\alpha=1$ and at low $R/\xi$. In the inset of the lower panel, we plot the same curve but for values $R/\xi=10^{-3},10^{-2},10^{-1}$. Each of these curves give the same distribution as $R/\xi=1$.}
\label{fig: TR}
\end{figure}

In Fig.~\ref{fig: TR} we evaluate this integral numerically for several values $\alpha$ and $R/\xi$, for $\theta$ in the domain $(-\pi/2,\pi/2)$. The upper panel of Fig.~\ref{fig: TR}, shows that at fixed distance from the origin, $\alpha=1$ corresponds to the broadest angular distribution of tension. Since $\alpha$ is a ratio of the effective spring constants acting on a node due to bending and stretching, we find that branching is maximal when the network equally supports a tensile load via bending and stretching. Due to the inversion symmetry $\alpha \to 1/\alpha$ of the function $M_{21}(\theta)$, networks that support tensile loads via mainly bending and via mainly stretching exhibit the same branching, provided their ratios of effective bending to stretching spring constants are inverse.

The lower panel of Fig.~\ref{fig: TR} shows that the angular distribution of tension becomes increasingly narrow at
distances $R \gg \xi$ away from the origin. Naturally, one would expect that at longer distances multiple branching events
would lead to a broader angular distribution of tension. This counterintuitive result is due to tension decay.  At long
distances from the origin, tension is determined primarily by the length of the tension chain, which favors paths along
the forward direction. At distances smaller than the decay length ($R < \xi$), the angular distribution is broad, and remains
fixed to its value at $R=\xi$ -- see the inset in the lower panel of Fig.~\ref{fig: TR}. In this limit, the tension distribution is highly
dependent on details of the network.

\section{Conclusions}
Experiments on the force propagation between cells in a filamentous extra-cellular matrix suggest that
forces are transmitted over long distances
along particular pathways.  This observation raises the question: do these force pathways arise
naturally in randomly structured filament networks, or
does the ECM contain particular structural heterogeneities, such as system-spanning
subnetworks of filament bundles, that prescribe these force-transmitting
paths?  In our work, we make a preliminary investigation of this question by addressing the first possibility.

Our simulations of random homogeneous and isotropic networks do indeed
show weakly branching force chains over scales of many mesh sizes in the network.  Based on the imperfect analogy between force propagation in
filament networks and force chains in granular media, such structures are not entirely surprising. We believe that the high-force pathways observed in
simulation are sufficiently similar to those seen in granular media, so as to warrant using the same terminology.  The principle distinction between the
two types of mechanical systems is that filament bending in the network
provides (typically) highly-compliant elastic degrees of freedom, which, in some sense, absorb the tensile loading
on a highly tensed filament.  In this way, filament-based force chains naturally terminate over some finite distance.

The simulations demonstrate that the point-force response of the network is highly heterogeneous.  It depends on both the point of loading and the direction
of that applied force.  This is, again, in qualitative agreement with the results of active microrheology experiments in fibrin gels.  In particular, we
observe that the microscopic geometry of the network near the point of loading has a large effect on the collective point-force response of the
system.  Based on the numerical data
presented here, we believe that the wide distribution of linear response observed in experiment is qualitatively consistent 
with our numerical results. In short, our distribution
appears to be narrower than that seen in experiment.  This suggests that the fibrin networks used in experiment are more structurally heterogeneous on the
microscale.

We also show the expected asymmetry
between pushing and pulling on a filament.  The importance of buckling in limiting the propagation of compressive loading is clear.  This effect may be
weakened in highly tensed networks, where the applied forces do not typically reach the threshold for Euler buckling.
Such tensed networks may, in fact,
behave more like continuum elastic solids with regard to their point-force response.  We do not, as yet, have numerical data to
test this supposition.  In addition, we observe in simulations that the collective point force response depends quite strongly on the direction of
pulling with respect to the local filament tangent and that there appear to be two distinct types of response corresponding to a
low-force regime and a high-force regime where the pulling changes the local structure of the network.

We also approached the problem of force propagation in the network through simple
analytic models of the system using harmonic springs (of two classes: bending and stretching) and rather simple assumptions of how forces
propagate through the network, {\em i.e.}, they do not typically form closed loops.  By invoking self-consistency for the collective
longitudinal compliance (which takes into account the motion of other nodes in the network) of the filament,  we obtain an analytic
prediction for the point-force response measured in our numerical experiments.

These calculations qualitatively agree with the simulated results in that
they predict the decay of tension along the filament of force application due to those tensile forces being transmitted into a combination of bending and
tensile loading on the filaments cross linked the filament of force application.  In one analytically tractable limit, the self-consistent calculation predicts
that tensions decay exponentially along the filament of force application.  Numerical solutions away from this point suggest that the
exponential behavior is more broadly applicable.

Using this result, we introduce the accordion approximation, which allows us to study this
exponential decay via transfer matrices.  It also allows us to study
the branching of force chains.  We learn a few things. First, the system produces the
largest angular distribution of force chain branches when bending and
stretching compliances are comparable.  Second, we find that branching occurs primarily at
small angles; in the language of optics, one may assert that
forces are strongly forward scattering.  Third and finally, we note that, due to the exponential
decay of tension from the main force paths, shorter and less
ramified force chains should dominate the distribution of these structures. This appears to be
qualitatively supported by the numerical experiments.   The second
point shows how force may propagate over long distances along essentially straight paths,
which are controlled primarily by the tension decay length.

To quantitatively compare the predicted and observed tension decay lengths,
we need to use parameters from the numerical simulations. 
Using the average filament segment length $\ell_c =0.56 \mu \text{m}$ and filament rigidity 
$\kappa = 5.6\times 10^{-2} \text{pN}\mu\text{m}^2$, and making the seemingly reasonable choice of 
$k_\perp' = \kappa/\ell_c^3 \approx 0.33 \text{pN}/\mu \text{m}$, we obtain a value that is much smaller than 
the average $k_\perp' \approx 8.2 \text{pN}/\mu\text{m}$ observed in the low force regime of our filament pulling simulations -- 
see Fig.~\ref{fig::simulation_effective_spring_constants}(D). 
If we also use the same analysis to determine $k'_{\parallel} \approx 3.4 \times 10^{2}\text{pN}/\mu\text{m}$, 
we obtain a prediction for the tension decay length $\xi \sim 18\mu$m, which is an order of magnitude greater than 
the $\xi =2.2 \mu\text{m}$ observed for the lowest magnitude pulling force simulation -- see Fig.~\ref{fig::simulation_box25x06x06}(D).  
If, on the other hand, we take our $k'_{\parallel}$
and $k'_{\perp}$ directly from the numerical data -- see Fig.~\ref{fig::simulation_effective_spring_constants}(C), we obtain a 
much more reasonable prediction of $\xi \sim 0.9\mu \text{m}$.
One might suggest that this discrepancy between the first estimate of the local bending spring constant and the measured one is attributable to
prestress in the numerically simulated networks. If the filaments are under tension $T$ then, $k_\perp = 4 T / l_c \approx 1 \text{pN} / \mu \text{m}$, but
this alone is not sufficient.  We speculate that the frozen-in filament curvature in our zero temperature simulations stiffen the filaments to
bending.

\section*{Acknowledgements}
We are grateful to Prof.~Carl Goodrich for fruitful discussions.  AJL, JK, and VS acknowledge partial support from NSF DMR-1709785.
VS acknowledges partial support from the Bhaumik Institute for Theoretical Physics.
MJG and WAW acknowledge partial support from BaCaTeC.



\balance


\bibliography{ForceChainreferences} 

\providecommand*{\mcitethebibliography}{\thebibliography}
\csname @ifundefined\endcsname{endmcitethebibliography}
{\let\endmcitethebibliography\endthebibliography}{}
\begin{mcitethebibliography}{38}
\providecommand*{\natexlab}[1]{#1}
\providecommand*{\mciteSetBstSublistMode}[1]{}
\providecommand*{\mciteSetBstMaxWidthForm}[2]{}
\providecommand*{\mciteBstWouldAddEndPuncttrue}
  {\def\EndOfBibitem{\unskip.}}
\providecommand*{\mciteBstWouldAddEndPunctfalse}
  {\let\EndOfBibitem\relax}
\providecommand*{\mciteSetBstMidEndSepPunct}[3]{}
\providecommand*{\mciteSetBstSublistLabelBeginEnd}[3]{}
\providecommand*{\EndOfBibitem}{}
\mciteSetBstSublistMode{f}
\mciteSetBstMaxWidthForm{subitem}
{(\emph{\alph{mcitesubitemcount}})}
\mciteSetBstSublistLabelBeginEnd{\mcitemaxwidthsubitemform\space}
{\relax}{\relax}

\bibitem[Liu \emph{et~al.}(1995)Liu, Nagel, Schecter, Coppersmith, Majumdar,
  Narayan, and Witten]{liu1995force}
C.-h. Liu, S.~R. Nagel, D.~Schecter, S.~Coppersmith, S.~Majumdar, O.~Narayan
  and T.~Witten, \emph{Science}, 1995, \textbf{269}, 513--515\relax
\mciteBstWouldAddEndPuncttrue
\mciteSetBstMidEndSepPunct{\mcitedefaultmidpunct}
{\mcitedefaultendpunct}{\mcitedefaultseppunct}\relax
\EndOfBibitem
\bibitem[Cates \emph{et~al.}(1998)Cates, Wittmer, Bouchaud, and
  Claudin]{Cates1998}
M.~E. Cates, J.~P. Wittmer, J.-P. Bouchaud and P.~Claudin, \emph{Phys. Rev.
  Lett.}, 1998, \textbf{81}, 1841--1844\relax
\mciteBstWouldAddEndPuncttrue
\mciteSetBstMidEndSepPunct{\mcitedefaultmidpunct}
{\mcitedefaultendpunct}{\mcitedefaultseppunct}\relax
\EndOfBibitem
\bibitem[Hurley \emph{et~al.}(2016)Hurley, Hall, Andrade, and
  Wright]{Hurley2016}
R.~C. Hurley, S.~A. Hall, J.~E. Andrade and J.~Wright, \emph{Phys. Rev. Lett.},
  2016, \textbf{117}, 098005\relax
\mciteBstWouldAddEndPuncttrue
\mciteSetBstMidEndSepPunct{\mcitedefaultmidpunct}
{\mcitedefaultendpunct}{\mcitedefaultseppunct}\relax
\EndOfBibitem
\bibitem[Head \emph{et~al.}(2005)Head, Levine, and
  MacKintosh]{head2005mechanical}
D.~Head, A.~Levine and F.~MacKintosh, \emph{Physical Review E}, 2005,
  \textbf{72}, 061914\relax
\mciteBstWouldAddEndPuncttrue
\mciteSetBstMidEndSepPunct{\mcitedefaultmidpunct}
{\mcitedefaultendpunct}{\mcitedefaultseppunct}\relax
\EndOfBibitem
\bibitem[Ma \emph{et~al.}(2013)Ma, Schickel, Stevenson, Sarang-Sieminski,
  Gooch, Ghadiali, and Hart]{Hart2013}
X.~Ma, M.~Schickel, M.~Stevenson, A.~Sarang-Sieminski, K.~Gooch, S.~Ghadiali
  and R.~T. Hart, \emph{Biophysical Journal}, 2013, \textbf{104},
  1410--1418\relax
\mciteBstWouldAddEndPuncttrue
\mciteSetBstMidEndSepPunct{\mcitedefaultmidpunct}
{\mcitedefaultendpunct}{\mcitedefaultseppunct}\relax
\EndOfBibitem
\bibitem[Rudnicki \emph{et~al.}(2013)Rudnicki, Cirka, Aghvami, Sander, Wen, and
  Billiar]{rudnicki2013nonlinear}
M.~S. Rudnicki, H.~A. Cirka, M.~Aghvami, E.~A. Sander, Q.~Wen and K.~L.
  Billiar, \emph{Biophysical journal}, 2013, \textbf{105}, 11--20\relax
\mciteBstWouldAddEndPuncttrue
\mciteSetBstMidEndSepPunct{\mcitedefaultmidpunct}
{\mcitedefaultendpunct}{\mcitedefaultseppunct}\relax
\EndOfBibitem
\bibitem[Missel \emph{et~al.}(2010)Missel, Bai, Klug, and Levine]{Missel2010}
A.~R. Missel, M.~Bai, W.~S. Klug and A.~J. Levine, \emph{Phys. Rev. E}, 2010,
  \textbf{82}, 041907\relax
\mciteBstWouldAddEndPuncttrue
\mciteSetBstMidEndSepPunct{\mcitedefaultmidpunct}
{\mcitedefaultendpunct}{\mcitedefaultseppunct}\relax
\EndOfBibitem
\bibitem[Foucard \emph{et~al.}(2015)Foucard, Price, Klug, and
  Levine]{foucard2015}
L.~Foucard, J.~Price, W.~Klug and A.~Levine, \emph{Nonlinearity}, 2015,
  \textbf{28}, R89\relax
\mciteBstWouldAddEndPuncttrue
\mciteSetBstMidEndSepPunct{\mcitedefaultmidpunct}
{\mcitedefaultendpunct}{\mcitedefaultseppunct}\relax
\EndOfBibitem
\bibitem[Majumdar \emph{et~al.}(2018)Majumdar, Foucard, Levine, and
  Gardel]{majumdar2018mechanical}
S.~Majumdar, L.~C. Foucard, A.~J. Levine and M.~L. Gardel, \emph{Soft matter},
  2018, \textbf{14}, 2052--2058\relax
\mciteBstWouldAddEndPuncttrue
\mciteSetBstMidEndSepPunct{\mcitedefaultmidpunct}
{\mcitedefaultendpunct}{\mcitedefaultseppunct}\relax
\EndOfBibitem
\bibitem[Kotlarchyk \emph{et~al.}(2011)Kotlarchyk, Shreim, Alvarez-Elizondo,
  Estrada, Singh, Valdevit, Kniazeva, Gratton, Putnam, and
  Botvinick]{kotlarchyk2011concentration}
M.~A. Kotlarchyk, S.~G. Shreim, M.~B. Alvarez-Elizondo, L.~C. Estrada,
  R.~Singh, L.~Valdevit, E.~Kniazeva, E.~Gratton, A.~J. Putnam and E.~L.
  Botvinick, \emph{PloS one}, 2011, \textbf{6}, year\relax
\mciteBstWouldAddEndPuncttrue
\mciteSetBstMidEndSepPunct{\mcitedefaultmidpunct}
{\mcitedefaultendpunct}{\mcitedefaultseppunct}\relax
\EndOfBibitem
\bibitem[Keating \emph{et~al.}(2017)Keating, Kurup, Alvarez-Elizondo, Levine,
  and Botvinick]{keating2017spatial}
M.~Keating, A.~Kurup, M.~Alvarez-Elizondo, A.~Levine and E.~Botvinick,
  \emph{Acta biomaterialia}, 2017, \textbf{57}, 304--312\relax
\mciteBstWouldAddEndPuncttrue
\mciteSetBstMidEndSepPunct{\mcitedefaultmidpunct}
{\mcitedefaultendpunct}{\mcitedefaultseppunct}\relax
\EndOfBibitem
\bibitem[Hu \emph{et~al.}(2020)Hu, Morris, Grosberg, Levine, and
  Botvinick]{Qingda2020}
Q.~Hu, T.~A. Morris, A.~Grosberg, A.~J. Levine and E.~Botvinick,
  \emph{Submitted}, 2020\relax
\mciteBstWouldAddEndPuncttrue
\mciteSetBstMidEndSepPunct{\mcitedefaultmidpunct}
{\mcitedefaultendpunct}{\mcitedefaultseppunct}\relax
\EndOfBibitem
\bibitem[Cyron and Wall(2010)]{Cyron2010}
C.~J. Cyron and W.~A. Wall, \emph{Physical Review E}, 2010, \textbf{82},
  66705\relax
\mciteBstWouldAddEndPuncttrue
\mciteSetBstMidEndSepPunct{\mcitedefaultmidpunct}
{\mcitedefaultendpunct}{\mcitedefaultseppunct}\relax
\EndOfBibitem
\bibitem[Cyron and Wall(2012)]{cyron2012}
C.~J. Cyron and W.~A. Wall, \emph{International Journal for Numerical Methods
  in Engineering}, 2012, \textbf{90}, 955--987\relax
\mciteBstWouldAddEndPuncttrue
\mciteSetBstMidEndSepPunct{\mcitedefaultmidpunct}
{\mcitedefaultendpunct}{\mcitedefaultseppunct}\relax
\EndOfBibitem
\bibitem[Cyron \emph{et~al.}(2013)Cyron, M{\"{u}}ller, Schmoller, Bausch, Wall,
  and Bruinsma]{Cyron2013phasediagram}
C.~J. Cyron, K.~W. M{\"{u}}ller, K.~M. Schmoller, A.~R. Bausch, W.~A. Wall and
  R.~F. Bruinsma, \emph{Europhysics Letters}, 2013, \textbf{102}, 38003\relax
\mciteBstWouldAddEndPuncttrue
\mciteSetBstMidEndSepPunct{\mcitedefaultmidpunct}
{\mcitedefaultendpunct}{\mcitedefaultseppunct}\relax
\EndOfBibitem
\bibitem[M{\"{u}}ller \emph{et~al.}(2014)M{\"{u}}ller, Bruinsma, Lieleg,
  Bausch, Wall, and Levine]{Mueller2014rheology}
K.~W. M{\"{u}}ller, R.~F. Bruinsma, O.~Lieleg, A.~R. Bausch, W.~A. Wall and
  A.~J. Levine, \emph{Physical Review Letters}, 2014, \textbf{112},
  238102\relax
\mciteBstWouldAddEndPuncttrue
\mciteSetBstMidEndSepPunct{\mcitedefaultmidpunct}
{\mcitedefaultendpunct}{\mcitedefaultseppunct}\relax
\EndOfBibitem
\bibitem[M{\"{u}}ller \emph{et~al.}(2015)M{\"{u}}ller, Meier, and
  Wall]{Mueller2015interpolatedcrosslinks}
K.~W. M{\"{u}}ller, C.~Meier and W.~A. Wall, \emph{Journal of Computational
  Physics}, 2015, \textbf{303}, 185--202\relax
\mciteBstWouldAddEndPuncttrue
\mciteSetBstMidEndSepPunct{\mcitedefaultmidpunct}
{\mcitedefaultendpunct}{\mcitedefaultseppunct}\relax
\EndOfBibitem
\bibitem[Maier \emph{et~al.}(2015)Maier, M{\"{u}}ller, Heussinger,
  K{\"{o}}hler, Wall, Bausch, and Lieleg]{Maier2015}
M.~Maier, K.~W. M{\"{u}}ller, C.~Heussinger, S.~K{\"{o}}hler, W.~A. Wall, A.~R.
  Bausch and O.~Lieleg, \emph{European Physical Journal E}, 2015, \textbf{38},
  1--7\relax
\mciteBstWouldAddEndPuncttrue
\mciteSetBstMidEndSepPunct{\mcitedefaultmidpunct}
{\mcitedefaultendpunct}{\mcitedefaultseppunct}\relax
\EndOfBibitem
\bibitem[Kachan \emph{et~al.}(2016)Kachan, M{\"{u}}ller, Wall, and
  Levine]{Kachan2016bundlingcasimir}
D.~Kachan, K.~W. M{\"{u}}ller, W.~A. Wall and A.~J. Levine, \emph{Physical
  Review E}, 2016, \textbf{94}, 032505\relax
\mciteBstWouldAddEndPuncttrue
\mciteSetBstMidEndSepPunct{\mcitedefaultmidpunct}
{\mcitedefaultendpunct}{\mcitedefaultseppunct}\relax
\EndOfBibitem
\bibitem[Slepukhin \emph{et~al.}(2019)Slepukhin, Grill, M{\"{u}}ller, Wall, and
  Levine]{Slepukhin2019}
V.~M. Slepukhin, M.~J. Grill, K.~W. M{\"{u}}ller, W.~A. Wall and A.~J. Levine,
  \emph{Physical Review E}, 2019, \textbf{99}, 042501\relax
\mciteBstWouldAddEndPuncttrue
\mciteSetBstMidEndSepPunct{\mcitedefaultmidpunct}
{\mcitedefaultendpunct}{\mcitedefaultseppunct}\relax
\EndOfBibitem
\bibitem[Meier \emph{et~al.}(2018)Meier, Grill, Wall, and Popp]{Meier2017b}
C.~Meier, M.~J. Grill, W.~A. Wall and A.~Popp, \emph{International Journal of
  Solids and Structures}, 2018, \textbf{154}, 124--146\relax
\mciteBstWouldAddEndPuncttrue
\mciteSetBstMidEndSepPunct{\mcitedefaultmidpunct}
{\mcitedefaultendpunct}{\mcitedefaultseppunct}\relax
\EndOfBibitem
\bibitem[{Institute for Computational Mechanics (Technical University of
  Munich)}(2020)]{BACI2020}
{Institute for Computational Mechanics (Technical University of Munich)},
  \emph{{BACI: A multiphysics simulation environment}}, 2020\relax
\mciteBstWouldAddEndPuncttrue
\mciteSetBstMidEndSepPunct{\mcitedefaultmidpunct}
{\mcitedefaultendpunct}{\mcitedefaultseppunct}\relax
\EndOfBibitem
\bibitem[Landau \emph{et~al.}(1986)Landau, Lifshitz, Kosevich, Sykes,
  Pitaevskii, and Reid]{Landau1986}
L.~Landau, E.~Lifshitz, A.~Kosevich, J.~Sykes, L.~Pitaevskii and W.~Reid,
  \emph{Theory of Elasticity: Volume 7}, Elsevier Science, 1986\relax
\mciteBstWouldAddEndPuncttrue
\mciteSetBstMidEndSepPunct{\mcitedefaultmidpunct}
{\mcitedefaultendpunct}{\mcitedefaultseppunct}\relax
\EndOfBibitem
\bibitem[Head \emph{et~al.}(2003)Head, Levine, and
  MacKintosh]{head2003distinct}
D.~Head, A.~Levine and F.~MacKintosh, \emph{Physical Review E}, 2003,
  \textbf{68}, 061907\relax
\mciteBstWouldAddEndPuncttrue
\mciteSetBstMidEndSepPunct{\mcitedefaultmidpunct}
{\mcitedefaultendpunct}{\mcitedefaultseppunct}\relax
\EndOfBibitem
\bibitem[Head \emph{et~al.}(2003)Head, Levine, and
  MacKintosh]{head2003deformation}
D.~A. Head, A.~J. Levine and F.~MacKintosh, \emph{Physical review letters},
  2003, \textbf{91}, 108102\relax
\mciteBstWouldAddEndPuncttrue
\mciteSetBstMidEndSepPunct{\mcitedefaultmidpunct}
{\mcitedefaultendpunct}{\mcitedefaultseppunct}\relax
\EndOfBibitem
\bibitem[Coppersmith \emph{et~al.}(1996)Coppersmith, Liu, Majumdar, Narayan,
  and Witten]{coppersmith1996model}
S.~Coppersmith, C.-h. Liu, S.~Majumdar, O.~Narayan and T.~Witten,
  \emph{Physical Review E}, 1996, \textbf{53}, 4673\relax
\mciteBstWouldAddEndPuncttrue
\mciteSetBstMidEndSepPunct{\mcitedefaultmidpunct}
{\mcitedefaultendpunct}{\mcitedefaultseppunct}\relax
\EndOfBibitem
\bibitem[Souslov \emph{et~al.}(2009)Souslov, Liu, and Lubensky]{Souslov2009}
A.~Souslov, A.~J. Liu and T.~C. Lubensky, \emph{Phys. Rev. Lett.}, 2009,
  \textbf{103}, 205503\relax
\mciteBstWouldAddEndPuncttrue
\mciteSetBstMidEndSepPunct{\mcitedefaultmidpunct}
{\mcitedefaultendpunct}{\mcitedefaultseppunct}\relax
\EndOfBibitem
\bibitem[Mao \emph{et~al.}(2010)Mao, Xu, and Lubensky]{Mao2010}
X.~Mao, N.~Xu and T.~C. Lubensky, \emph{Phys. Rev. Lett.}, 2010, \textbf{104},
  085504\relax
\mciteBstWouldAddEndPuncttrue
\mciteSetBstMidEndSepPunct{\mcitedefaultmidpunct}
{\mcitedefaultendpunct}{\mcitedefaultseppunct}\relax
\EndOfBibitem
\bibitem[Kane and Lubensky(2014)]{kane2014topological}
C.~Kane and T.~Lubensky, \emph{Nature Physics}, 2014, \textbf{10}, 39--45\relax
\mciteBstWouldAddEndPuncttrue
\mciteSetBstMidEndSepPunct{\mcitedefaultmidpunct}
{\mcitedefaultendpunct}{\mcitedefaultseppunct}\relax
\EndOfBibitem
\bibitem[MacKintosh \emph{et~al.}(1995)MacKintosh, K{\"a}s, and
  Janmey]{mackintosh1995}
F.~C. MacKintosh, J.~K{\"a}s and P.~A. Janmey, \emph{Physical review letters},
  1995, \textbf{75}, 4425\relax
\mciteBstWouldAddEndPuncttrue
\mciteSetBstMidEndSepPunct{\mcitedefaultmidpunct}
{\mcitedefaultendpunct}{\mcitedefaultseppunct}\relax
\EndOfBibitem
\bibitem[Mao \emph{et~al.}(2013)Mao, Stenull, and Lubensky]{mao2013effective}
X.~Mao, O.~Stenull and T.~C. Lubensky, \emph{Physical Review E}, 2013,
  \textbf{87}, 042601\relax
\mciteBstWouldAddEndPuncttrue
\mciteSetBstMidEndSepPunct{\mcitedefaultmidpunct}
{\mcitedefaultendpunct}{\mcitedefaultseppunct}\relax
\EndOfBibitem
\bibitem[Mao \emph{et~al.}(2013)Mao, Stenull, and Lubensky]{mao2013elasticity}
X.~Mao, O.~Stenull and T.~C. Lubensky, \emph{Physical Review E}, 2013,
  \textbf{87}, 042602\relax
\mciteBstWouldAddEndPuncttrue
\mciteSetBstMidEndSepPunct{\mcitedefaultmidpunct}
{\mcitedefaultendpunct}{\mcitedefaultseppunct}\relax
\EndOfBibitem
\bibitem[Heussinger and Frey(2006)]{Heussinger2006}
C.~Heussinger and E.~Frey, \emph{Phys. Rev. Lett.}, 2006, \textbf{97},
  105501\relax
\mciteBstWouldAddEndPuncttrue
\mciteSetBstMidEndSepPunct{\mcitedefaultmidpunct}
{\mcitedefaultendpunct}{\mcitedefaultseppunct}\relax
\EndOfBibitem
\bibitem[Zhou \emph{et~al.}(2018)Zhou, Zhang, and Mao]{zhou2018}
D.~Zhou, L.~Zhang and X.~Mao, \emph{Physical review letters}, 2018,
  \textbf{120}, 068003\relax
\mciteBstWouldAddEndPuncttrue
\mciteSetBstMidEndSepPunct{\mcitedefaultmidpunct}
{\mcitedefaultendpunct}{\mcitedefaultseppunct}\relax
\EndOfBibitem
\bibitem[Wilhelm and Frey(2003)]{Frey:03}
J.~Wilhelm and E.~Frey, \emph{Phys. Rev. Lett.}, 2003, \textbf{91},
  108103\relax
\mciteBstWouldAddEndPuncttrue
\mciteSetBstMidEndSepPunct{\mcitedefaultmidpunct}
{\mcitedefaultendpunct}{\mcitedefaultseppunct}\relax
\EndOfBibitem
\bibitem[Gradshteyn and Ryzhik(2014)]{gradshteyn2014table}
I.~S. Gradshteyn and I.~M. Ryzhik, \emph{Table of integrals, series, and
  products}, Academic press, 2014\relax
\mciteBstWouldAddEndPuncttrue
\mciteSetBstMidEndSepPunct{\mcitedefaultmidpunct}
{\mcitedefaultendpunct}{\mcitedefaultseppunct}\relax
\EndOfBibitem
\bibitem[M{\'e}zard \emph{et~al.}(1987)M{\'e}zard, Parisi, and
  Virasoro]{mezard1987spin}
M.~M{\'e}zard, G.~Parisi and M.~Virasoro, \emph{Spin glass theory and beyond:
  An Introduction to the Replica Method and Its Applications}, World Scientific
  Publishing Company, 1987, vol.~9\relax
\mciteBstWouldAddEndPuncttrue
\mciteSetBstMidEndSepPunct{\mcitedefaultmidpunct}
{\mcitedefaultendpunct}{\mcitedefaultseppunct}\relax
\EndOfBibitem
\bibitem[Kardar(2007)]{kardar2007statistical}
M.~Kardar, \emph{Statistical physics of fields}, Cambridge University Press,
  2007\relax
\mciteBstWouldAddEndPuncttrue
\mciteSetBstMidEndSepPunct{\mcitedefaultmidpunct}
{\mcitedefaultendpunct}{\mcitedefaultseppunct}\relax
\EndOfBibitem
\end{mcitethebibliography}
\bibliographystyle{rsc} 
\end{document}